\documentclass{aa}
\usepackage{amsmath}
\usepackage{graphicx}
\usepackage{txfonts}

\newcommand{\gtsimeq}{\raisebox{-0.6ex}{$\,\stackrel
        {\raisebox{-.2ex}{$\textstyle >$}}{\sim}\,$}}
\defcitealias{BB2014}{BB2014}
\defcitealias{BBC2015}{BBC2015}

\begin{document}

\title{Ionisation in Turbulent Magnetic Molecular Clouds}
\subtitle{I. Effect on Density and Mass-to-Flux Ratio Structures }

\author{Nicole D. Bailey\inst{1}, Shantanu Basu\inst{2}, \and Paola Caselli\inst{1}}
\institute{Max-Planck-Institut f$\ddot{\rm u}$r extraterrestrische Physik, Giessenbachstraße, D-85748 Garching, Germany\\
\email{ndbailey@mpe.mpg.de (NDB), caselli@mpe.mpg.de (PC)}
\and
Department of Physics and Astronomy, University of Western Ontario, 1151 Richmond Street, London, Ontario, Canada, N6A 3K7 \\
\email{basu@uwo.ca}
}

\date{Received - February 8, 2016 ; accepted - December 30, 2016}

\abstract
{Previous studies show that the physical structures and kinematics of a region depend significantly on the ionisation fraction. These studies have only considered these effects in non-ideal magnetohydrodynamic simulations with microturbulence. The next logical step is to explore the effects of turbulence on ionised magnetic molecular clouds and then compare model predictions with observations to assess the importance of turbulence in the dynamical evolution of molecular clouds.}
{In this paper, we extend our previous studies of the effect of ionisation fractions on star formation to clouds that include both non-ideal magnetohydrodynamics and turbulence. We aim to quantify the importance of a treatment of the ionisation fraction in turbulent magnetised media and investigate the effect of the turbulence on shaping the clouds and filaments before star formation sets in. In particular, here we investigate how the structure, mass and width of filamentary structures depend on the amount of turbulence in ionised media and the initial mass-to-flux ratio.}
{To determine the effects of turbulence and mass-to-flux ratio on the evolution of non-ideal magnetised clouds with varying ionisation profiles, we have run two sets of simulations. The first set assumes different initial turbulent Mach values for a fixed initial mass-to-flux ratio. The second set assumes different initial mass-to-flux ratio values for a fixed initial turbulent Mach number. Both sets explore the effect of using one of two ionisation profiles: Step-Like (SL) or Cosmic Ray only (CR-only). We compare the resulting density and mass-to-flux ratio structures both qualitatively and quantitatively via filament and core masses and filament fitting techniques (Gaussian and Plummer profiles.)}
{We find that even with almost no turbulence, filamentary structure still exists although at lower density contours. Comparison of simulations show that for turbulent Mach numbers above 2, there is little structural difference between the SL and CR-only models, while below this threshold the ionisation structure significantly affects the formation of filaments. This holds true for both sets of models. Analysis of the mass within cores and filaments show that the mass decreases as the degree of turbulence increases. Finally, observed filaments within the Taurus L1495/B213 complex are best reproduced by models with supercritical mass-to-flux ratios and/or at least mildly supersonic turbulence, however, our models show that the sterile fibres observed within Taurus may occur in highly ionised,subcritical environments.
}
{From the analysis of the simulations, we conclude that in the presence of low turbulent velocities, the ionisation structure of the medium still plays a role in shaping the structure of the cloud, however, above Mach 2, the differences between the two profiles become indistinguishable. However, differences may be present in the underlying velocity structure. Kinematics studies will be the focus of the next paper in this series. Regions with fertile fibres likely indicate a trans- or supercritical mass-to-flux ratio within the region while sterile fibres are likely subcritical and transient. 
}


\maketitle

\section{Introduction}

Observations of star forming regions from telescopes such as Herschel \citep{Andre2010, Andre2014, Arz2011, Motte2010}, and IRAM \citep{Hacar2013, Henshaw2014, Peretto2014, TH2015, Beuther201509, Beuther201512}, among others \citep[e.g.,][]{Kirk2013,Li2015}, have shown the ubiquity of filaments in star forming regions, however the conditions for the formation of these structures are not yet well known. Observations and theory \citep[][among others]{FP2000, Padoan2001, Alves2008, NL2008, Heitsch2013, Arz2013, Hacar2013, Henshaw2014, Tomisaka2014, Kirk2015, SW2015, MB2015, Li2015, FA2015} indicate that turbulence and magnetic fields  both have a role in the formation of these structures, however to what degree and in what form (e.g., driven or decaying turbulence) is still up for debate. In addition, the degree of ionisation within the molecular cloud needs to be taken into account. In the past couple of decades, observations of star forming regions have revealed a plethora of neutral and ionic molecules (e.g., CO, NH$_{3}$, N$_{2}$H$^{+}$, etc.) that exist at and help probe various density thresholds. In a magnetised medium, the ions are bound to the field lines, however, depending on the radiation field and the ionisation fraction, neutral particles may be able to slip past the field lines to create density enhancements which may collapse into clumps and filaments. The ability for neutral particles to slip past the field lines is dependent on the degree of ambipolar diffusion within a region, however, the majority of simulations only consider a flux frozen medium (i.e., a medium where the neutral particles are coupled to the magnetic field lines due to frequent collisions with ions). Ambipolar diffusion of the neutral fluid relative to the ions is a necessary ingredient to
understand both core formation and turbulence in partially ionised molecular clouds. 
For example, the preferred core fragmentation scale is a sensitive function of
ionisation fraction and mass-to-flux ratio \citep{CB2006, BB2012}. \citet{BB2013} showed how this can lead to a much broader core mass function than pure hydrodynamic fragmentation. Furthermore, turbulent flows in  a strongly magnetised medium can induce rapid ambipolar diffusion in compressed regions, leading to localised star formation with low overall efficiency 
\citep[e.g.,][]{NL2005, BCDW2009}. Here we extend such models using 
planar models of molecular clouds threaded by a large scale magnetic field,
including turbulence, ambipolar diffusion, and a step-like ionisation profile
that reflects ionisation by background ultraviolet starlight in the low column density
regions and cosmic ray induced ionisation in higher column density regions. These models have a background magnetic field that is perpendicular to the plane of the sheet, consistent with the presence of a dynamically important magnetic field, self-gravity, and turbulence that is Alfv\'enic or sub-Alf\'enic. These conditions allow settling of matter along the field lines and the formation of a flattened layer before further evolution occurs. Here we do not explore the opposite limit of an initially weak magnetic field that can be swept up in a sheet by turbulence or expanding supernova shells and be aligned parallel to a sheet. For example \citet{Nagai1998} performed a linear stability analysis of such a sheet and found that it can form elongated structures either perpendicular or parallel to the magnetic field direction in the cases of low or high ratio of external pressure to self-gravitational pressure respectively.  Previous three dimensional simulations with dynamically important magnetic fields  \citep{KB2008, KB2011} have shown similar results as the corresponding thin sheet results of \citet{NL2005} and \citet{BCDW2009}. The thin sheet models are actually more useful in studying initially near-critical mass-to-flux ratio clouds, since the critical mass-to-flux ratio is known analytically in a thin sheet geometry \citep{CB2006} but not identifiable analytically for three-dimensional models with finite extent along the magnetic field direction. Some of the outstanding questions in this field are the values of the initial mass-to-flux ratio in molecular clouds and the level of initial turbulence and whether it is freely decaying or continues to be driven.

With the above issues in mind, our aim is to self consistently explore all of these effects. In a previous set of papers, we looked at the effects of the ionisation profile on the structure and kinematics of core collapse within non-ideal magnetic molecular clouds \citep[][hereafter BB2014 and BBC2015, respectively]{BB2014,BBC2015}. To further this analysis, we include the effects of turbulence in our simulations in order to see how/if the density, mass-to-flux ratio structures, and kinematics can change. We assess if the ionisation structure is playing a role in the dynamical evolution and structure of a cloud just before star formation. To do so, we explore the effects of changing the turbulent Mach number or initial mass-to-flux ratio in two separate sets of models (Set 1 and Set 2, respectively). In this paper we focus on the density and mass-to-flux ratio structures found within the two sets of simulations while a second paper will look at how the addition of turbulence affects the kinematics when different ionisation profiles are considered. The rest of this paper will be structured as follows. Section 2 describes the numerical code used to produce the simulations. Section 3 describes the model parameters investigated as well as the results of the simulations. Section 4 looks at the physical properties of filaments and cores found within each model and compares this analysis to other studies. Finally, in Sections 5 and 6 we present our discussion and summary, respectively.

\section{Numerical Code}
For our simulations, we use the \citet{BC2004} IDL code to model a weakly ionised, magnetic, self-gravitating, isothermal molecular cloud. Specifically, we are using the modified version \citep{BB2014} that includes density dependent ionisation profiles which regulate the degree of ambipolar diffusion within the molecular cloud. We explore the resulting kinematics, density, mass-to-flux ratio and velocity structures of simulated clump-core complexes within partially ionised, isothermal magnetic turbulent interstellar molecular clouds. The model assumes planar clouds with periodic boundary conditions in the $x$- and $y$- directions and a local vertical half thickness Z. This so-called ``thin-sheet'' approximation allows us to explore a significant parameter space while minimising computational resources and has been shown to provide results that are similar to three-dimensional non-ideal MHD simulations \citep{KB2011,NL2005,BCDW2009}.Three-dimensional effects of e.g., turbulent support along the magnetic field direction \citep{kud03} and ionisation fraction variation along the vertical direction are not accounted for in this thin sheet model. A full description of the assumptions, non-axisymmetric equations and formulations and possible short-comings of this model can be found in \citet{BC2004, CB2006, BCDW2009, BCW2009}, however we highlight those essential for the analysis within this paper. 

\subsection{Magnetic Fields}
The model assumes a magnetic field that threads the cloud perpendicular to the $xy$ plane and includes the effects of ambipolar diffusion. The timescale for collisions between neutral particles and ions is 
\begin{equation} 
\tau_{ni} = 1.4 \left(\frac{m_i +m_{H_2}}{m_i} \right) \frac{1}{n_i\langle\sigma w \rangle_{iH_2}}, 
\label{tni}
\end{equation} 
where $m_{i}$ is the ion mass, $m_{H_2}$ is the mass of molecular hydrogen, $n_{i}$ is the number density of ions, and $\langle\sigma w\rangle_{iH_2}$ is the neutral-ion collision rate. Typical ions within a molecular cloud include singly ionised Na, Mg, and HCO which have an average mass of 25 amu. Assuming collisions between H$_{2}$ and HCO$^+$, the neutral-ion collision rate is $1.69\times 10^{-9}$ cm$^{-3}$ s$^{-1}$ \citep{MM1973}. The factor of 1.4 in Equation~\ref{tni} accounts for the fact that the inertia of helium is neglected in the calculation of the slowing-down time of neutrals by collisions with ions \citep{MC1999, CB2006}. 

Collapse within a molecular cloud is regulated by the normalised mass-to-flux ratio of the background reference state,
\begin{equation}
\mu_{0} \equiv 2\pi G^{1/2}\frac{\sigma_{n,0}}{B_{\rm ref}},
\end{equation}
where $(2\pi G^{1/2})^{-1}$ is the critical mass-to-flux ratio for gravitational collapse in the adopted model \citep{CB2006}, $\sigma_{n,0}$ is the initial column density and $B_{\rm ref}$ is the magnetic field strength of the background reference state. Regions with mass-to-flux ratios greater than, approximately equal to, and less then one are known to be supercritical, transcritical and subcritical, respectively. In the limit where $\tau_{ni} \rightarrow 0$, the medium is defined to be flux-frozen, that is, frequent collisions between the neutral particles and ions couple the neutrals to the magnetic field. Under these conditions, subcritical regions are supported by the magnetic field and only supercritical regions may collapse within a finite time frame. Non-zero values of $\tau_{ni}$ are inversely dependent on the ion number density and therefore on the degree of ionisation for a fixed neutral density.

\subsection{Turbulence}
\label{turb}
Turbulence within our simulations is incorporated via an initial ``turbulent'' velocity field added to the background reference state of uniform column density and vertical magnetic field strength $B_{\rm ref}$. The velocity field is generated in Fourier space using the same method described in \citet{BCDW2009}. For a physical grid containing $N$ zones in both the $x-$  and $y-$ directions, there exists a corresponding wavenumber in Fourier space, $k_{x}$ and $k_{y}$, respectively. For each ($k_{x}$,$k_{y}$) pair, a Fourier velocity amplitude is assigned from a Gaussian distribution and scaled by the power spectrum $v_{k}^{2} \propto k^{n}$, where $k = (k_{x}^{2}+k_{y}^{2})^{1/2}$. The resulting Fourier velocity field is then transformed back into physical space. The distributions of $v_{x}$ and $v_{y}$ are chosen independently in this manner and each rescaled such that the rms amplitude is equal to the chosen turbulent velocity amplitude $v_{a}$. The turbulent velocity field is only added at the beginning of each simulation and allowed to decay. For our simulations, we test the effect of varying both the turbulent velocity amplitude and the initial mass-to-flux ratio within the context of varying ionisations profiles.

To enable comparison between models, we construct the Gaussian distributions for each model using the same realisation (i.e., the same random number generator seed). However, within this realisation, we ensure that the $x$ and $y$ velocity fields are generated using two distinct seed values. By maintaining the same realisation, we are able to distinguish which physical changes are due to changes in physical parameters as opposed to the random nature of the initial velocity distribution. The exact choice of random seed will affect the run time of the simulation, therefore quoted simulation times within this paper should be taken as representative but not exact values.

\subsection{Ionisation Profiles}

The modified version of the IDL code includes density dependent ionisation profiles. For our analysis, we utilise two different ionisation profiles which describe the ionisation fraction ($\chi_{i}$) as a function of density. These profiles set the neutral-ion collision time within each pixel since $\tau_{ni} \propto 1/\chi_{i}$. The first profile used is a step-like ionisation profile that includes both the ultraviolet (UV) and cosmic ray (CR) regimes. The form of this step-like ionisation function is given by \begin{equation} 
\log \chi_{i} = \left\{ \begin{array}{ll}
\log \chi_{i,0} + 0.5(\log \chi_{i,c} - \log \chi_{i,0})\times& \\
~~~~~~\left(1 + \tanh\frac{A_{V}-A_{V, \rm crit}}{A_{V,d}}\right) & \mbox{$A_{V}\le A_{V,CR}$} \\
\log[ 1.148\times10^{-7}(1 +\tilde{P}_{\rm ext})^{-1/2}~\times  &\\
~~~~~~\left(\frac{T}{10~\rm K}\right)^{1/2}\left(\frac{2.75~ \rm mag}{A_{V}}\right)] & \mbox{$A_{V} > A_{V,CR}$ },
\end{array}
\right.
\label{eqn:ionmodel}
\end{equation}
where $\tilde{P}_{\rm ext}$ is the dimensionless external pressure and $A_{V,CR} = 6.365$ mag is the location of the transition from the UV regime to the CR regime. The step function parameters are set to $\log \chi_{i,0} = -4.0$, $\log \chi_{i,c} = -7.362$, $A_{V,\rm crit} = 4.0$ mag, and $A_{V,d} = 1.05$ mag. This ionisation profile is based upon those presented by \citet{Ruffle1998}, where we have chosen our high and low ionisation levels ($\log \chi_{i,0}$ and $\log \chi_{i,c}$, respectively) to depict a rough average of the different profiles presented. The location of the step was chosen to correspond with the typical $A_{V}$ where dust shielding sets in, thus eliminating the effects of the UV radiation. Naturally, other step-like profiles can be constructed in this way, however we have chosen to focus on this one to facilitate comparison with our previous analysis and simulations \citep[see][]{BB2012, BB2014, BBC2015}. With the inclusion of this ionisation profile, the ionisation fraction regime (and neutral-ion collision time) within the cloud depend on the column density/visual extinction of each region and therefore changes as the cloud evolves. The derived neutral-ion collision time for each pixel then depends on which ionisation regime it is in. For the UV regime, the neutral-ion collision time is computed via 
\begin{eqnarray}
\nonumber\tau_{ni}/t_{0} = &0.262& \left(\frac{T}{10~\rm K}\right)^{1/2}\left(\frac{0.01~\rm g~cm^{-2}}{\sigma_{n,0}}\right)\times \\
&& \left(\frac{10^{-7}}{\chi_{i}}\right)(1+\tilde{P}_{\rm ext})^{-1},
\label{eqn:tauni}
\end{eqnarray}
where $t_{0}$ is the dimensionless time scale ($t_{0} = c_{s}/2\pi G \sigma_{n,0}$).
For the CR regime, we set the dimensionless neutral-ion collision time to $\tau_{ni,0}/t_{0} = 0.2$. 

 The second ionisation profile corresponds to a cosmic ray (CR) only profile which is achieved by  assuming a dimensionless neutral-ion collision time of $\tau_{ni,0}/t_{0} = 0.2$ for all densities.

\subsection{Dimensionless and Dimensional Values}
Our model is characterised by several dimensionless free parameters including a dimensionless form of the initial neutral-ion collision time ($\tau_{ni,0}/t_{0}~\equiv~2\pi G\sigma_{n,0}\tau_{ni,0}/c_{s}$) and a dimensionless external pressure ($\tilde{P}_{\rm ext} \equiv 2 P_{\rm ext}/\pi G \sigma^{2}_{n,0}$).  Here, $c_{s} = (k_{B} T/m_{n})^{1/2}$ is the isothermal sound speed; $k_{B}$ is the Boltzmann constant, $T$ is the temperature in Kelvin, and $m_{n}$ is the mean mass of a neutral particle ($m_{n} = 2.33$ amu). We normalise column densities by $\sigma_{n,0}$, length scales by $L_{0} = c_{s}^{2}/2\pi G \sigma_{n,0}$ and time scales by $t_{0} = c_{s}/2\pi G \sigma_{n,0}$. Based on these parameters, typical values of the units used and other derived quantities are 

\begin{eqnarray}
\nonumber\sigma_{n,0} &=& \frac{3.63\times 10^{-3}}{(1+\tilde{P}_{\rm ext})^{1/2}}\left(\frac{n_{n,0}}{10^3 \rm ~cm^{-3}}\right)^{1/2}\left(\frac{T}{10 ~\rm K}\right)^{1/2} \rm g~cm^{-2},\\ 
&&\\
c_{s} &=& 0.188\left(\frac{T}{10 ~\rm K}\right)^{1/2} \rm km~s^{-1},\\
t_{0} &=& 3.98\times 10^5\left(\frac{10^3 \rm~ cm^{-3}}{n_{n,0}}\right)^{1/2}(1 + \tilde{P}_{\rm ext})^{1/2}~\rm yr,\label{time}\\ 
\nonumber L_{0} &=& 7.48\times 10^{-2} \left(\frac{T}{10 ~\rm K}\right)^{1/2}\times\\
&&\left(\frac{10^3 \rm ~cm^{-3}}{n_{n,0}}\right)^{1/2}(1 + \tilde{P}_{\rm ext})^{1/2}~\rm pc,\label{length}
\end{eqnarray}
where $n_{n,0}$ is the initial neutral number density. For our analysis, we assume a dimensionless external pressure $\tilde{P}_{\rm ext} = 0.1$ ($P_{\rm ext} \approx 10^{3}~\rm cm^{-3}~K$), a neutral number density of 1.1 $\times 10^{3}~\rm cm^{-3}$, and a temperature $T = 10$ K.

\section{Simulations}

\subsection{Model Parameters}

\label{model}

The models and analyses presented in this paper form an extension to the analyses presented in \citetalias{BB2014} and \citetalias{BBC2015} to include turbulence. We assume an initially diffuse cloud with an initial background column density which corresponds to a visual extinction $A_{V,0} = 1$ mag. From the prescription of \citet{Pineda2010} \citep[see also][]{BB2012} and assuming a mean molecular weight of 2.33 amu, the resulting conversion between visual extinction and mass column density is 
\begin{equation} 
\sigma_{n} =3.638\times 10^{-3} (A_{V}/\rm mag)~\rm g~cm^{-2}.
\label{av2sigma}
\end{equation}
All simulations begin with an initial turbulent velocity perturbation assuming a power spectrum exponent $n = -4$, such that $v_{k}^{2} \propto k^{-4}$ and are performed on a 512 $\times$ 512 periodic box. The box size is 64$\pi L_{0}$, which translates to a size of 15.16 pc, or a pixel size of 0.0296 pc, for $T$ = 10 K and $\sigma_{n,0} = 3.638\times10^{-3}$\rm~g~cm$^{-2}$. This results in a total mass of 4000 M$_\odot$ within the simulation region. All simulations terminate when any pixel within the region attains a density enhancement that is 10 times the initial column density ($\sigma_{n,0}$). This threshold was chosen as it corresponds to the runaway gravitational collapse of cloud into a dense core. \citet{BCDW2009} verified, with high resolution runs, that collapse does indeed continue past this value. In addition, we chose this stopping conditions because  we are interested in how the physical conditions of the environment affect the ability of a molecular cloud to achieve runaway collapse and not the subsequent evolution of the high density regions formed. For our simulations, we are looking at both the effects of changing the turbulent Mach number and the degree of magnetic support in the context of different ionisation profiles. To achieve this, we present the results of two sets of simulations. The first set (Models I - X) assumes the same initial mass-to-flux value while varying the initial Mach number. For the second set (Models XI - XVI), we have fixed the initial Mach number and vary the initial mass-to-flux ratio.  The initial parameters for both model sets are listed in Table~\ref{models}. Here, $v_{a}/c_{s}$ is the turbulent velocity amplitude scaled by the sound speed (see Section~\ref{turb}), which is equivalent to the Mach number, $t_{r}$ refers to the total run time of the simulation, and $\sigma_{n,max}$, $\mu_{max}$ and $\mu(\sigma_{n,max})$ refer to the maximum density, maximum mass-to-flux ratio and mass-to-flux ratio at the location of the maximum density within the simulation at the final time, respectively. The abbreviations quoted for the ionisation profile indicate whether the simulations assumes a step-like (SL) or cosmic-ray-only (CR) ionisation profile. These two profiles correspond to an initial neutral-ion collision time $\tau_{ni,0}/t_{0} = 0.001$ and $\tau_{ni,0}/t_{0} = 0.2$, respectively. The characteristic free-fall time for our system is 1.01 Myr. From the linear analysis of \citet{BB2012}, the ambipolar diffusion time is equivalent to the fragmentation time scale for $\mu_{0} = 0.5$ and is dependent on the initial ionisation fraction of the medium. This corresponds to values of 1.56 Gyr and 7.88 Myr for the SL and CR-only models, respectively. For the initial mass-to-flux ratio value used in Set 1 ($\mu_{0} = 1.1$), the characteristic times found through linear analysis are 7.2 Myr and 4.1 Myr for the SL and CR-only models, respectively.

The simulations track the evolution of various physical properties of the cloud including the density enhancement, velocity, magnetic field strength, mass-to-flux ratio and the ionisation. 
\begin{table*}
\small
\caption{Simulation Parameters}
\centering
\begin{tabular}{lccccccccc}
\hline\hline
Model & $\mu_{0}$ & $v_{a}/c_{s}$ &  $\chi_{i,0}$ Profile\tablefootmark{a} & $t_{\rm r}/t_{0}$ & $t_{\rm r}$ (Myr) & $\sigma_{n,max}/\sigma_{n,0}$\tablefootmark{b} & $\mu_{max}$\tablefootmark{b} & $\mu(\sigma_{n,max})$\tablefootmark{b}\\
\hline
\multicolumn{9}{c}{Set 1: Variation in Mach Number}\\
\hline
I    &  1.1  &  0.03  &  SL & 81.3 & 31.9 & 10.18 & 1.18 & 1.18\\
II   &  1.1  &  0.03  &  CR & 44.1 & 17.3 & 10.28 & 1.46 & 1.46\\
III  &  1.1  &  1.00  &  SL & 11.3 &  4.4 & 10.46 & 1.25 & 1.20\\
IV   &  1.1  &  1.00  &  CR &  7.7 &  3.0 & 10.24 & 1.35 & 1.33\\
V    &  1.1  &  2.00  &  SL &  3.8 &  1.5 & 10.78 & 1.19 & 1.18\\ 
VI   &  1.1  &  2.00  &  CR &  3.7 &  1.5 & 10.58 & 1.32 & 1.23\\
VII  &  1.1  &  3.00  &  SL &  2.4 & 0.94 & 10.02 & 1.15 & 1.14\\
VIII &  1.1  &  3.00  &  CR &  2.4 & 0.94 & 10.60 & 1.33 & 1.20\\
IX   &  1.1  &  4.00  &  SL &  1.8 & 0.71 & 11.54 & 1.16 & 1.14\\
X    &  1.1  &  4.00  &  CR &  1.7 & 0.67 & 10.12 & 1.30 & 1.20\\ 
\hline
\multicolumn{9}{c}{Set 2: Variation in Initial Mass-to-Flux Ratio}\\
\hline
XI   &  0.5  &  2.00  &  SL & $>$ 156.5 & $>$ 61.4 & 3.08 & 0.50 & 0.50 \\
XII  &  0.5  &  2.00  &  CR & 51.5      & 20.2 & 10.11 & 1.62 & 1.62 \\
XIII &  0.8  &  2.00  &  SL & $>$ 656.6 & $>$ 256.1  & 2.34 & 0.81 & 0.80 \\   
XIV  &  0.8  &  2.00  &  CR & 17.8      & 7.0  & 10.74 & 1.68 & 1.66 \\
XV   &  2.0  &  2.00  &  SL &  2.7      & 1.0  & 10.50 & 2.05 & 2.05 \\
XVI  &  2.0  &  2.00  &  CR &  2.7      & 1.0  & 11.10 & 2.23 & 2.16 \\
\hline\hline
\end{tabular}
\tablefoot{
\tablefoottext{a}{SL and CR refer to the (S)tep-(L)ike and (C)osmic (R)ay only profiles, respectively. }\\
\tablefoottext{b}{These values are taken at the final time of the respective simulation. }
}
\label{models}
\normalsize
\end{table*}
Given the wealth of information available from these simulations, we are splitting the discussion of these parameters into two separate papers. In this paper, we will focus on the structures formed within the simulations by looking at the density enhancements and mass-to-flux ratio maps. The second paper will focus on the kinematic information revealed through the velocity maps and synthetic spectra of the simulations. The following subsections present the effect of varying the set-specific initial conditions on the density enhancement and mass-to-flux ratio maps, paying particular attention to the difference between the two different ionisation profiles.

\subsection{Effect of Varying the Initial Mach Number}

\subsubsection{Density Enhancement Structures}
\label{density}

\begin{figure*}
\centering
\includegraphics[width = 0.3\textwidth]{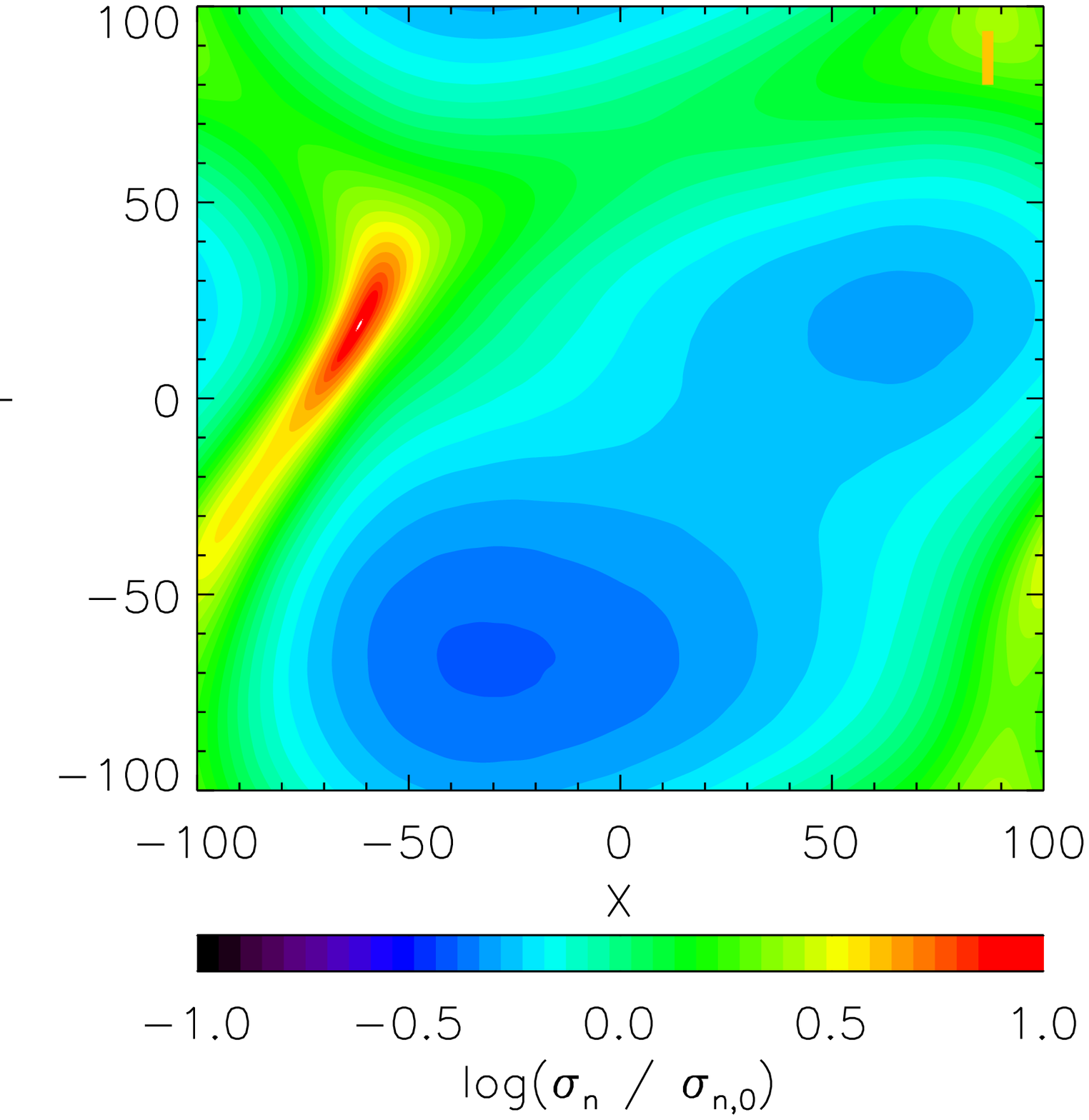} 
\includegraphics[width = 0.3\textwidth]{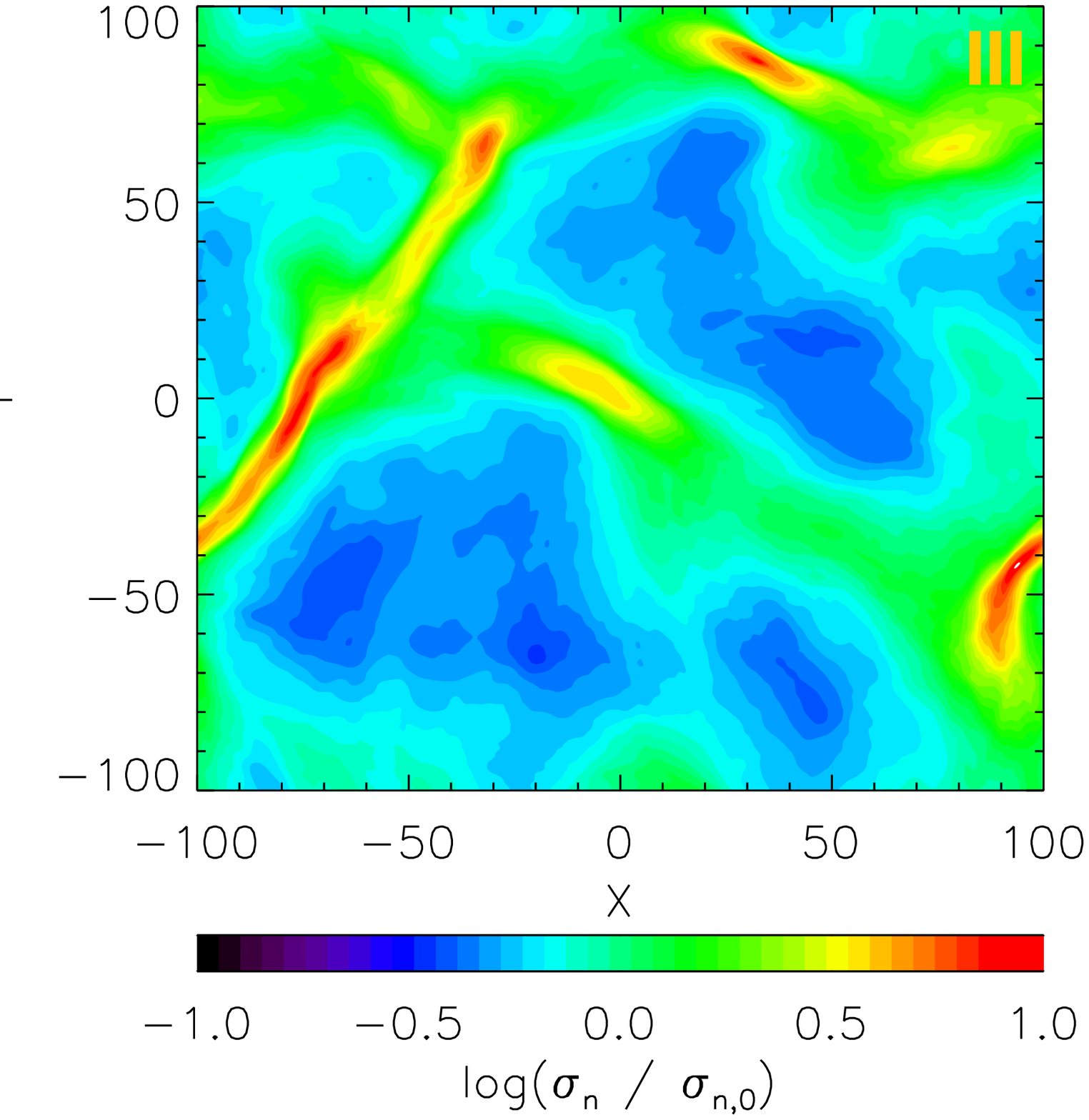} 
\includegraphics[width = 0.3\textwidth]{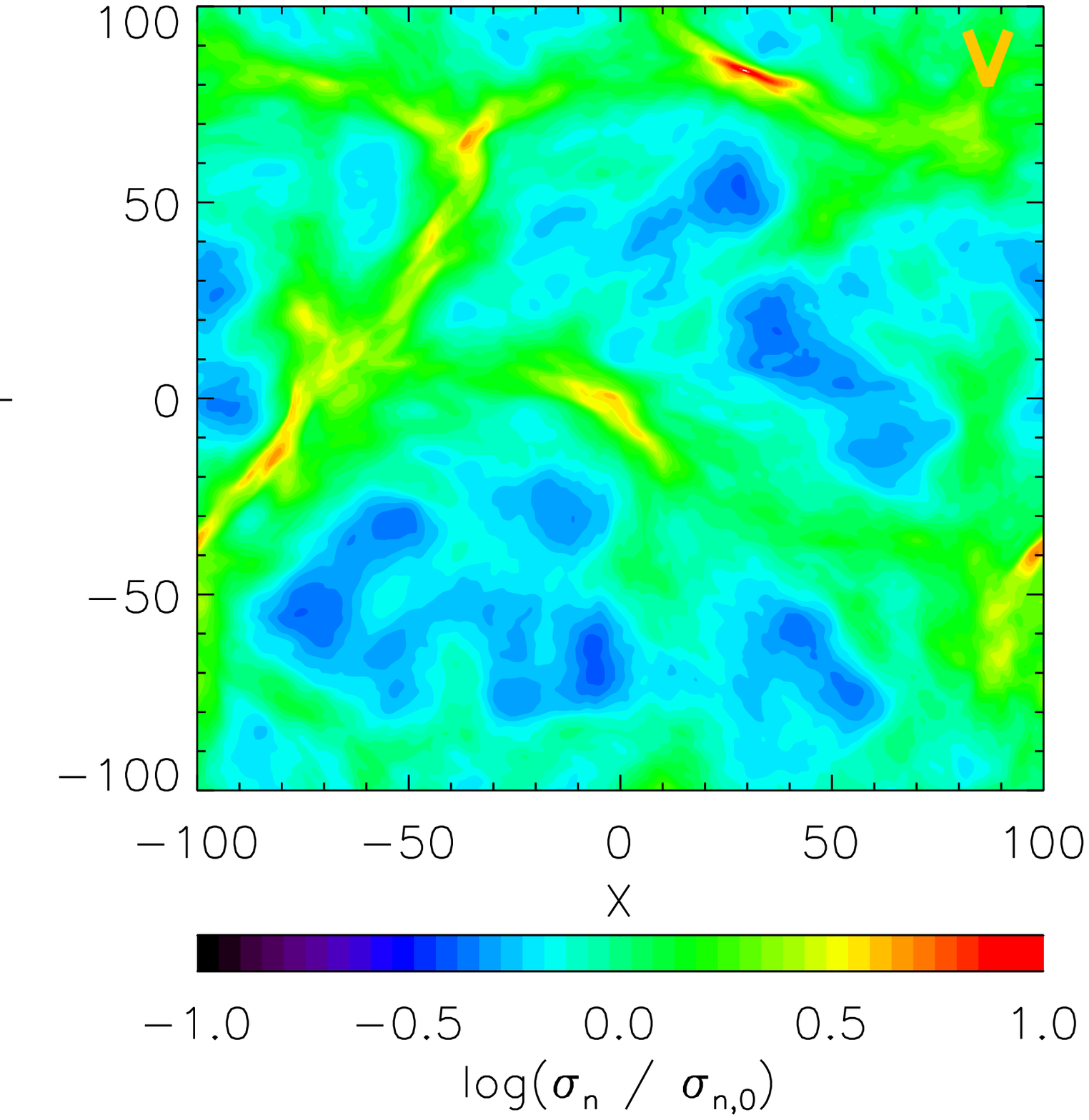}\\ 
\includegraphics[width = 0.3\textwidth]{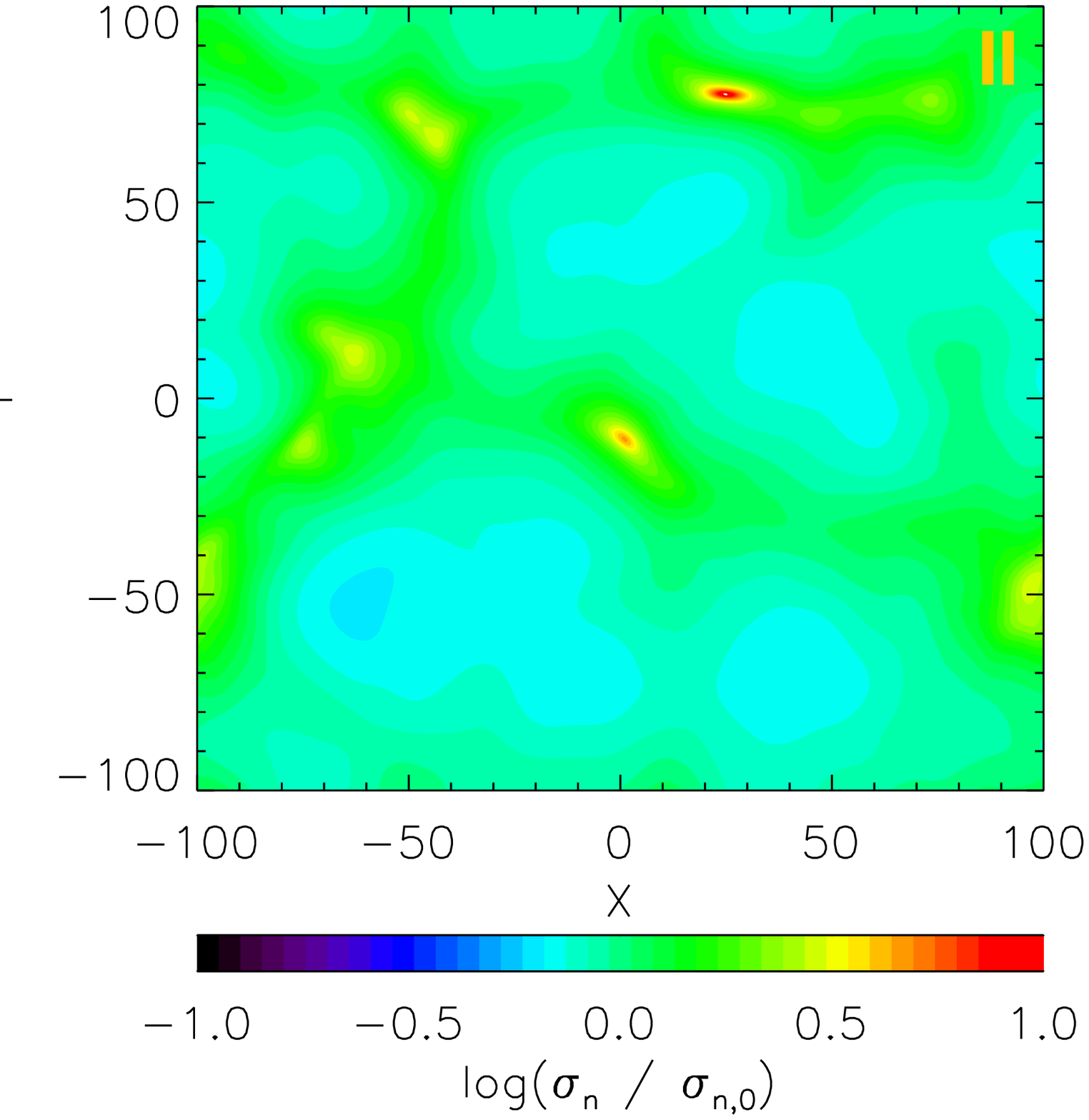} 
\includegraphics[width = 0.3\textwidth]{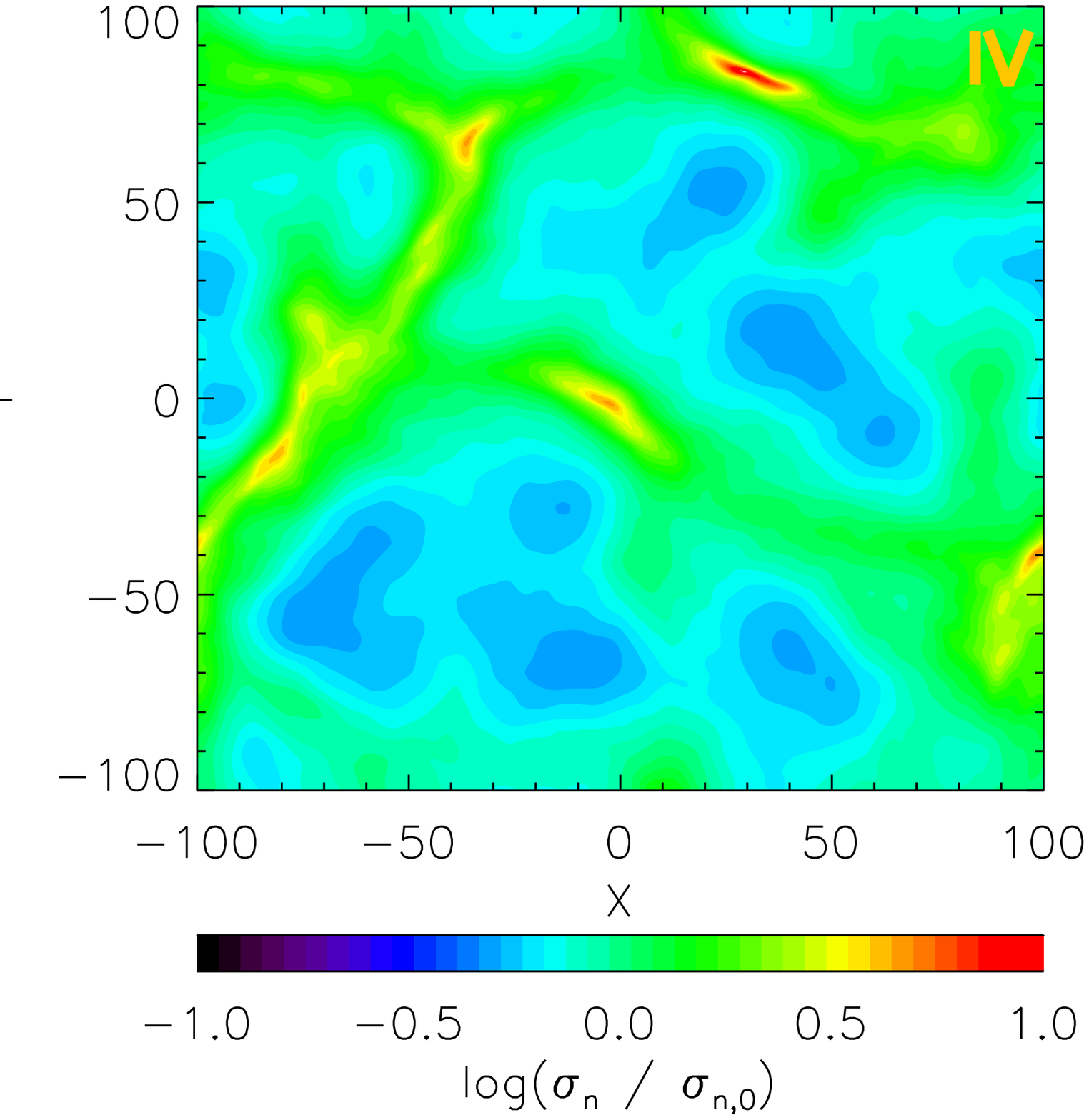} 
\includegraphics[width = 0.3\textwidth]{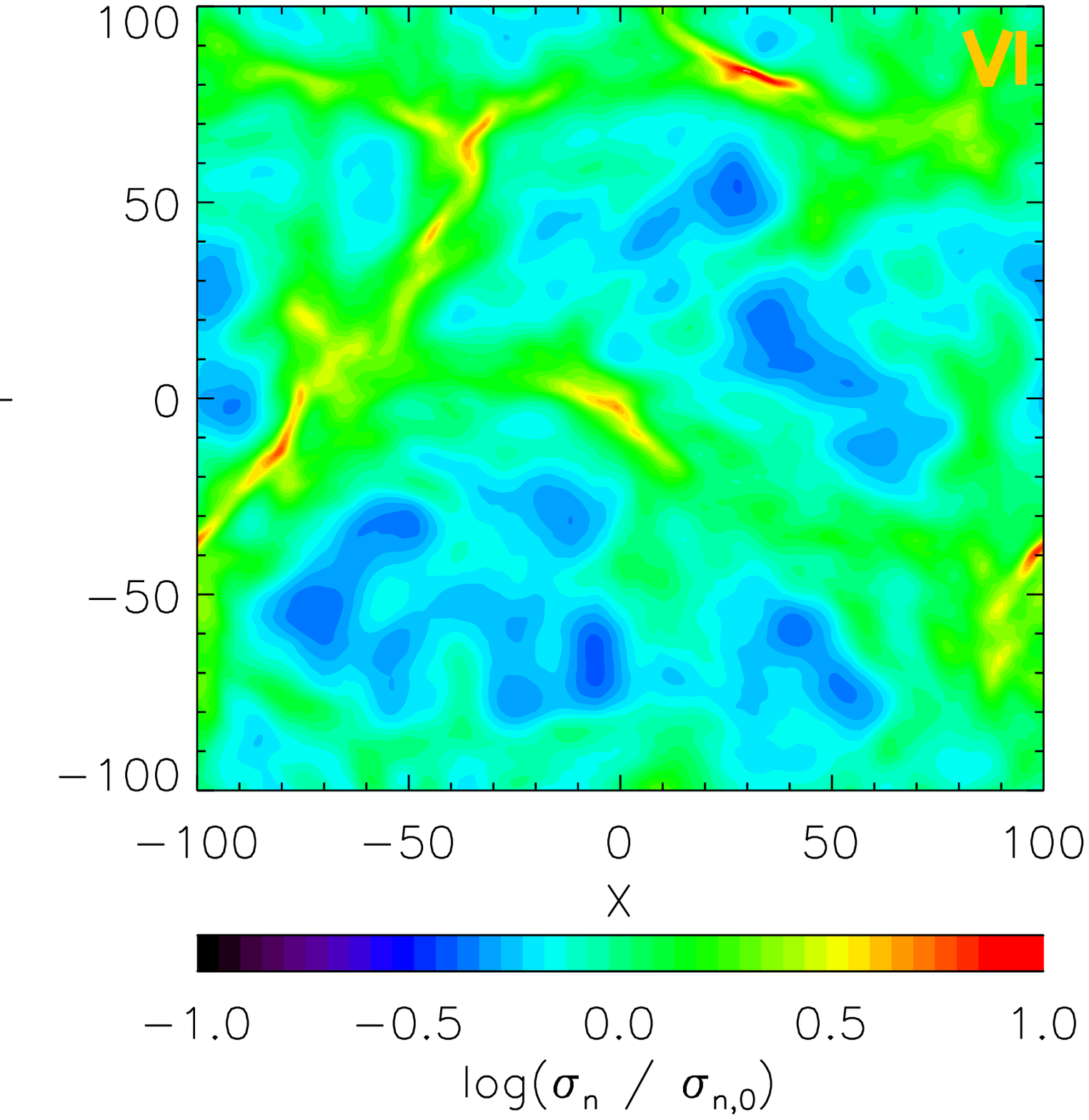} 
\caption{Column density enhancement maps for representative models with Mach number = 0.03 (left panels), 1 (centre panels), and 2 (right panels). Colour bar shows the logarithm of the density enhancement. Roman numerals in the top right of each panel indicate the model number of the simulation (see Table~\ref{models}). Top row: models with the Step-Like ionisation profile. Bottom Row: models with CR-only ionisation profile.  }
\label{sigma}
\end{figure*}

In this section, we look at the effect of changing the turbulent Mach number on the density enhancement structures formed within a collapsing molecular cloud (Models I - X). We are specifically interested in the differences due to the underlying ionisation profile. Figure~\ref{sigma} shows a sample of the density enhancement maps (with respect to the initial density) for three values of the turbulent Mach number (0.03 (left), 1.0 (centre), and 2.0 (right)) for both the step-like (top row) and CR-only (bottom row) ionisation profiles. Models VII - X do not show significant differences compared to models V and VI and therefore are not shown, however, these models do exhibit shorter collapse timescales than the regions with lower turbulence (see Table~\ref{models}). The colour scheme depicts the logarithm of the density enhancement. Looking across the three columns, we see that the assumed ionisation profile results in the largest difference in the simulations with the least amount of turbulence (Models I and II) and the smallest, virtually indistinguishable difference in the simulations with the strongest turbulence (Models V and VI). This indistinguishability is also observed in models for Mach 3 and 4 (not shown). Focussing on the SL models (top row), we see that the degree of fragmentation increases from a single monolithic filament/clump in Model I ($v_{a}/c_{s} = 0.03$) to multiple thin filaments in Model V ($v_{a}/c_{s} = 2.0$). Conversely, the CR-only models (bottom row) show a decrease in the degree of fragmentation from single isolated cores in Model II to thin filaments in Model VI. Finally, we note that our ``non-turbulent" simulations (Models I and II) are very similar to the corresponding micro-turbulent simulations in \citetalias{BB2014} \& \citetalias{BBC2015}. 

The formation of either a filament or a core is a direct consequence of the state of the medium. As shown in in \citet{BB2012,BB2014}, the fragmentation length scale of a cloud depends on the mass-to-flux ratio and the ionisation profile. The transcritical nature of the these simulations puts them in the regime where a change in the ionisation fraction can drastically change the fragmentation length scale. As shown in Figure 2 (right panel) of \citet{BB2012}, for a nearly flux frozen medium ($\tau_{n,i} = 0.001$, i.e., models with SL ionisation profile) the fragmentation length scale is much larger than in a medium with a smaller ionisation fraction ($\tau_{n,i} = 0.2$, i.e., models with CR-only ionisation profile). The thinning out of the structures found in Models I and II to those observed in Models V and VI is due to the turbulence present in the medium.

\begin{figure*}
\centering
\includegraphics[width = 0.3\textwidth]{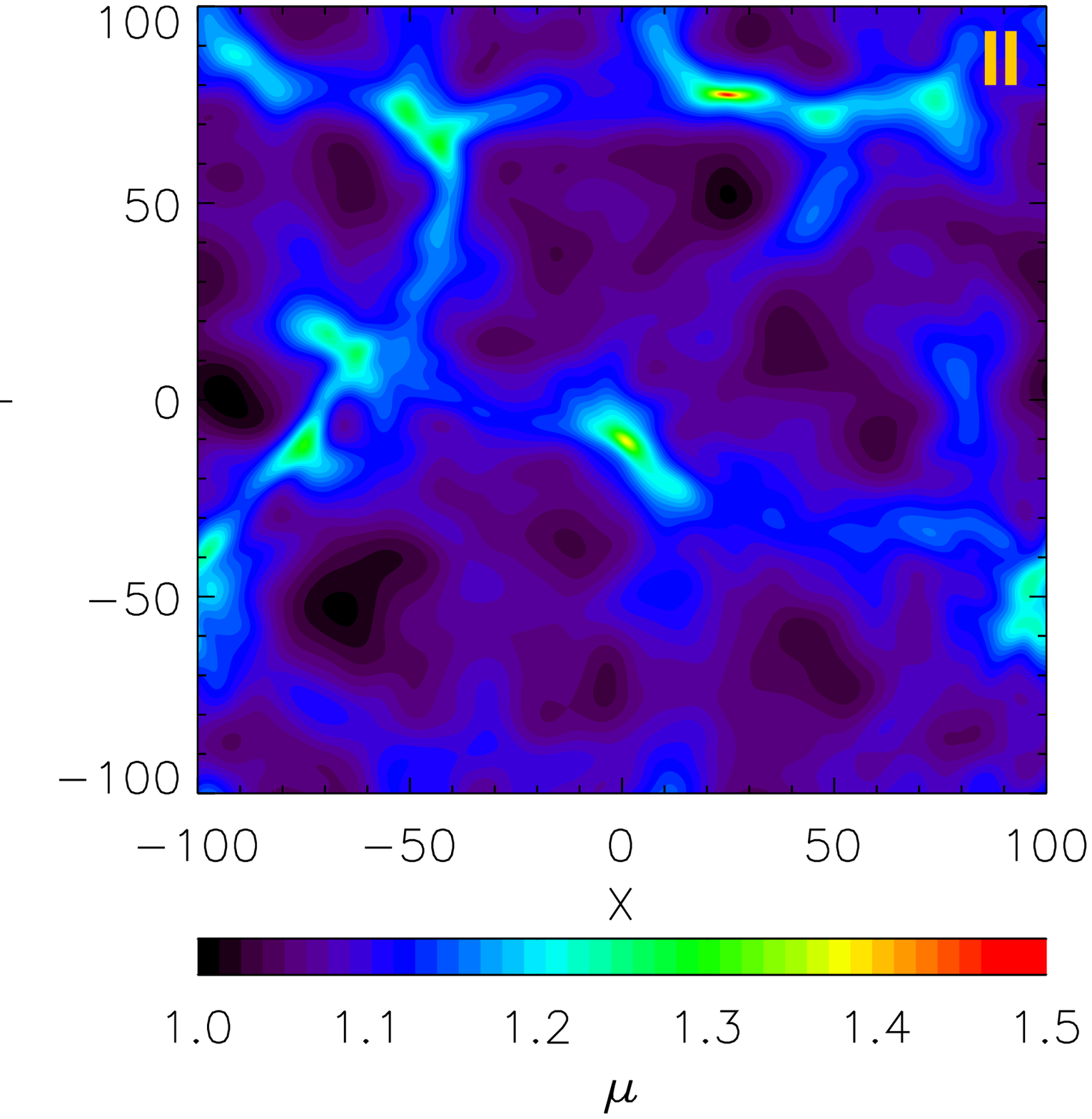} 
\includegraphics[width = 0.3\textwidth]{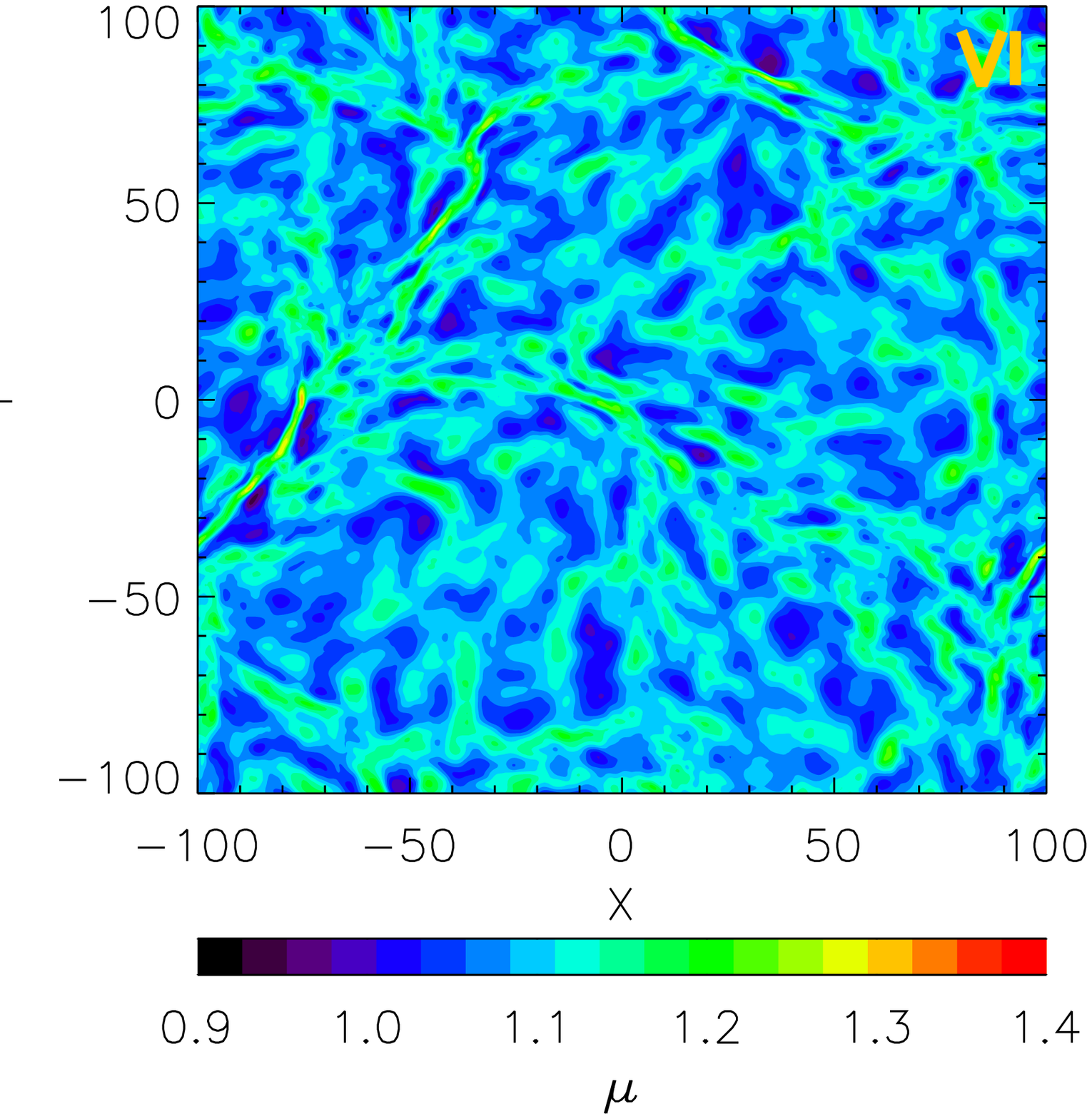} 
\includegraphics[width = 0.3\textwidth]{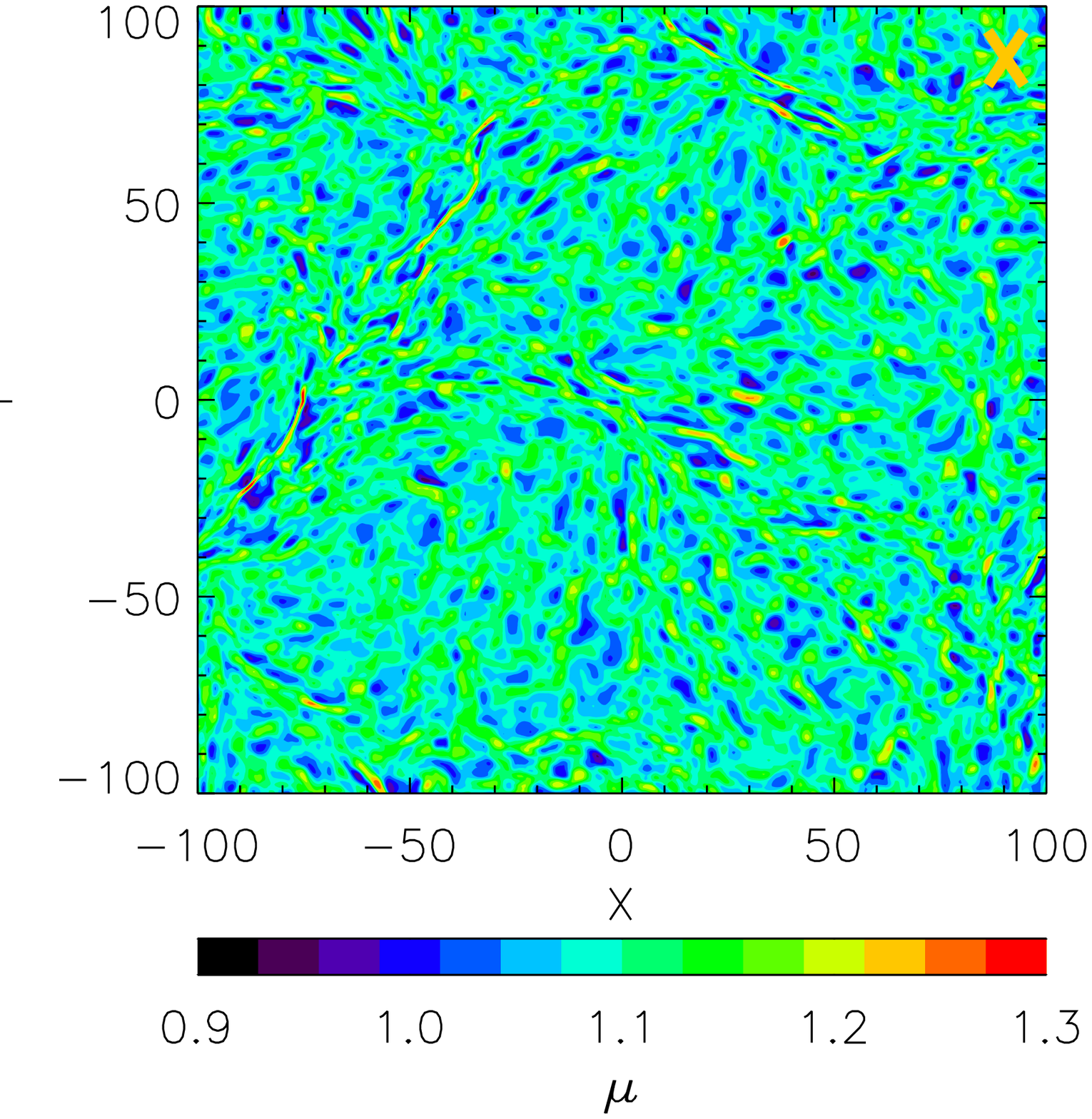} 
\caption{Mass-to-flux ratio maps for representative CR models with Mach number = 0.03 (left), 2 (center), and 4 (right). Colour bar shows the mass-to-flux ratio on a linear scale. }
\label{mu}
\end{figure*}

\begin{figure}
\centering
\includegraphics[width = 0.45\textwidth]{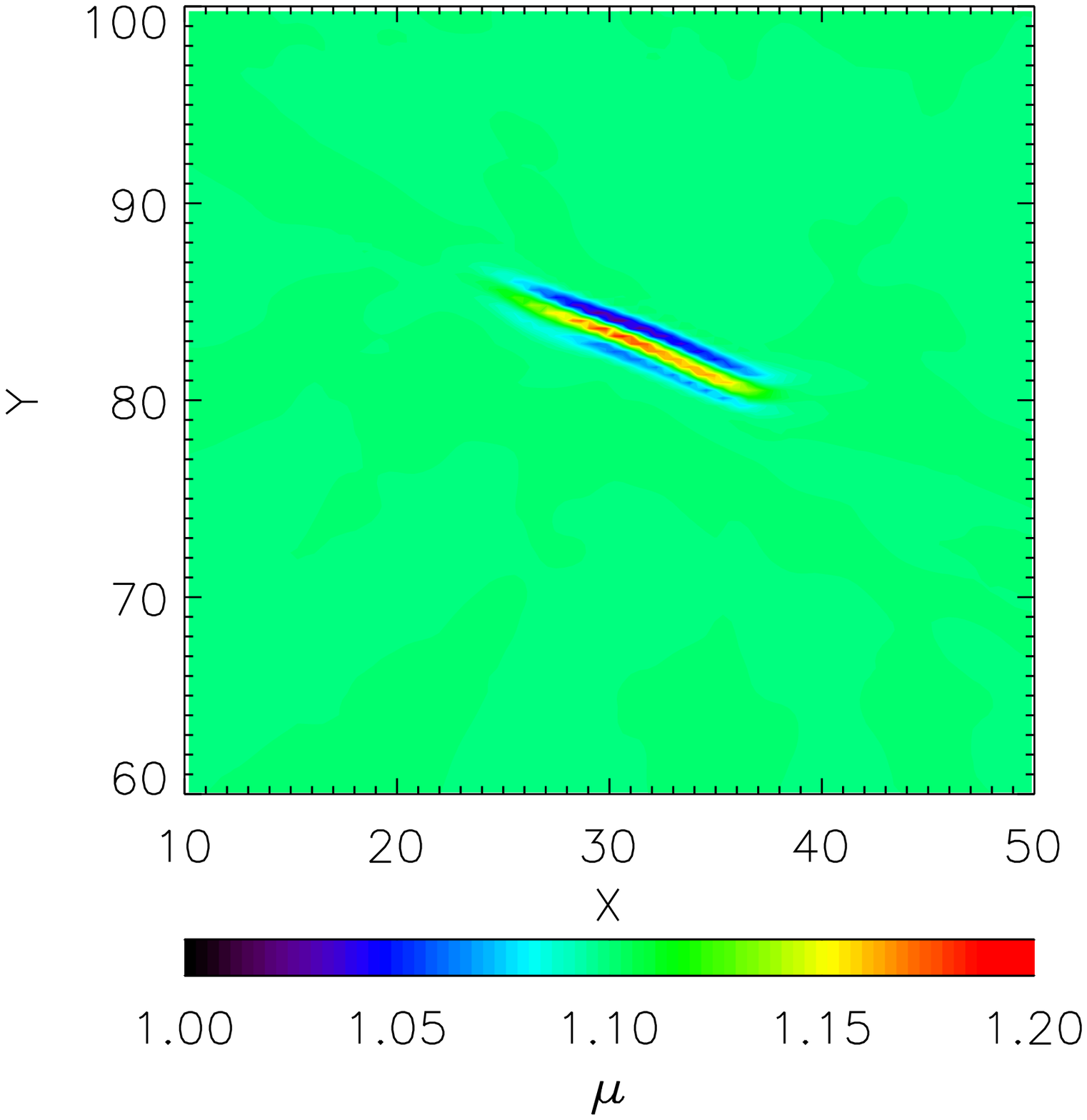}\\ 
\includegraphics[width = 0.35\textwidth, angle=-90]{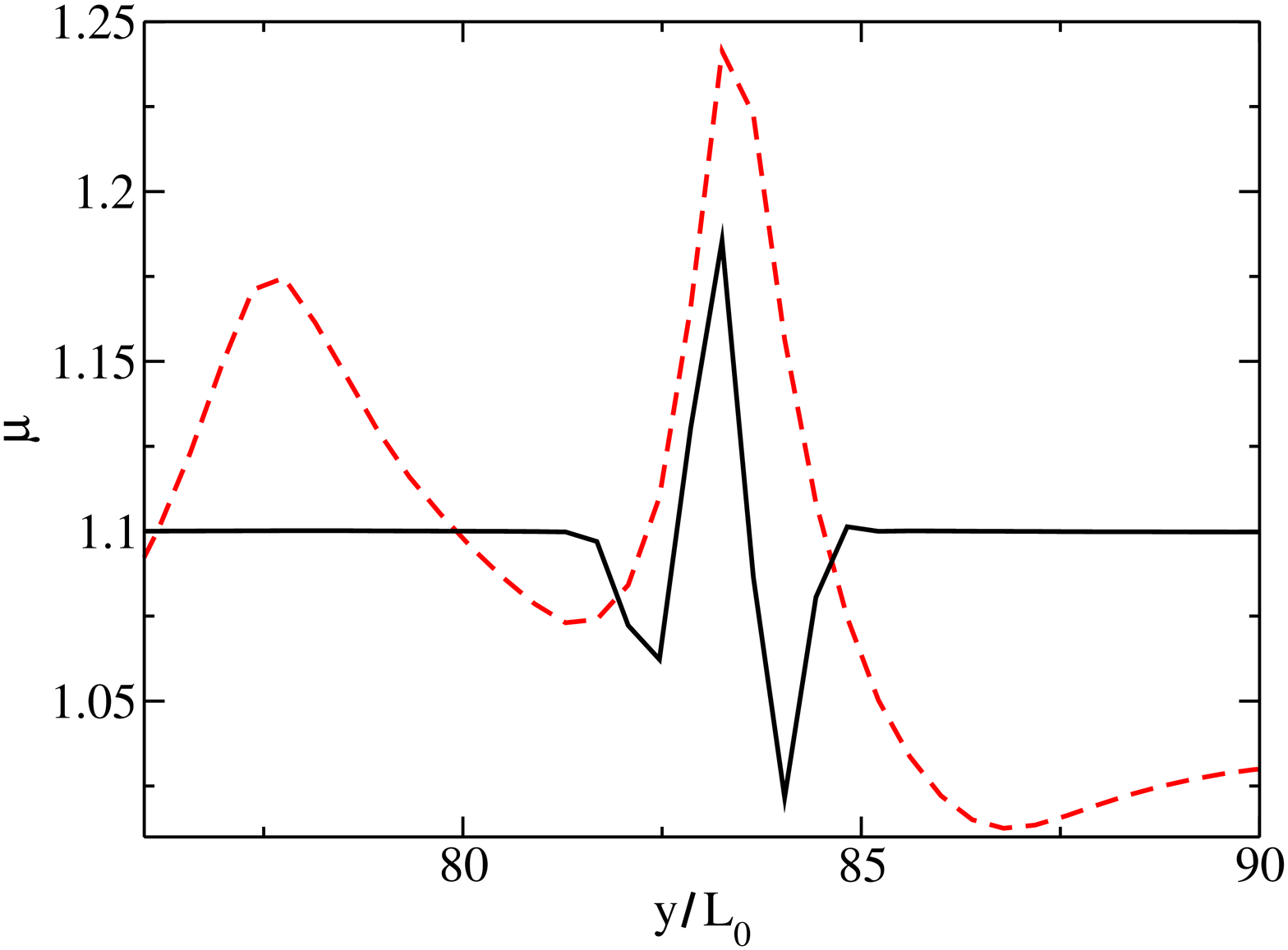}
\caption{Top: Representative mass-to-flux ratio map for Filament V-4 (See Table~\ref{Filamentmasses}). Bottom: Mass-to-flux ratio profiles through the core for Filament V-4 (SL, black, solid line) and Filament VI-4 (CR, red, dashed line). The centres of the cores are at $y/L_{0}$ = 83.28 for Filament V-4  and $y/L_{0}$ = 83.25 for Filament VI-4. The profiles are taken along the $y$-axis. }
\label{muprofile}
\end{figure}

\subsubsection{Mass-to-flux Ratio Structure}
\label{mtfstructure}
In this section, we focus on the effect that varying the initial turbulent Mach number and ionisation profile have on the distribution of mass-to-flux ratio within the first set of simulations. The effect of varying the initial mass-to-flux ratio will be explored in the following section. Figure~\ref{mu} shows representative mass-to-flux ratio maps for CR-only models with Mach number equal to 0.03, 2 and 4 (Models II, VI and X, respectively). From these maps, we see that the structure of the mass-to-flux ratio within the regions changes on smaller and smaller scales as the degree of turbulence increases. This is also evident in the corresponding SL simulations (not shown), however due to the nearly flux frozen nature of these simulations, the range of mass-to-flux ratios present within the region are not significantly different from the initial value ($\mu_{0} = 1.1$) and therefore do not show up within a linearly scaled map. As indicated in Table~\ref{models}, there is a significant difference between the maximum mass-to-flux ratio values found in the SL and CR-only models; the average maximum value for SL models is 1.186 while the average maximum value for the CR-only models is 1.352. Looking at the low density regions, we see that the structure within the CR-only models exhibits $\sim$0.1 differences between the high and low mass-to-flux ratio regions (green vs. blue) while in the SL simulations, the differences are on the order of 0.003 (not shown). The CR-only models are able to achieve a higher peak mass-to-flux ratio due to the lower ionisation fraction within these simulations as compared to the SL models. This allows for more neutrals to slip past the magnetic field lines to increase the density while locally keeping the magnetic field strength the same and thus increasing the mass-to-flux ratio.

The top panel of Figure~\ref{muprofile} shows a representative map of the mass-to-flux ratio structure within the dense region of a SL model (Model V). Looking at the clump/core regions in detail we see that the models exhibit mass-to-flux ratio structures similar to those described in \citet{BB2014} (Model A), i.e., a high mass-to-flux ratio corresponding to the highest density region surrounded by an extremely low mass-to-flux ratio on either side. Specifically for the core depicted in Figure~\ref{muprofile}, we see that the high density central region has a mass-to-flux ratio of $\sim 1.18$, while the low density outer regions have a mass-to-flux ratio of only $\sim 1.04$. The bottom panel of Figure~\ref{muprofile} shows profiles of the mass-to-flux ratio across the core as a means to further illustrate this interesting structure. The black, solid line shows the profile for the core in the top panel, while the red, dashed line shows the profile for the corresponding core in Model VI. As shown by these two profiles, this characteristic structure/profile is evident within all the models, i.e., a spike at the location of the core and depressions on the outskirts as described above. This is a clear signature of flux redistribution by ambipolar diffusion. The profile is more pronounced in the SL models however due to the smaller variations in the background mass-to-flux ratio. In contrast, the feature can be much broader within the CR-only simulations. In this case, the flux redistribution occurs over a larger region than in the SL models. Initial analysis of the kinematics show that Model V (SL) and Model VI (CR-only) have similar velocity structures near the core, while Model III (SL) has a steeper velocity gradient near the core than in Model IV (CR-only), thus suggesting that the SL and CR simulations present distinguishable velocity structures for apparently similar density enhancement structures. Further analysis of the kinematics will be presented in our next paper.

\begin{figure}
\centering
\includegraphics*[width=0.38\textwidth, angle = -90]{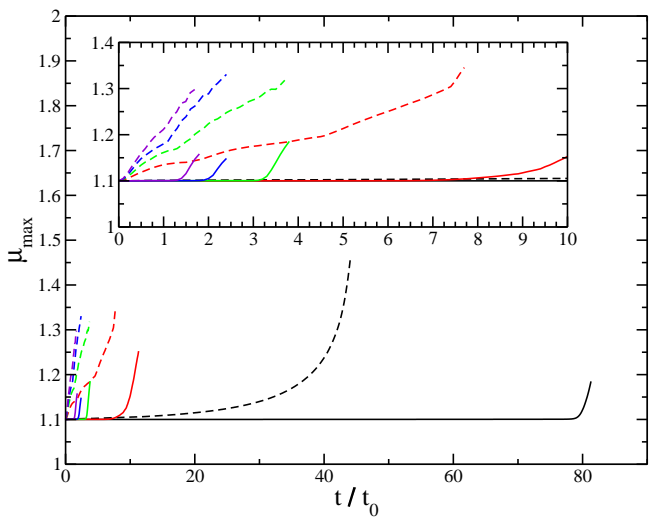}\\
\caption{Maximum mass-to-flux ratio as a function of dimensionless time ($t/t_{0}$) for all models. Solid lines indicate simulations with SL ionisation profiles while dashed lines indicate CR-only ionisation profiles. Colours indicate different Mach values (Black = 0.03, Red = 1, Green = 2, Blue = 3, and Purple = 4). The main plot shows the full time range while the inset zooms in on the range from 0 to $10t_{0}$. }
\label{muvst}
\end{figure}

Figure~\ref{muvst} shows the maximum mass-to-flux ratio within the simulation region as a function of time for the ten models in Set 1 ($\mu_{0} = 1.1$). Solid lines indicate simulations with SL ionisation profiles while dashed lines indicate CR-only ionisation profiles. Colours indicate different Mach values (Black = 0.03, Red = 1, Green = 2, Blue = 3, and Purple = 4). The inset shows a zoom-in of the first $10t_{0}$. Note that the location of the maximum mass-to-flux ratio ($\mu_{max}$) does not necessarily correspond to the location of the maximum density enhancement within the simulation. This is evident by the fact that the $\mu_{max}$ values quoted in Table~\ref{models} do not necessarily equal the mass-to-flux ratio at the location of the maximum density enhancement $\mu(\sigma_{n,max})$. This mismatch between mass-to-flux ratio values can occur throughout the evolution of the region. 

\begin{figure*}
\centering
\includegraphics[width = 0.3\textwidth]{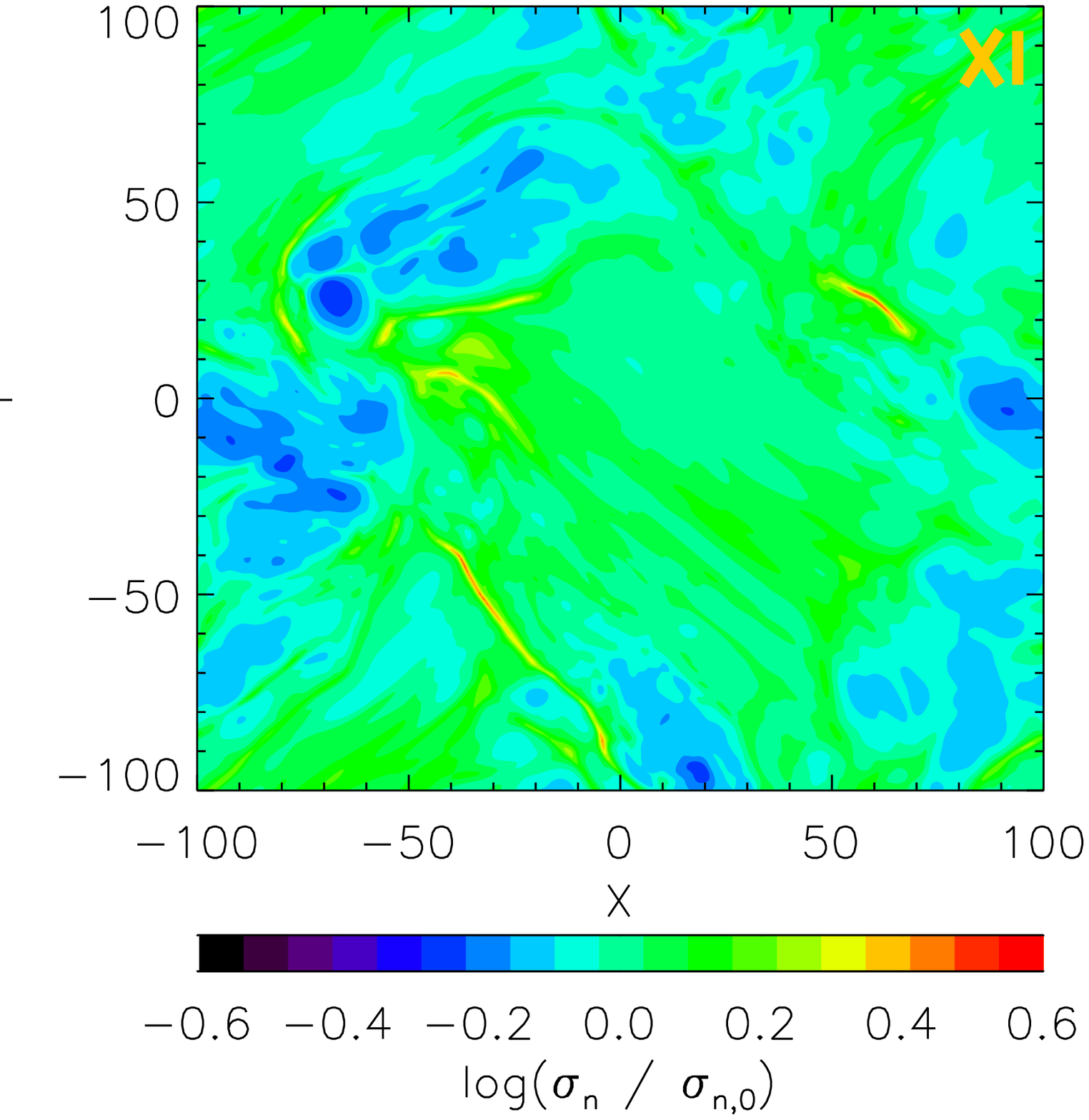} 
\includegraphics[width = 0.3\textwidth]{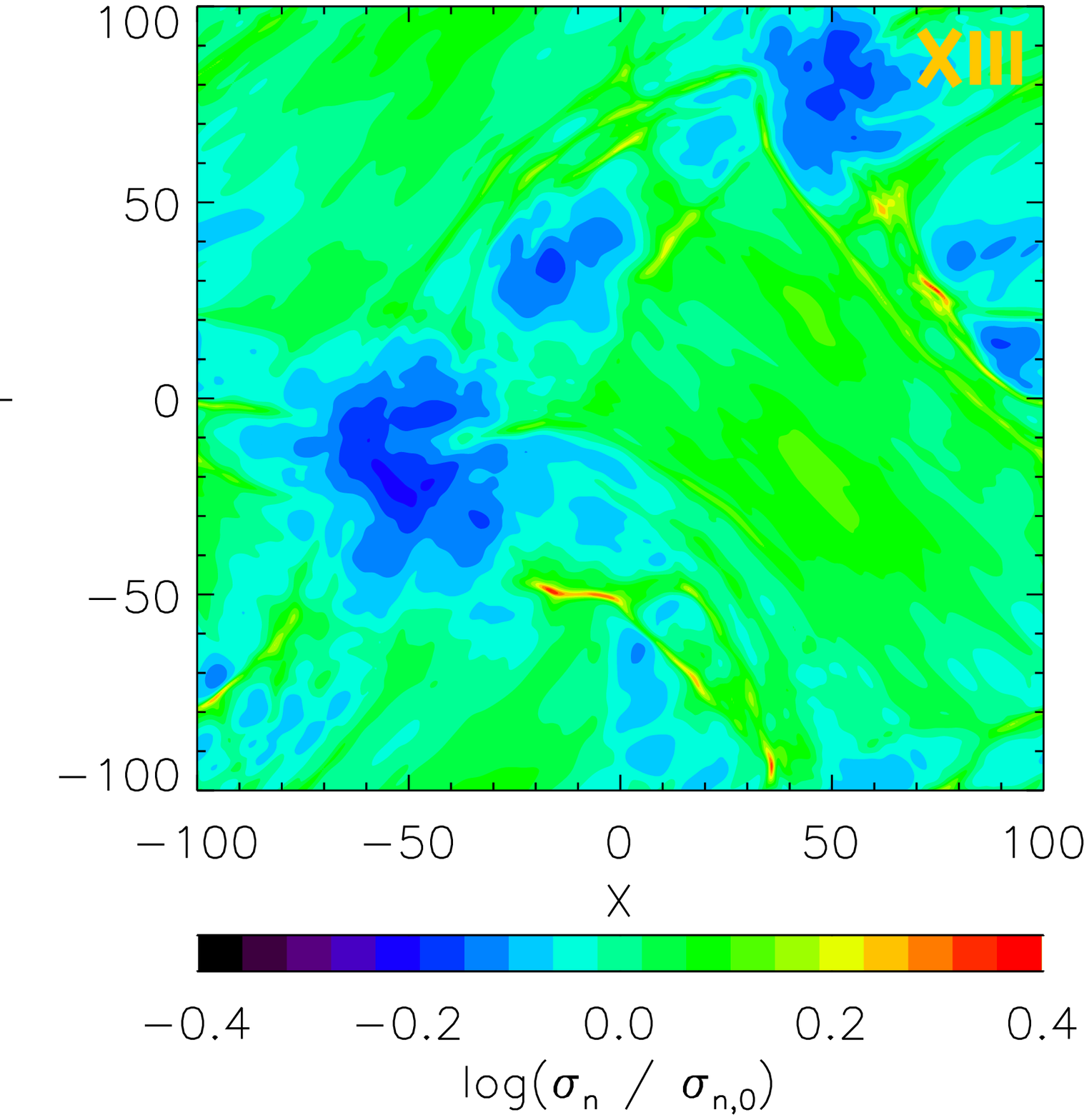} 
\includegraphics[width = 0.3\textwidth]{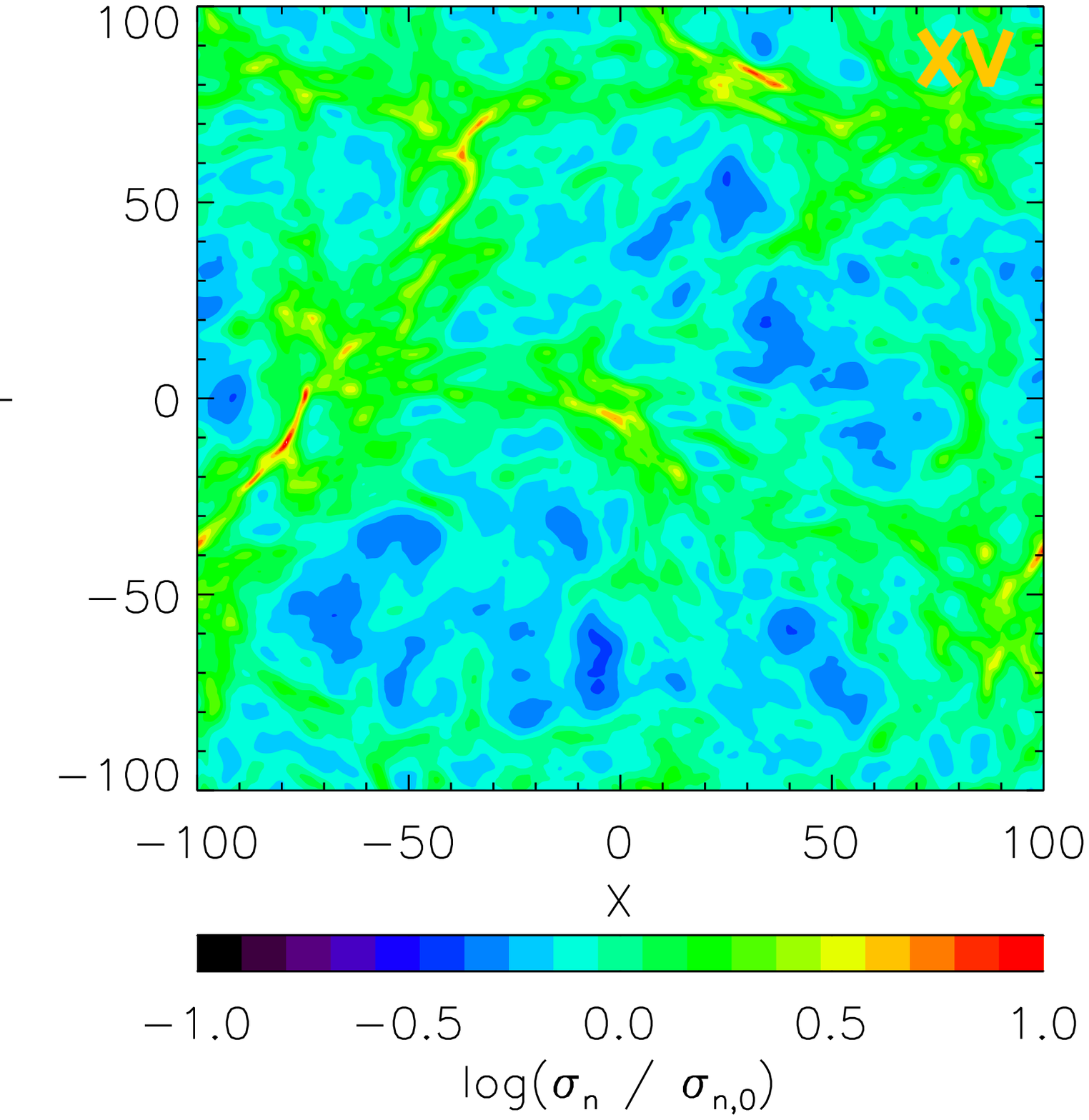}\\ 
\includegraphics[width = 0.3\textwidth]{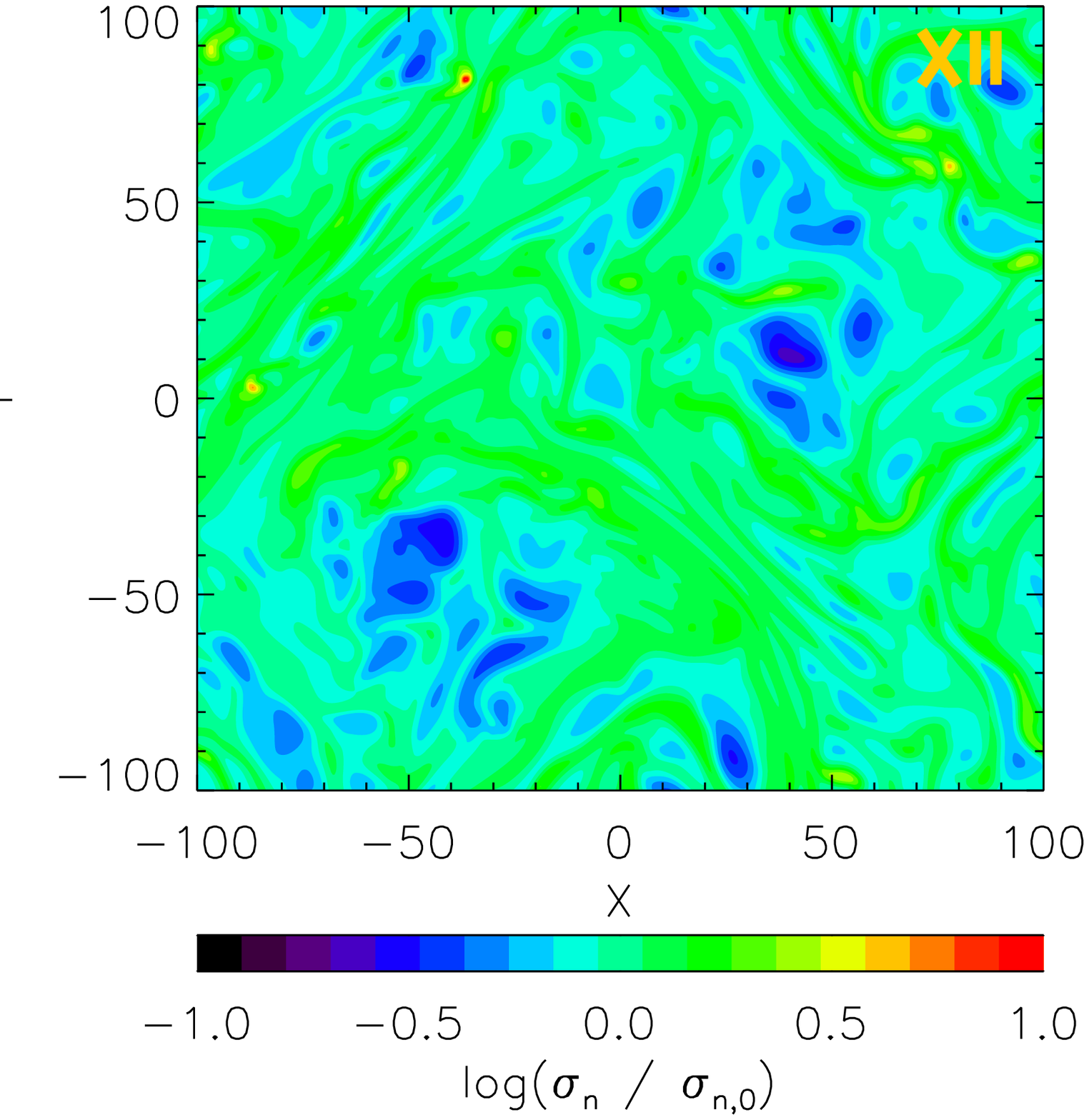} 
\includegraphics[width = 0.3\textwidth]{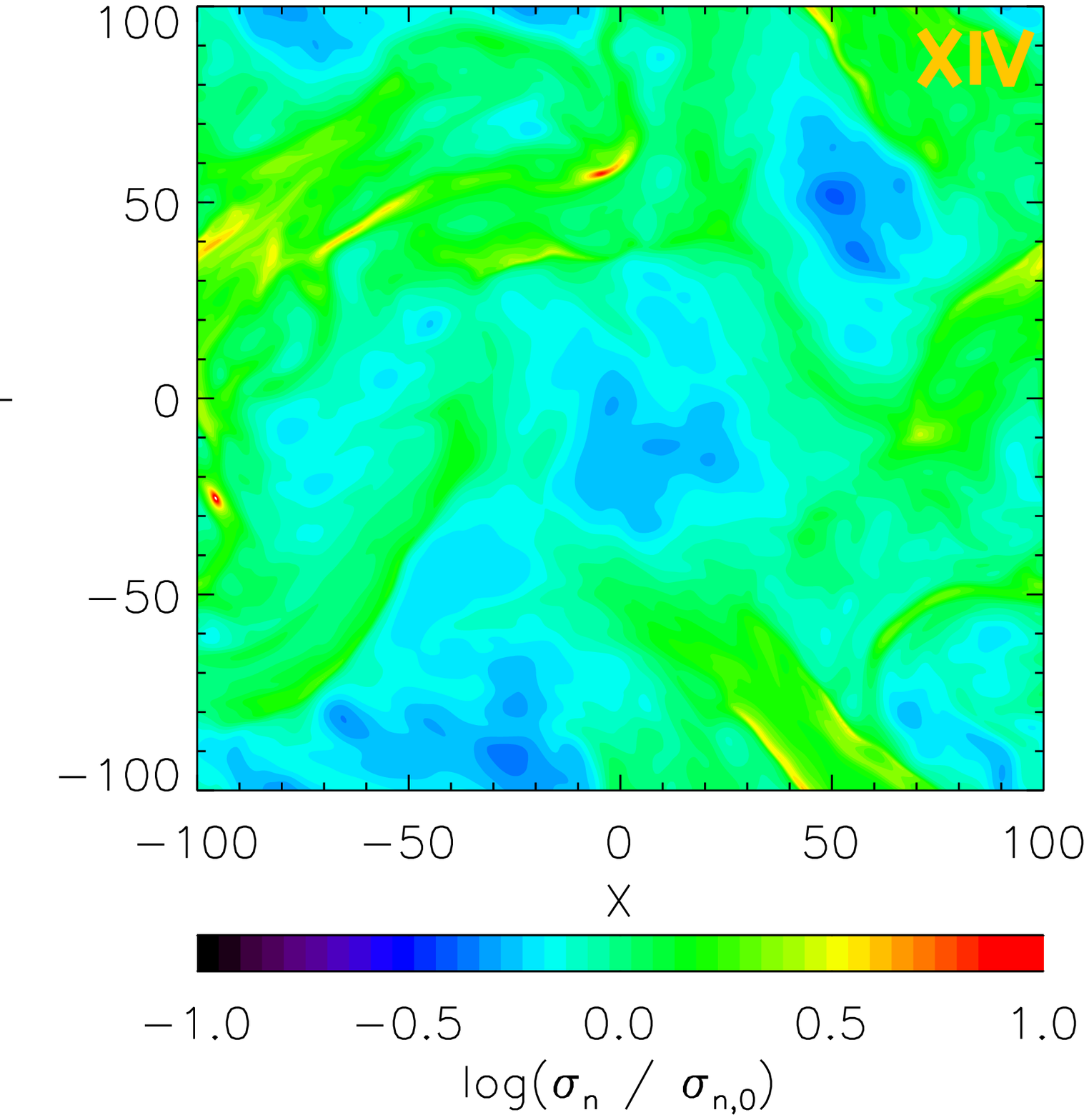} 
\includegraphics[width = 0.3\textwidth]{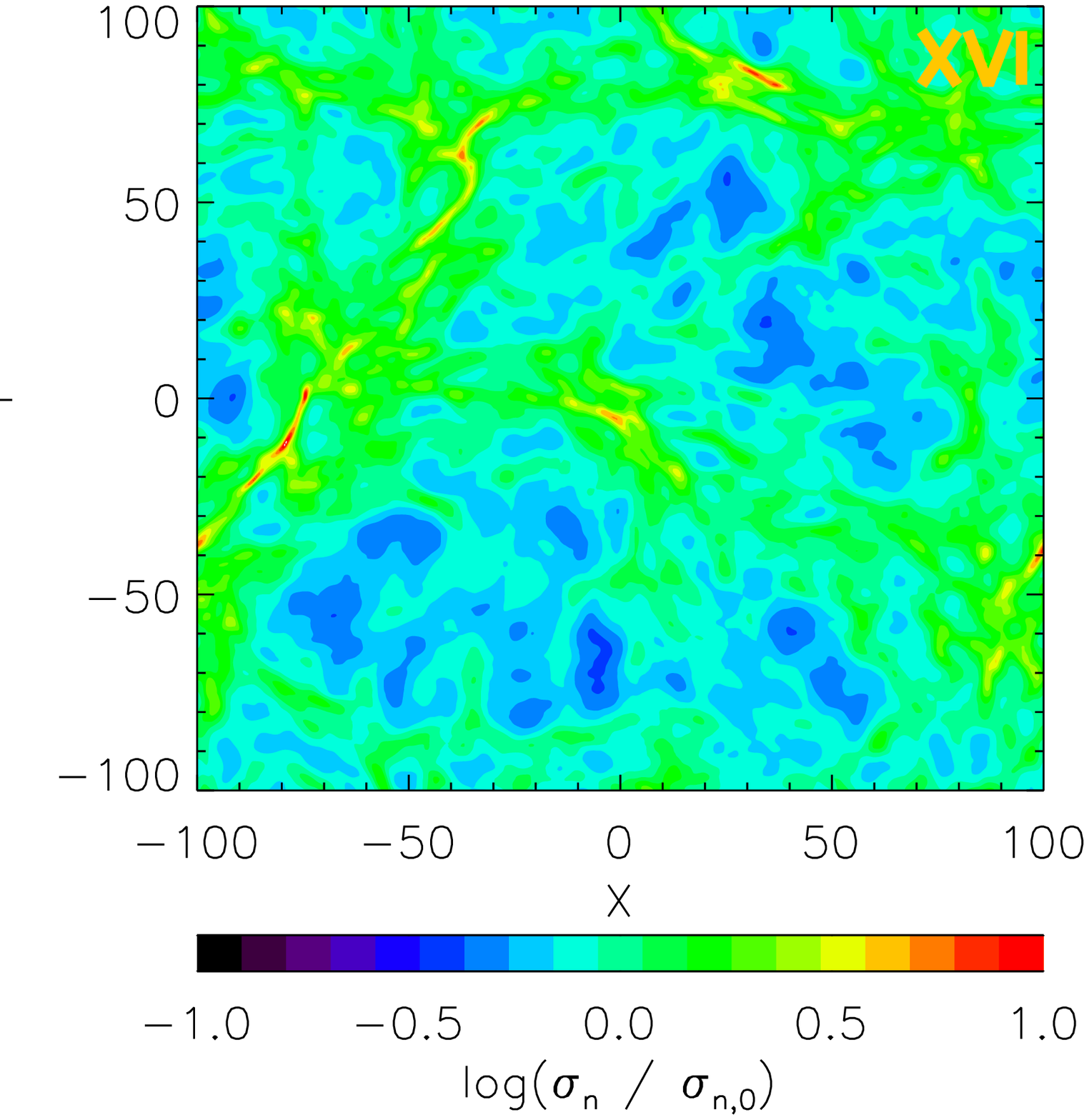} 
\caption{Column density enhancement maps for models with initial mass-to-flux ratio of 0.5 (left panels), 0.8 (centre panels), and 2.0 (right panels). Colour bar shows the logarithm of the density enhancement. Roman numerals in the top right of each panel indicate the model number of the simulation (see Table~\ref{models}). Top row: models with the Step-Like ionisation profile. Bottom Row: models with CR-only ionisation profile. Note the different limits on the colour bars for Models XI and XIII. }
\label{sigma_varymu}
\end{figure*}

\begin{figure*}
\centering
\includegraphics[width = 0.3\textwidth]{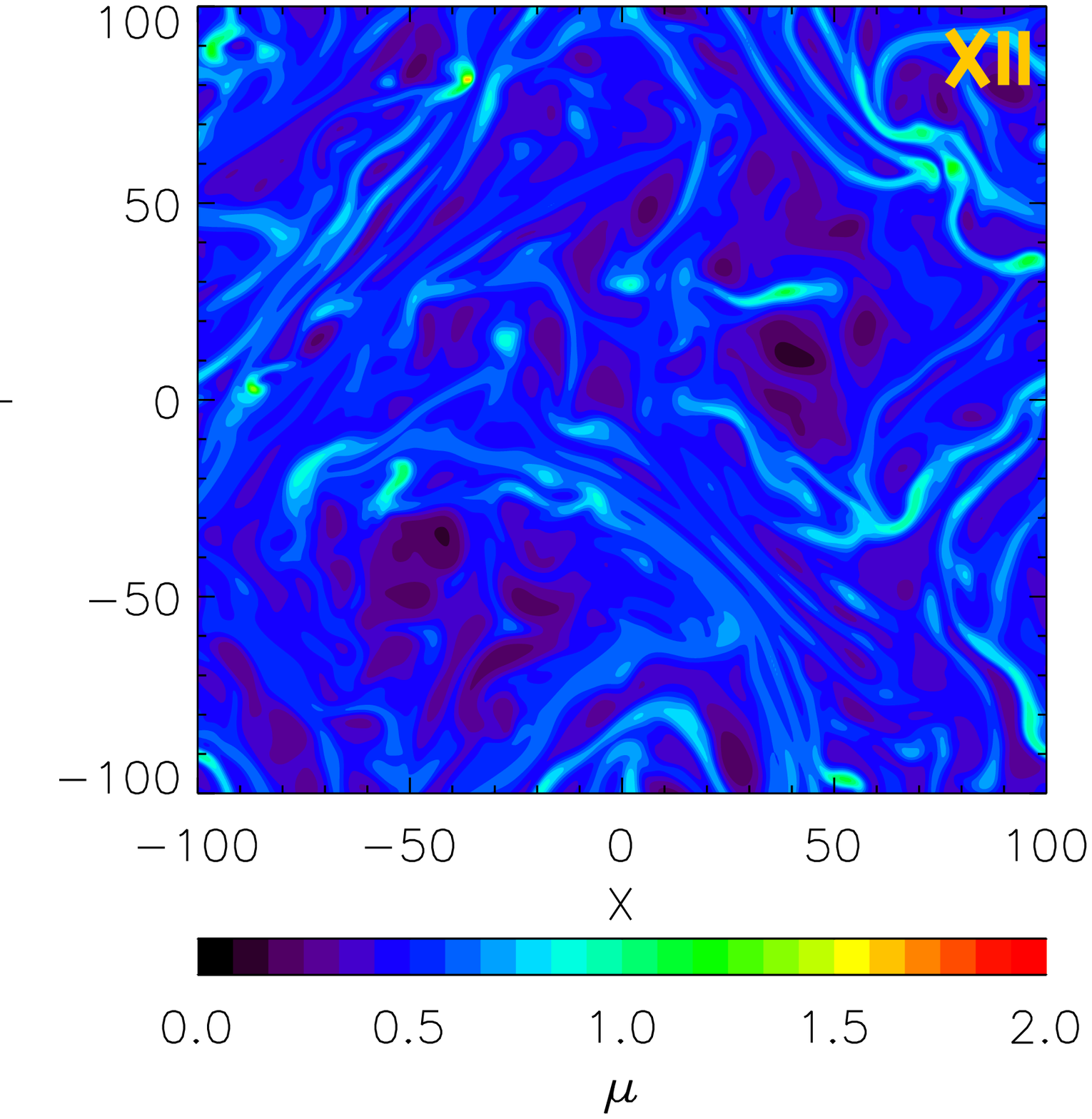} 
\includegraphics[width = 0.3\textwidth]{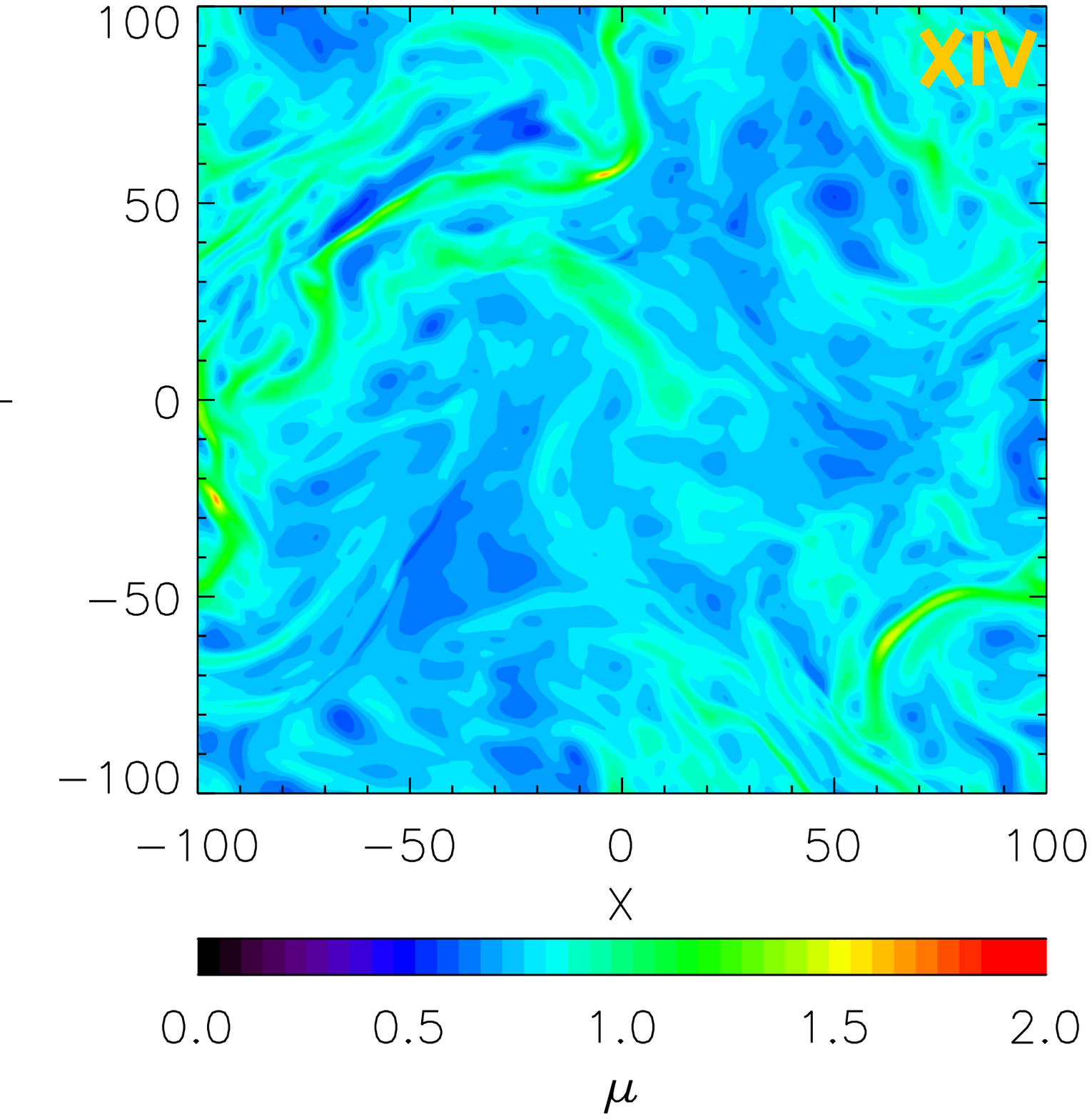} 
\includegraphics[width = 0.3\textwidth]{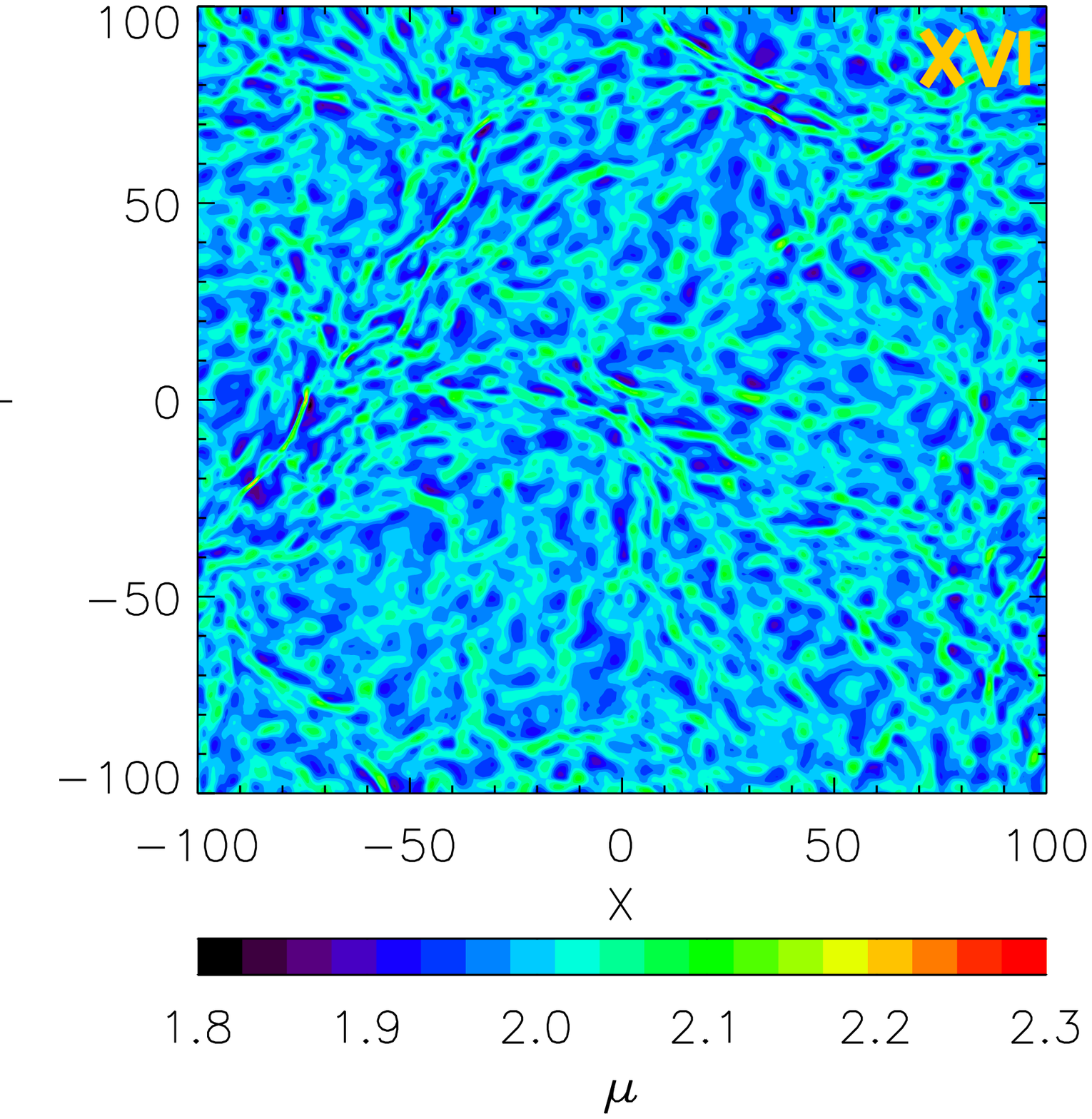} 
\caption{Mass-to-flux ratio maps for CR-only models with initial mass-to-flux ratio of 0.5 (left panels), 0.8 (centre panels), and 2.0 (right panels). Colour bar shows the mass-to-flux ratio on a linear scale. Roman numerals in the top right of each panel indicate the model number of the simulation (see Table~\ref{models}).}
\label{mtf_varymu}
\end{figure*}

Looking at the set as a whole, it is clear that the increase in the Mach number results in a marked decrease in the evolutionary time. In addition we note that for low value Mach numbers (i.e., those in Models I-IV), the SL simulation takes up to two times longer to evolve compared to the corresponding CR simulation, however for Mach numbers greater or equal to 2, the evolutionary times for the two different profiles are more or less identical. These temporal comparisons can also be seen in Table~\ref{models}. Focussing on the evolution of the mass-to-flux ratio itself, we see that all SL profile models exhibit a constant mass-to-flux ratio for the majority of the run with a sharp increase at near the end. This mimics the evolution of the maximum density and is a result of the nearly flux frozen nature of the medium at lower densities. In contrast, the CR models show a steady, mostly linear increase over the entire duration of the simulation. Again this is a direct consequence of the increased ambipolar diffusion throughout the simulation region in these models. Finally, we can see that the CR simulations result in greater final maximum mass-to-flux ratios than the SL models. These different mass-to-flux ratios should imply different dynamical histories. Therefore, as mentioned before, we expect the velocity structure to be different between the two ionisation profiles which can cause variations in simulated observations of molecular cloud tracers (Bailey et al., in prep) and enable us to quantify the effects of the ionisation structure and magnetic fields via molecular line observations.

\subsection{Effect of Mass-to-Flux ratio on Turbulent Simulations}

In addition to the models discussed in the previous sections (Set 1: Models I-X) we also ran a second set of models (Set 2: Models XI - XVI) which look at the evolution of the medium for different initial mass-to-flux ratios while keeping the initial turbulent Mach number fixed ($v_{a}/c_{s} = 2.0$). For this set, we have explored both sub- and supercritical initial mass-to-flux ratios which we can compare to the transcritical runs from Set 1 (Models V and VI for SL and CR, respectively). The initial parameters and resulting values are noted in Table~\ref{models}. In the following subsections, we look at the effect of changing the initial mass-to-flux ratio on the density enhancement structures and mass-to-flux ratio structures. Note that Models XI and XIII did not collapse into high density structures, therefore the values in Table~\ref{models} reflect those at the time when these models were terminated manually.

\subsubsection{Density Enhancement Structures}

Linear analysis from \citet{BB2012} shows that although fragmentation of a subcritical region may be possible, it can take up to infinitely longer than media with trans- or supercritical initial mass-to-flux ratios, however this analysis does not take into account the possible effects of turbulence on the fragmentation of a subcritical region. Looking at the results of the SL-simulations, we see that the addition of a Mach 2 turbulent component does not add enough energy to create regions where ambipolar diffusion can overcome the magnetic field and allow a subcritical medium (Models XI and XIII) to evolve to a point where runaway gravitational collapse can occur. For Model XIII, an initial run which assumed our fiducial stopping condition seemed to suggest that runaway collapse could occur even though the medium is sub-critical. However, a subsequent run with an increased stopping condition ($\sigma/\sigma_{n,0} = 20$) revealed that although a maximum density enhancement of 12.82 was achieved at $t/t_{0} = 11$, runaway collapse was not achieved. Instead, these dense regions were still subcritical and transient in nature. Figure~\ref{sigma_varymu} shows density enhancement maps at the final time for each simulation in Set 2, as quoted in Table~\ref{models}. The top row shows the SL models while the bottom row shows the CR-only models. The initial mass-to-flux ratio values increase from left to right ($\mu = 0.5, 0.8$ and 2.0, respectively). The colour scale shows the logarithm of the density enhancement. Looking at the subcritical SL models which did not result in runaway gravitational collapse (Model XI and Model XIII, top left and middle respectively), we see that the medium has become quite smooth with some evidence of low density filaments. These filaments tend to be transient and do not persist long enough for cores to form within them. For the supercritical SL case (Model XV), the medium fragments into dense filaments as expected. In this case, the higher mass-to-flux ratio allows this fragmentation to occur on a shorter timeframe than in the corresponding model from Set 1 (Model V). Looking at the bottom row, we see that with the CR-only ionisation profile, all three models are able to fragment into dense cores/filaments. Interestingly, change in the resulting structures via the increase in mass-to-flux ratio mimics the evolution seen with increasing turbulence, i.e. the smallest mass-to-flux ratio yields small isolated compact cores while larger mass-to-flux ratios result in a stretching of these single cores into filamentary structures with evidence of multiple cores within \citep[see also][]{BCW2009, BCDW2009}.

\subsubsection{Mass-to-flux ratio Structures}

Figure~\ref{mtf_varymu} shows mass-to-flux ratio maps at the final time for the Set 2 CR-only simulations as quoted in Table~\ref{models}. As with the SL models from Set 1, the variations in the mass-to-flux ratios for SL models in this set are not significantly different from the initial value and therefore do not show up well on a linear scale. The initial mass-to-flux ratio value increases from left to right. The colour scale shows the mass-to-flux ratio on a linear scale. Note that the minimum and maximum scale for each panel is not the same. The maximum mass-to-flux ratio values present at the displayed times can be found in Table~\ref{models}. Looking at differences between the SL and CR simulations, we see that the CR-only simulations result in larger changes from the initial mass-to-flux ratio values than the SL only simulations. This is due to the SL simulations being held at near flux-frozen conditions for the majority of the run. Note that the large difference between the final mass-to-flux ratio values for the $\mu = 0.5$ (Models XI and XII) and $\mu = 0.8$ (Models XIII and XIV) is due to the fact that the SL simulations were not able to collapse and therefore were always held at near flux frozen conditions. Looking at the bottom row of Figure~\ref{mtf_varymu}, the CR-only simulations show that as the mass-to-flux ratio increases, the amount of structure within the mass-to-flux ratio map increases. However, the relative amplitude of variations is greater for the lower mass-to-flux ratio models. Looking back at Model VI ($\mu_0 = 1.1$), we see that the degree of structure present in that simulation fits in with the pattern in the three CR-only simulations in Set 2. Analysis of the profiles across the cores within the models of this set shows that, as with the other set, both the SL and CR-only simulations show the same characteristic shapes with the SL models again exhibiting spikier profiles due to the smaller variation in mass-to-flux ratio between the low and high density regions. We note that the fluctuations in mass-to-flux ratio are introduced in our initial conditions, and we are tracking the difference in their subsequent evolution.

\section{Physical properties of Filaments and Cores}

The structures visible within a star forming region are highly dependent on the sensitivity of the instrument used to observe them. Data from the \textit{Herschel} space telescope, the Institut de Radioastronomie Millim\'{e}trique (IRAM) 30m telescope, the James Clerk Maxwell Telescope (JCMT) and, now even more from the Atacama Large Millimetre/submillimetre Array (ALMA) and the Jansky Very Large Array (JVLA), show that complex structures are typically observed in the direction of star forming regions. These structures can take on many different forms including subfragmentation of large clump regions into several cores \citep[i.e.,][]{Sadavoy2012}, ubiquitous filaments in star forming regions \citep{Andre2010,Hacar2013,Henshaw2014}, bundles of fibres within filaments \citep{Hacar2013,TH2015} and even filaments within dense cores \citep{Pineda2011}. Our simulations show formation of both filaments and dense cores.

\begin{figure}
\centering
\includegraphics[width = 0.48\textwidth]{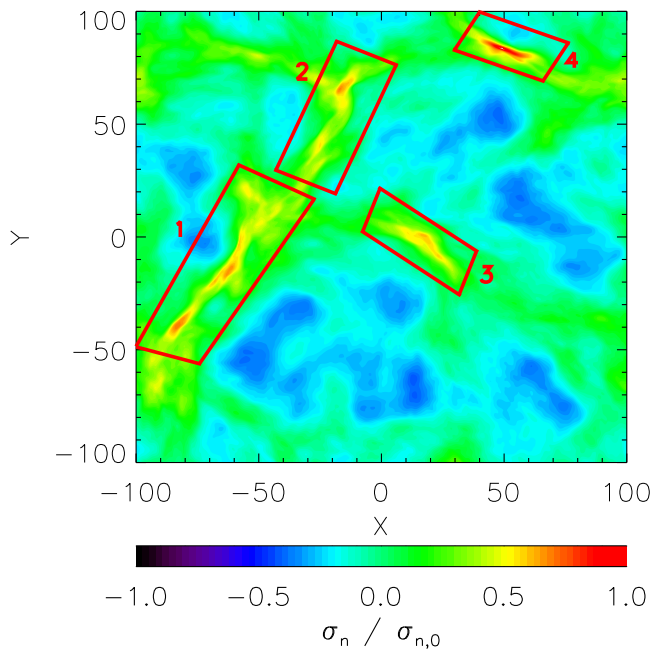}
\caption{Representative map based on Model V showing the definition of filament regions. Note that this map is shifted by 50 pixels (20 map units) in the $x$ direction compared to those in Figure~\ref{sigma} to ensure that the filaments contained within region 1 are not split at the periodic boundary. To avoid confusion, all quoted locations of structures will be based upon the original locations in Figure~\ref{sigma}.}
\label{filamentmap}
\end{figure}

As shown in Figures~\ref{sigma}~\&~\ref{sigma_varymu}, the majority of the models exhibit the formation of dense structures in four main regions. Figure~\ref{filamentmap} shows a representative map of these four visually recognisable regions which contain high density filaments within our simulations.\footnote{The filament in Model I spans both regions 1 and 2. We denote this filament as I-1,2 to highlight this fact.}. This map is constructed using Model V, however please note that this map has been shifted by 50 pixels (20 map units) in the $x$ direction in order to ensure that the filaments contained within region 1 are not split at the periodic boundary of the box. To avoid confusion however, all quoted locations of regions will be based upon the original unshifted figures (i.e., those shown in Figure~\ref{sigma}.) The following sections will examine the physical properties of the filaments and embedded cores within the four regions defined above. Specifically we will determine the masses of the cores and filaments and analyse the profiles across the filaments to determine the widths via two different fitting functions.

\subsection{Filament and Core Masses}

To determine the mass within the filaments and dense structures in the simulations we utilise {\sc CLUMPFIND2D} \citep{Williams1994}. This is a set of IDL routines which find the extent of structures within observations or simulated data. Structures are determined by assuming linearly spaced contours based upon user definitions and connecting pixels at each contour level that are within one resolution element of each other \citep{Williams1994}. For our simulations we have defined the outer edges of the filaments to be at $A_{V}=3$ mag and dense structures to be regions above $A_{V}=7$ mag. Recall that for our simulations, $A_{V}$ refers to the visual extinction perpendicular to the 2D plane. Output from these routines include the ``intensity'' of the pixels within the identified structure. In the case of our simulations, this intensity is the sum of the column density over all pixels within the defined structure. From these data we then determine the enclosed mass of the filaments and dense structures. Finding filaments and cores within observational data usually requires more sophisticated techniques such as morphological component analysis (MCA) or use of programs such as getfilaments \citep{Menshchikov2013} or DisPerSe \citep{Sousbie2013}.  Comparison of these techniques by \citet{Konyves2015} showed they all give approximately the same output. In our case, however, we do not require such sophisticated methods as our models are not affected by the noise and background fluctuations that are present in astronomical observations. 

\begin{table}
\small
\caption{Filament Masses}
\centering
\begin{tabular}{llc}
\hline\hline
Filament & Filament range\tablefootmark{b} & Mass\\
Designation\tablefootmark{a} & $(L/L_{0})$ & ($M_{\odot}$)\\
\hline\hline
\multicolumn{3}{c}{Set 1}\\
\hline\hline
I-1,2\tablefootmark{c} & Y: -42.9 - 45.7 & 442\\
\hline
III-1 & Y: -69.7 -- 20.1 & 380 \\ 
III-2 &  Y:  20.9 -- 72.4 & 105 \\
III-3 &  X: -18.1 -- 5.9  & 54.9 \\ 
III-4 &  X:  19.3 -- 48.8 & 86.9 \\
\hline
IV-1a & Y: -46.1 -- -33.9 & 19.2\\  
IV-1b & Y: -22.0 --  13.4 & 43.6\\ 
IV-2  & Y:  59.1 --  72.0 & 24.4\\ 
IV-3  & X: -10.2 --   3.5 & 24.4\\ 
IV-4  & X:  20.5 --  42.9 & 44.2\\
\hline
V-1a & Y: -44.9 -- -29.9 & 21.4\\
V-1b  & Y: -23.2 --   3.5 & 35.1\\ 
V-2 & Y:  56.7 --  72.4 & 22.6\\ 
V-3 & X: -11.8 --   8.3 & 29.1\\ 
V-4 & X:  19.3 --  42.9 & 45.8\\
\hline
VI-1a & Y: -45.7 -- -31.1 & 18.6\\ 
VI-1b & Y: -23.6 --   3.9 & 31.9\\
VI-2a & Y:  37.8 --  46.9 & 5.94\\ 
VI-2b  & Y:  56.7 --  74.4 & 21.4\\
VI-3 & X: -12.2 --   8.6 & 23.2\\ 
VI-4 & X:  17.3 --  43.3 & 43.8\\
\hline
VII-1a & Y: -46.5 -- -29.9 & 19.7\\  
VII-1b  & Y: -24.8 --   3.5 & 34.4\\
VII-2a & Y:  37.8 --  48.8 & 7.78\\ 
VII-2b  & Y:  53.5 --  74.4 & 23.5\\
VII-3 & X: -12.6 --   6.3 & 20.3\\ 
VII-4 & X:  13.4 --  44.1 & 44.8\\
\hline
VIII-1a & Y: -46.5 -- -30.7 & 19.7\\
VIII-1b  & Y: -26.0 --   3.5 & 34.1\\    
VIII-2a & Y:  37.8 --  48.8 & 8.39\\ 
VIII-2b  & Y:  52.0 --  74.4 & 20.0\\
VIII-3 & X: -14.2 --   6.7 & 22.3\\ 
VIII-4 & X:  12.6 --  43.7 & 44.4\\
\hline
IX-1 & Y: -46.9 --   4.3 & 67.1\\ 
IX-2 & Y:  37.8 --  74.4 & 36.1\\ 
IX-3 & X: -14.6 --   7.1 & 23.6\\ 
IX-4 & X:  12.2 --  43.7 & 47.6\\
\hline
X-1a & Y: -46.9 -- -30.3 & 17.7\\ 
X-1b  & Y: -25.6 --   4.3 & 30.5\\ 
X-2a & Y: 38.6 -- 50.0  & 7.93\\ 
X-2b & Y: 51.2 -- 74.4  & 19.4\\
X-3 & X: -13.8 --   4.3 & 16.0 \\ 
X-4 & X:  11.4 --  42.9 & 39.9 \\
\hline\hline
\multicolumn{3}{c}{Set 2}\\
\hline\hline
XV-1a & Y: -43.7 -- -33.1  & 10.2\\
XV-1b & Y: -23.2 -- 3.1 & 22.9\\
XV-2a & Y: 40.2 -- 48.4 & 4.27\\
XV-2b & Y: 52.8 -- 73.6 & 13.4\\
XV-3  & X: -5.12 -- 3.94 & 7.17\\
XV-4a & X: 16.1 -- 20.5 & 1.43\\
XV-4b & X: 27.2 -- 39.8 & 11.9\\
\hline 
XVI-1a & Y: -44.5 -- -33.5 & 10.5\\
XVI-1b & Y: -23.2 -- 3.54 & 23.3\\
XVI-1c & Y: 10.2 -- 14.6 & 3.05 \\
XVI-2a & Y: 43.3 -- 49.2 & 4.58\\
XVI-2b & Y: 52.0 -- 73.6 & 14.2\\
XVI-3  & X: -5.12 -- 4.33 & 7.32\\
XVI-4a & X: 16.1 -- 20.9 & 1.53\\
XVI-4b & X: 27.6 -- 39.8 & 11.7\\
\hline\hline
\end{tabular}
\tablefoot{
\tablefoottext{a}{The filament designation is formatted as (Model number - Filament region). For filaments regions with multiple distinct filaments, a lowercase letter is added to indicate this multiplicity. }\\
\tablefoottext{b}{The filament range refers to our defined extent of the filament along the X or Y direction oriented closest to the major axis of the filament. }\\
\tablefoottext{c}{The filament in this simulation exists within both regions 1 and 2 (See Figures~\ref{sigma}~\&~\ref{filamentmap}).}
}
\label{Filamentmasses}
\normalsize
\end{table}

Table~\ref{Filamentmasses} shows the results of this analysis for the filaments within each of the models. For some models, the regions show evidence of multiple filamentary structures. This occurs when the density falls below our defined lower threshold such that the lowest defined contours enclose more than one filamentary structure. Note that we have not performed this analysis on Models II, XI, XII, XIII and XIV since they either do not exhibit filamentary structure within the region\footnote{We use the definition of a filament from \citet{Panopoulou2014} which states that filaments must have a greater than 3:1 aspect ratio.} or the maximum density within the simulation at the final time lies below our density definition of a filament. As shown in Table~\ref{Filamentmasses}, the mass contained within a filament tends to decrease dramatically as the degree of turbulence increases. This continues until a Mach value of 2. For higher turbulence values, the mass contained seems to remain relatively constant. Comparing the two ionisation profiles, we see that for simulations with Mach $<$ 2, the filaments within the SL models tend to have larger masses than the CR-only simulations. Above Mach 2, the masses contained within the filaments tend to be equal across the two different ionisation profiles. Comparing the two model sets (Models V \& VI vs. Models XV \& XVI) we see that the regions with higher mass-to-flux ratios have smaller filament masses. This is due to both the shorter collapse time and smaller collapse length scales associated with regions containing supercritical mass-to-flux ratios.

Table~\ref{coremasses} shows the results of this analysis for the cores (i.e., coherent regions with $A_{V}>7$ mag). Looking at the locations of the resulting cores, we see that the majority of the cores tend to form in regions 1 and 4. This indicates that although the other two regions have formed filaments, the density is not high enough to have started to form cores yet. Note that the cores in the Set 2 models do not coincide with any of the defined filament regions. From the resulting masses, we again see that, as the degree of turbulence increases, the mass contained within the cores decreases. Note that these masses are for the full core regions found within each simulation at the end of the run, however this does not mean that each core will collapse into the same mass star. Each core still requires further evolution before a star can be formed which could include further fragmentation into several smaller cores. Comparing the two ionisation profiles, we see that on average, the cores evolving within the SL models accumulate more mass than those in the CR-only models. This is in contrast to the findings of \citet{CO2014} who found that their core masses did not depend on the degree of ambipolar diffusion. Such a discrepancy could be due to the differences in model assumptions.

Comparing the masses from Set 2 to Models V and VI, we see that in all cases, the models with sub- and supercritical mass-to-flux ratios exhibit smaller filament and core masses. The smaller masses found within the Set 2 models is entirely due to the sub- or supercritical initial mass-to-flux ratio. As shown in the linear analysis of \citet{BB2012}, structures formed within regions with magnetic fluxes that are significantly larger or smaller than the mass result in smaller fragmentation length scales than in those regions where the mass and magnetic flux are comparable.  
 
\begin{table}
\small
\caption{Core Masses}
\centering
\begin{tabular}{lccc}
\hline\hline
Region & Peak location & Peak & Mass\\
Designation\tablefootmark{a} & (x,y) & ($A_{V}$) & ($M_{\odot}$)\\
\hline\hline
\multicolumn{4}{c}{Set 1}\\
\hline\hline
I-1,2a & -61.8, 18.9 & 10.18 &  39.66\\
\hline
II-4a & 24.8, 77.6 & 10.28& 3.66\\
\hline 
III-1a & 94.1, -42.5& 10.46 & 13.27\\ 
III-1b & -66.0, 12,2& 8.40& 9.76\\ 
III-1c & -76.8, -4.72& 8.27 & 10.63\\ 
III-4b & 32.3, 86.2& 7.84 & 1.53\\
\hline
IV-4a  &29.5, 83.5& 10.24 & 6.25\\  
\hline
V-4a   & 29.5, 83.6 &10.78 & 6.86\\  
\hline
VI-4a  & 29.5, 83.6 &10.58 & 7.92\\ 
\hline
VII-1a  & 100, -37.4 & 7.58 & 1.98\\  
VII-4a  & 30.7, 83.5& 10.02 & 6.76\\
\hline
VIII-1a  & 99.2, -38.6 &8.64 & 3.20\\  
VIII-1b  & -80.3, -13.0 & 7.50 & 1.98 \\ 
VIII-4a  & 31.1, 83.5& 10.60& 3.67\\ 
VIII-4b  & 33.9, 82.3 & 10.52& 3.97\\
\hline
IX-1a  & 99.6, -38.2& 9.69 & 6.38 \\ 
IX-1b  & -79.9, -11.4& 8.61 & 4.48\\ 
IX-1c  & -76.0, -0.79 & 7.70 &  0.79\\ 
IX-4a  & 31.5, 83.5& 11.55 & 9.90 \\
\hline
X-1a  & -79.5, -11.0& 8.30 & 2.75\\  
X-1b  & 100, -38.2 & 8.05& 2.75\\ 
X-1c  & -75.6, 0.39 & 7.58 &1.01 \\ 
X-4a  & 33.5, 82.7 & 10.04&2.14 \\
X-4b & 31.1, 83.6  & 9.17& 1.10\\
\hline\hline
\multicolumn{4}{c}{Set 2}\\
\hline\hline
XII-a & -36.6, 81.5 & 10.11 & 1.05\\
\hline
XIV-a & -95.6 -25.6 & 10.74 & 1.53\\
\hline
XV-1a & -79.1, -11.0 & 10.50 & 2.29\\
XV-1b & -74.8, 0.39 & 8.53 & 0.98\\
\hline
XVI-1a & -79.5, -11.8 & 11.10 & 2.73\\
XVI-1b & -74.8, 0.39& 9.21 & 1.16\\
\hline\hline
\end{tabular}
\tablefoot{
\tablefoottext{a}{The roman numeral indicates the model, the number indicates the filament region within which the core exists (if applicable) while the letter denotes the particular core listed in order of density. Cores in Set 2 models do not correspond to one of the defined filament regions in Figure~\ref{filamentmap}. }
}
\label{coremasses}
\normalsize
\end{table}

\begin{figure*}
\centering
\includegraphics[width = 0.25\textwidth, angle = -90]{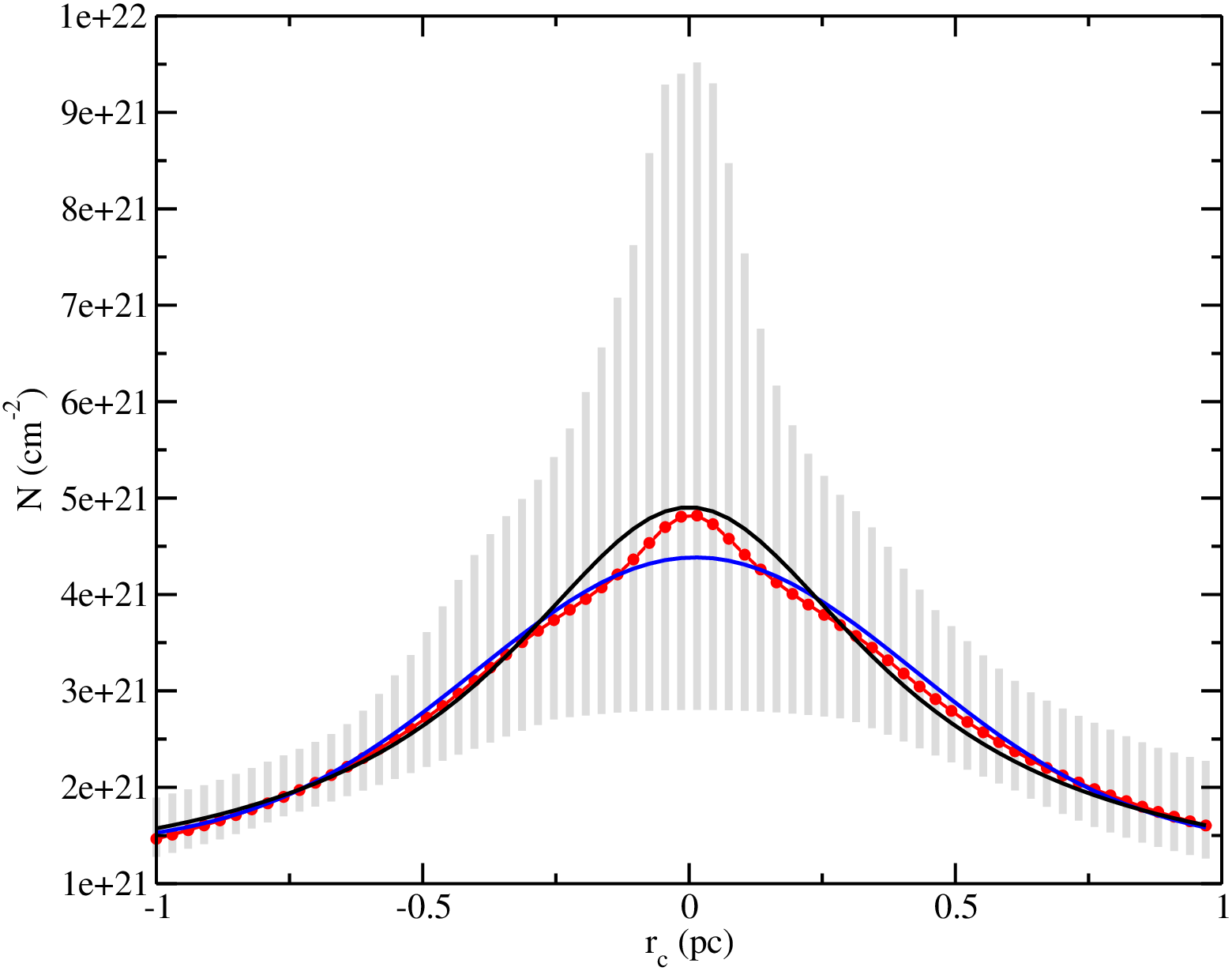}
\includegraphics[width = 0.25\textwidth, angle = -90]{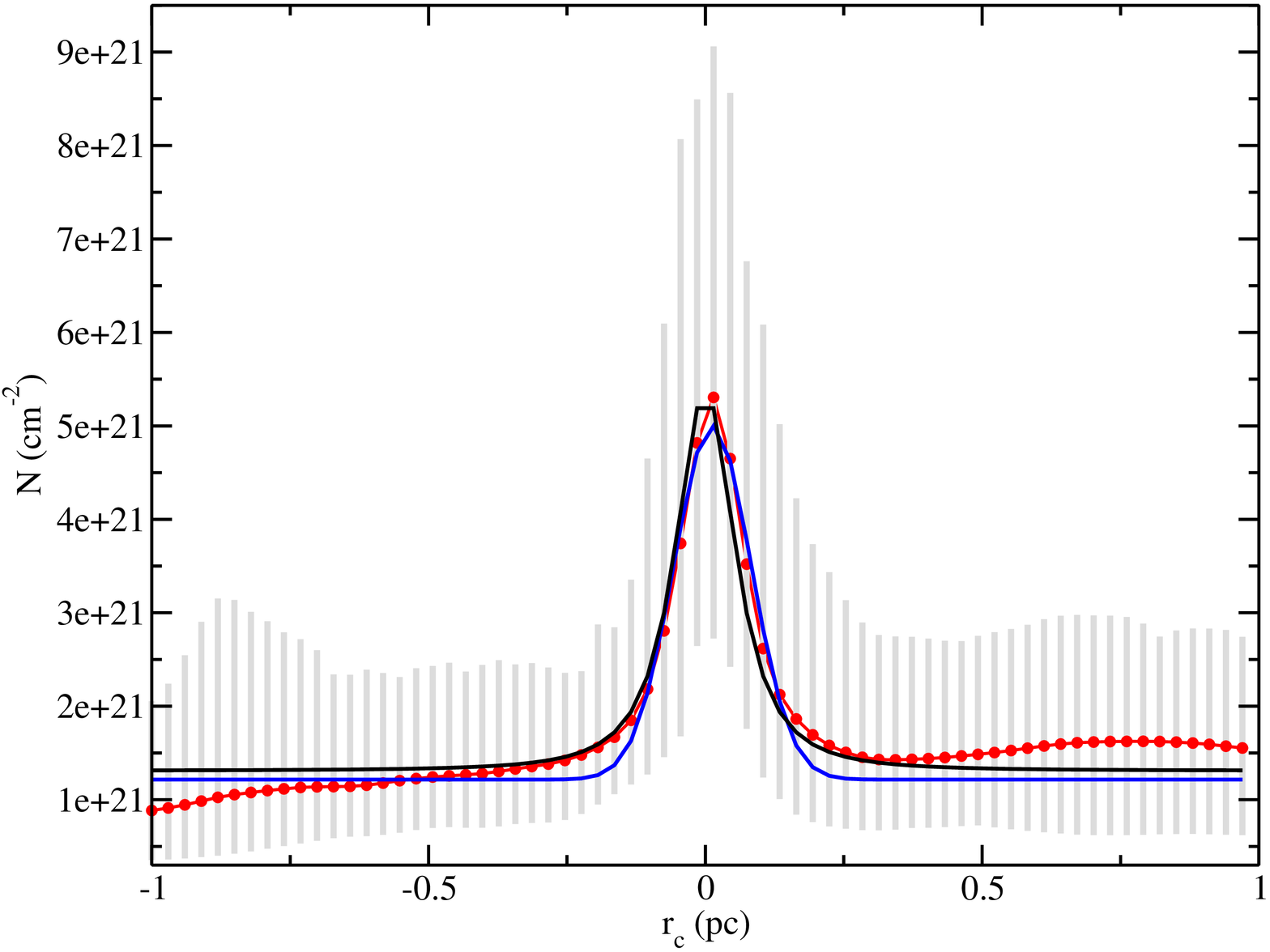}
\includegraphics[width = 0.25\textwidth, angle = -90]{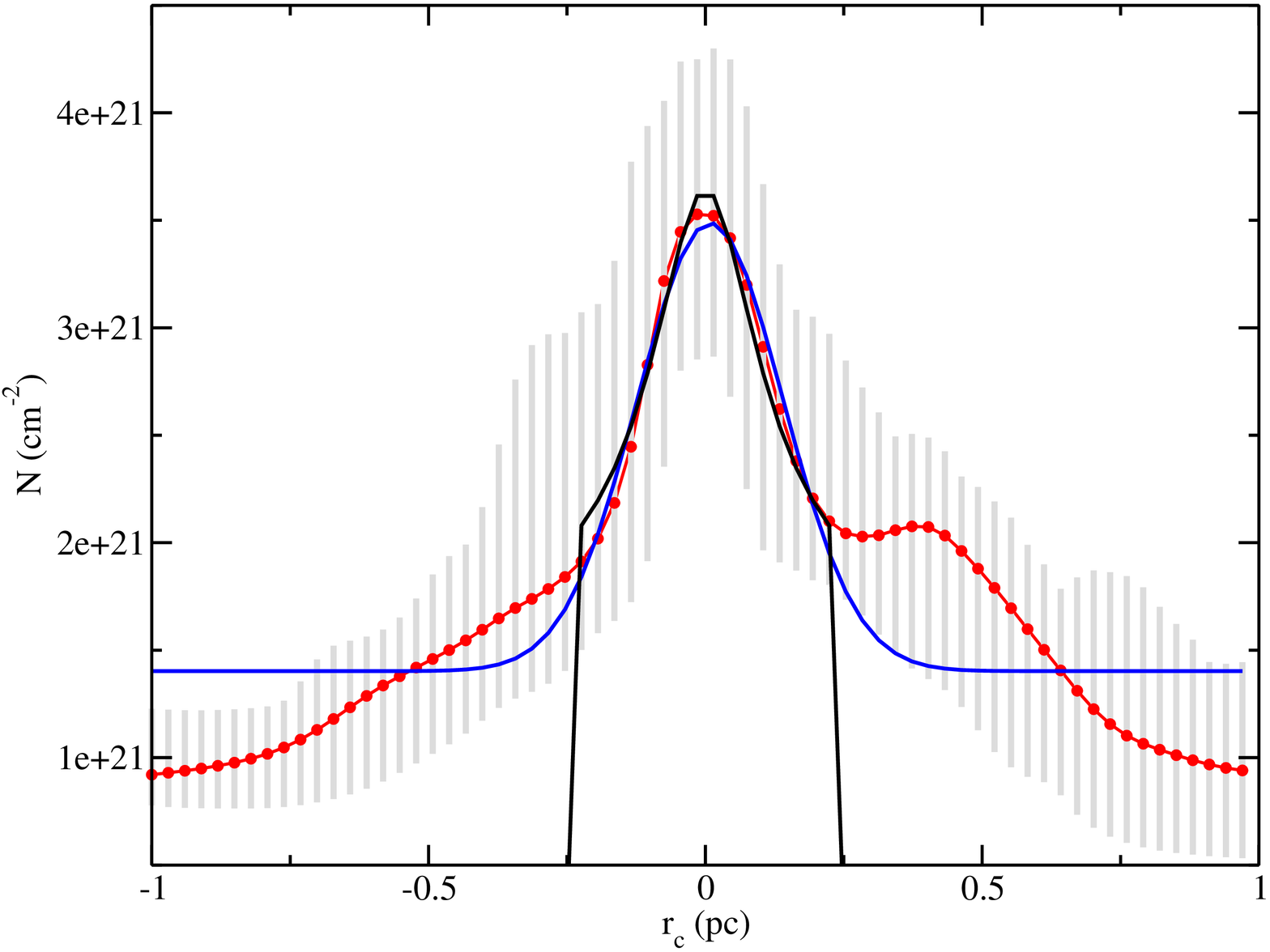}
\caption{Example filament profiles. Red curve shows the averaged data. Grey bars indicate the error range based upon the maximum and minimum density values at that particular radius. Black and blue curves show the best fit Plummer and Gaussian profiles respectively. Left and middle panels show examples of filament profiles for the two Mach number extremes (Left: Model I, Mach 0.03. Middle: Model IX, Mach 4, Filament 1) while the right panel shows an example of a model with complicated wing structures and truncated Plummer profile (Model VII, Mach 3, Filament 3). }
\label{profiles}
\end{figure*}

\subsection{Filament width as a function of Mach Number}
\label{fits}

The discovery of filaments pretty much everywhere within star forming regions has promoted significant efforts in determining their widths and overall structure within different environments.
Initial studies of the widths of filaments discovered through sub-millimetre dust continuum observations of the Gould Belt with the \textit{Herschel} space telescope tend to show a characteristic width of 0.1 pc \citep{Arz2011}. Studies of other regions such as the Taurus Molecular cloud (TMC) \citep{Malinen2012} and B211 \citep{Pal2013} also revealed this seemingly canonical value. Studies of Galactic cold core regions as chosen from the Cold Clump Catalogue of \textit{Planck} Objects (C3PO) by \citet{Juvela2012} found that some regions (e.g., G94.15+6.50 and G298.31-13.05) conformed to a characteristic width of 0.1 pc, however other regions revealed much larger widths of up to 1 pc. Other studies also find regions with filaments that do not exhibit a width of 0.1 pc. A search for filamentary structure in $^{13}$CO data of Taurus reveal a broad distribution of profile widths peaking at 0.4 pc \citep{Panopoulou2014}, observations of filaments within the Cygnus X region find a mean profile width between 0.26 and 0.34 pc \citep{Hennemann2012} and \textit{Herschel} observations of the Galactic plane via the Hi-GAL project find widths ranging from 0.1 to 2.5 pc \citep{Schisano2014}. Numerical studies have also found a wide range of filament widths. For example, \citet{Smith2014} finds a range of FWHM between 0.2 pc and 0.35 pc while \citet{Kirk2015} find filament FWHMs that range from 0.05 pc to 0.12 pc. \citet{Ntormousi2016} found a difference in the peak of the filament width distribution depending on whether the region assumed ideal or non-ideal MHD, however, without the inclusion of self-gravity in their models, they were unable to conclude whether this was due to the different filament selection or the physics inherent to filament formation. Theoretical considerations by \citet{Hocuk2015} of the Jeans length within a non-magnetic medium, where the gas and dust temperature is followed in detail, suggests that the length scale is on the order of 0.1 pc, while studies of both non-self-gravitating and self-gravitating filaments tends to suggest that dissipation via ion-neutral friction may be a promising mechanism to explain the structure of self-gravitating filaments \citep{Hennebelle2013,HA2013}. In addition to this, recent studies of the L1495/B213 regions in Taurus tend to suggest that large scale filaments may be composed of thread-like bundles \citep{Hacar2013}.  Given the discrepancy between all these different studies, it is only natural for us to analyse our own filaments to determine how the width varies with the Mach number and if our simulations agree/disagree with previous studies.

To determine the width of the filamentary structures within our simulations we utilise the fitting techniques of previous studies \citep[e.g.,][]{Arz2011,Smith2014,Kirk2015}. Within these techniques, one constructs an average profile across the filament and then fits this profile with a proper function from which a width can be extracted. The average profile is constructed by taking several density profile cuts along the length of the filament. The maximum of each profile is then shifted to coincide with each other and the average profile computed. For our simulations, we will examine the average profiles for the filaments found via CLUMPFIND2D listed in Table~\ref{Filamentmasses}. Our profiles, are constructed by taking cuts across the filament for each pixel along the length of the filament. We assume that the maximum of each profile within the average occurs at $r_{c}$ = 0, (where $r_{c}$ is the radius from the centre of the filament) and the profile is followed out to 1 pc from this maximum on both sides. We chose this range to both ensure that the entire width of the filament is represented within the average profile and to enable comparisons with previous studies that also adopt this profile extent.

As described by \citet{Arz2011}, \citet{Smith2014} and \citet[][among others]{Kirk2015}, the average profiles are fit by either a Plummer profile or a Gaussian profile. For our analysis, we will also fit these two profiles. We utilise the 2D surface density Plummer profile from \citet{Smith2014},
\begin{equation}
\Sigma(r) = \frac{A_{p}n_{c}R_{\rm flat}}{[1+(r/R_{\rm flat})^{2}]^{(p-1)/2}} + B~[\rm cm^{-2}],
\label{plummer}
\end{equation}
where $n_{c}$ is the central density, $R_{\rm flat}$ is the radius within which the profile is flat, $p$ is the slope of the power law decrease outside this radius and $B$ is the background density. Note that $p = 2$ corresponds to Bonnor-Ebert model clouds \citep[e.g.][]{DappBasu2009}. The $A_{p}$ value in the relation above is a finite constant normalisation factor based on an assumed distribution of filament inclinations with respect to the observer as given by 
\begin{equation}
A_{p} = \frac{1}{\cos~i} \int^{\infty}_{\infty} \frac{du}{(1 + u^{2})^{p/2}},
\end{equation} 
where $i$ is the inclination of the filament on the plane of the sky, generally assumed to be 0 \citep{Arz2011}. \citet{Arz2011} show that non-zero inclinations can affect the derived density by up to 60\% but have no effect on the shape of the profile, however given the two dimensional nature of our simulations, it is natural to assume an inclination angle of zero. For the cases where p = 2 or 4, with $i = 0$, $A_{p} = \pi/2$, which is the same value used by \citet{Smith2014}. For our analysis, we will also adopt this value for $A_{p}$ but still allow the value of $p$ to vary. Errors in the average profile are assumed to be equal to the maximum and minimum density values of the individual profiles at each distance from the maximum density of the filament. When these errors are taken into account in computing the profiles, the wings of the profile can sometimes prevent finding a good fit. To combat this problem, in these situations we truncate the average profile to only include the main density peak centred around $r_{c}$. This ensures that we are only fitting the primary filament within each region and that the widths are not contaminated by other nearby filaments.  

Our Gaussian function takes the form
\begin{equation}
\Sigma(r) = A_{g}~\exp\left(\frac{-(r-r_{0})^{2}}{2\sigma^{2}}\right),
\label{gaussian}
\end{equation}
where $A_{g}$ is the Gaussian amplitude, $r_{0}$ is the mean of the Gaussian (which by definition should be zero for all of the profiles) and $\sigma$ is the dispersion of the Gaussian. 

\subsubsection{Results}

\begin{table}[!ht]
\small
\caption{Filament Profile Parameters}
\centering
\begin{tabular}{lcccc}
\hline\hline
& Gaussian  & \multicolumn{3}{c}{Plummer}\\
\cline{2-5}
Filament & FWHM & $p$ & $R_{\rm flat}$ & $n_{c}$\\
Designation & (pc) &  & (pc) & ($\times 10^{3}$ cm$^{-3}$)\\
\hline\hline
\multicolumn{5}{c}{Set 1}\\
\hline\hline
I-1,2  & 0.97 & 2.83 & 0.41 & 2.0\\
\hline
III-1 & 0.63 & 3.41 & 0.25 & 4.2\\ 
III-2 & 0.73 & 2.68 & 0.30 & 2.2 \\
III-3 & 0.79 & 2.50 & 0.35 & 2.1 \\ 
III-4 & 0.62 & 3.96 & 0.25 & 3.9\\
\hline
IV-1a & 0.42 & 3.16 & 0.15 & 3.5\\  
IV-2  & 0.53 & 2.32 & 0.11 & 5.0\\ 
IV-3  & 0.69 & 2.59 & 0.22 & 2.9\\ 
IV-4  & 0.29 & 2.40 & 0.11 & 9.5\\
\hline
V-1a & 0.35 & 2.35 & 0.10 & 6.4\\
V-2  & 0.44 & 3.74 & 0.26 & 2.1\\ 
V-3  & 0.68 & 2.54 & 0.15 & 3.6\\ 
V-4  & 0.22 & 3.06 & 0.11 & 10.2\\
\hline
VI-1a & 0.25 & 2.77 & 0.008 & 10.2\\ 
VI-1b & 0.30 & 2.90 & 0.13 & 4.9\\
VI-2a & 0.34 & 2.50 & 0.11 & 4.5\\ 
VI-2b & 0.32 & 2.24 & 0.10 & 5.0\\
VI-3  & 0.55 & 2.89 & 0.20 & 2.7\\ 
VI-4  & 0.21 & 3.86 & 0.12 & 6.4\\
\hline
VII-1a & 0.21 & 3.96 & 0.10 & 8.6\\  
VII-1b & 0.24 & 2.80 & 0.12 & 6.0\\
VII-2a & 0.25 & 3.57 & 0.10 & 5.1\\
VII-2b & 0.33 & 2.78 & 0.10 & 5.8\\
VII-3  & 0.31 & 2.37 & 0.10 & 4.5\\ 
VII-4  & 0.24 & 2.66 & 0.07 & 12.7\\
\hline
VIII-1a & 0.18 & 3.99 & 0.09 & 10.7\\
VIII-1b & 0.19 & 2.13 & 0.05 & 19.1\\    
VIII-2a & 0.22 & 3.74 & 0.10 & 6.2\\
VIII-2b & 0.24 & 3.37 & 0.10 & 6.1\\
VIII-3  & 0.29 & 2.60 & 0.10 & 4.8\\ 
VIII-4  & 0.22 & 2.75 & 0.06 & 15.0\\
\hline
IX-1    & 0.16 & 3.71 & 0.08 & 10.8\\
IX-2    & 0.22 & 3.64 & 0.10 & 7.1\\ 
IX-3    & 0.26 & 2.90 & 0.12 & 4.9\\ 
IX-4    & 0.21 & 3.30 & 0.09 & 9.9\\
\hline
X-1a    & 0.16 & 2.56 & 0.05 & 17.8\\  
X-1b    & 0.14 & 2.97 & 0.05 & 19.1\\
X-2a    & 0.18 & 3.10 & 0.07 & 10.4\\ 
X-2b    & 0.22 & 3.67 & 0.12 & 5.0\\ 
X-3     & 0.23 & 3.49 & 0.09 & 5.2\\ 
X-4     & 0.18 & 1.98 & 0.06 & 14.6\\
\hline\hline
\multicolumn{5}{c}{Set 2}\\
\hline\hline
XV-1a & 0.18 & 2.84 & 0.08 & 9.4 \\
XV-1b & 0.13 & 2.56 & 0.03 & 25.5\\
XV-2a & 0.24 & 3.49 & 0.10 & 5.5 \\
XV-2b & 0.22 & 2.53 & 0.07 & 8.0\\
XV-3  & 0.23 & 3.44 & 0.10 & 5.0\\
XV-4a & 0.17 & 3.52 & 0.13 & 4.2\\
XV-4b & 0.11 & 3.34 & 0.11 & 7.5\\
\hline
XVI-1a & 0.17 & 2.87 & 0.08 & 9.4 \\
XVI-1b & 0.13 & 2.39 & 0.04 & 20.2\\
XVI-1c & 0.19 & 2.46 & 0.07 & 8.1 \\
XVI-2a & 0.21 & 3.09 & 0.08 & 7.0 \\
XVI-2b & 0.20 & 3.70 & 0.1 & 4.9 \\
XVI-3  & 0.22 & 3.54 & 0.1 & 5.0 \\
XVI-4a & 0.15 & 3.18 & 0.07 & 6.5 \\
XVI-4b & 0.17 & 3.22 & 0.07 & 9.4 \\
\hline\hline
\end{tabular}
\label{Filamentwidths}
\normalsize
\end{table}

\begin{figure}
\centering
\includegraphics*[width = 0.37\textwidth, angle=-90]{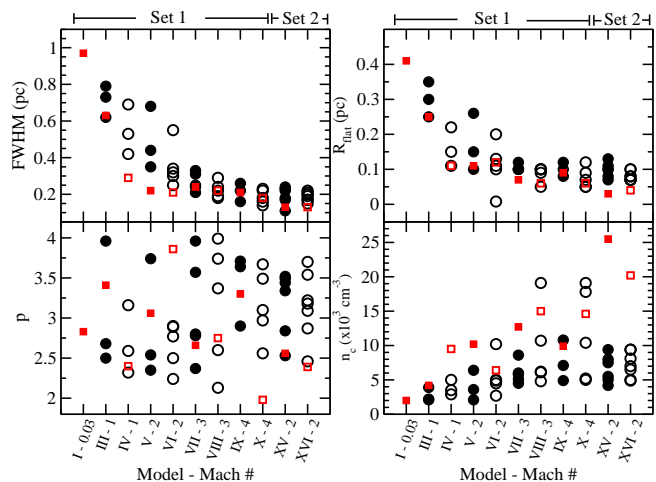}
\caption{Fitted values as a function of model for both Gaussian and Plummer profiles. Red squares denote the filaments which contain the densest region which stopped the simulations. Closed symbols depict SL models while open symbols depict CR- only models. X-axis labels indicate the model number (roman numerals) and associated Mach number. Top left: Gaussian FWHM. Bottom left: Plummer $p$ parameter. Top right: Plummer $R_{\rm flat}$ parameter. Bottom right: Central filament density as calculated through Plummer profile. Recall that the resolution limit for the models is 0.0296 pc (1 pixel).}
\label{fittedvalues}
\end{figure}

For this analysis, we attempt to fit the above two profiles to all of the filaments identified in Table~\ref{Filamentmasses}. Figure~\ref{profiles} shows representative examples of three of the fits. The left and middle panels show the examples of fits for the lowest and highest Mach values, Model I (Mach 0.03) and Model IX (Mach 4) respectively. The right hand panel shows an example of a profile with a complicated wing structure (Model VII, Mach 3, Filament 3), for which we show an example of a truncated Plummer profile. In all panels, the red curve shows the average profile, the grey bars indicate the error range based upon the maximum and minimum density values at that particular radius from the centre of the filament. The blue curve depicts the Gaussian fit while the black curve depicts the Plummer profile. The results of all the fits can be found in Table~\ref{Filamentwidths}. Note that filaments IV-1b and V-1b are missing from the table because it was not possible to find a satisfactory fit for either the Gaussian function or Plummer profile. Figure~\ref{fittedvalues} shows the results from Table~\ref{Filamentwidths} in a visual form. Closed symbols depict the SL models while open symbols depict CR-only models. The red squares denote the filament which contains the region with the maximum density within the simulation. X-axis labels indicate the model number (roman numerals) and associated Mach number.  The top left plot shows the FWHM of the Gaussian fit as a function of model. The other three panels show the Plummer profile parameters: $p$ (bottom left), $R_{\rm flat}$ (top right) and $n_{c}$ (bottom right). Focussing on the upper left panel first, we find that there is a scatter in the FWHM between the filaments within each model, however the overall FWHM tends to decrease as a function of Mach number. Above Mach 2, the scatter within a single model decreases significantly. Looking at the filaments which contain the highest density region (red squares) we see that these filaments tend to have the lowest FWHM amongst all the filaments within a particular model for Mach values less than 2. Above Mach 2, these regions no longer exhibit the smallest FWHM.

Looking at the three panels for the Plummer profile parameters, we see that there does not seem to be a discernible trend in the fitted $p$ value, however we see that for increasing Mach number, the flat region of the Plummer profile ($R_{\rm flat}$) decreases in size while the central density of the filament tends to increase. Looking at the right panels, we see that the filaments which contain the densest regions within the simulations tend to have the smallest central flat Plummer profile regions (upper right panel) and for the most part, the largest central densities (lower right panel). Finally, comparing the two sets of models (Models V \& VI vs Models XV \& XVI) we see that the larger mass-to-flux ratio results in overall smaller FWHM and $R_{\rm flat}$ values. Looking at the $p$ values, we see that their range is still scattered, however the filaments that contain the densest region of the simulation have the smallest $p$ values. Finally, the models with the higher mass-to-flux ratios have larger central filament densities, particularly for the densest regions in the simulations (red squares).

\subsubsection{Comparison to Previous Studies}

\noindent\textbf{Computational}
\\

In the past couple of years since the discovery of the seemingly characteristic width of 0.1 pc for filaments within star forming regions \citep{Arz2011}, several studies investigating the origin of this value \citep[e.g.,][among others]{Smith2014, Kirk2015,Federrath2016} have been published. Here we will compare the results of our filament analysis to these three studies. 

Before comparing, however, we first note the similarities and differences between all six simulations. Table~\ref{comp} shows the important defining features of each of the six set ups in an easy to compare manner. For papers which include various physical set ups, (e.g., driven and decaying turbulence, HD and MHD etc.) we will only compare, if possible, to the simulations and results that closest match our set up (i.e., simulations with decaying turbulence and/or magnetic fields). As shown in Table~\ref{comp}, the most notable differences between the four different studies are the absence of magnetic fields in \citet{Smith2014}, the absence of ambipolar diffusion in all three \citep{Smith2014,Kirk2015, Federrath2016}, the absence of a treatment for chemistry in \citet{Federrath2016} and the two-dimensional nature of our simulations compared to the others. In addition, note that both our simulations and those of \citet{Smith2014} assume an initial cloud that is 7.5 times larger than the clouds in the other two.

\begin{table*}
\small
\caption{Filament study comparison computational features}
\centering
\begin{tabular}{lcccc}
\hline\hline
 Study\tablefootmark{a} &   This work  & S14  & K15  & F16 \\
 \hline
Code Used\tablefootmark{b} &  IDL code &  {\sc AREPO} & {\sc FLASH} v2.5 & {\sc FLASH} v4 \\
Dimensions                   &   2  &     3   &    3    &   3  \\
MHD                              &   Y  & N & Y & Y \\
Ambipolar diffusion        & Y &  N & N  & N\\
Initial Mach \#                & 0.03 - 4 &  12 & 6.01 & 5\tablefootmark{c}\\
B field strength ($\mu$G) & 2.14 or 1.18 & 0 & 56.7 or 120.5 & 10\\
Gravity                          & Y & Y & Y &  Y\\
Chemistry                      & Y\tablefootmark{d}  & Y\tablefootmark{e}& Y\tablefootmark{f} & N \\
Mass (M$_{\odot}$)      & 4000 & 10000 & 500 or 2000  & 388 \\
Temperature (K)            &10    & 10\tablefootmark{g}  & 10   & 10\\
Box Size (pc)                &15  & 15\tablefootmark{h}& 2 & 2 \\
Resolution                     & 512$^{2}$ & ...\tablefootmark{i} & 8195$^{3}$\tablefootmark{j}  & 1024$^{3}$\\
Pixel size (AU)              & 5987  & 1013\tablefootmark{k} & 50\tablefootmark{l} & 395\\
\hline\hline
\end{tabular}
\tablefoot{
\tablefoottext{a}{Paper citations for study labels: S14 = \citet{Smith2014}, K15 = \citet{Kirk2015}, and F16 = \citet{Federrath2016};}
\tablefoottext{b}{References for codes are \citet[][IDL code]{BC2004}, \citet[][ARPEO]{Springel2010}, and \citet[][FLASH v2.5, FLASH v4]{Fryxell2000}; }
\tablefoottext{c}{driven;}
\tablefoottext{d}{via ionisation profiles;}
\tablefoottext{e}{Hydrogen chemistry of \citet{GM2007a,GM2007b} together with a highly simplified treatment of CO formation and destruction introduced in \citet{NL1997};} 
\tablefoottext{f}{gas cooling by dust and molecular lines from \citet{Banerjee2006};}
\tablefoottext{g}{Average value used to calculate Jeans length;}
\tablefoottext{h}{Setup is a 7 pc radius sphere of cold gas embedded within a larger 65 pc periodic box of tenuous warm medium with temperature of several thousand Kelvin \citep{Smith2014};}
\tablefoottext{i}{Code uses Adaptive Mesh Refinement (AMR) for which resolution is not based on cell sized but rather mass within a cell. Base resolution based on this criteria is $10^{-2}$ M$_{\odot}$ ;}
\tablefoottext{j}{Code uses AMR. This value is based on smallest cell size;}
\tablefoottext{k}{Value quoted here is the radius with their standard refinement scheme;}
\tablefoottext{l}{Refers to smallest pixel size within the simulation}.
}

\label{comp}
\normalsize
\end{table*}

The two parameters related to the filament width are the FWHM and $R_{\rm flat}$ for the Gaussian and Plummer profiles, respectively. As defined in \citet{Arz2011} the FWHM from the Gaussian profile approximately corresponds to 3 times $R_{\rm flat}$ for $p = 2$. By fitting the Gaussian function to the profiles, \citet{Smith2014} found average FWHM values ranging from $\sim$ 0.2 pc to 0.35 pc depending on the radius from the centre of the filament considered (0.35 pc vs 1.0 pc) while \citet{Kirk2015} find FWHM values between 0.05 pc and 0.12 pc.  For our simulations, we find that the FWHM decreases as function of Mach number with a range between 0.11 pc and 1 pc. \citet{Federrath2016} does not quote FWHM values, however, they do find that they are consistent with the 3$R_{\rm flat}$ definition. Looking at this criteria for our simulations, as well as those by \citet{Smith2014} and \citet{Kirk2015}, we see that the majority of the models do not satisfy this relation. This is due to the fact that in these analyses, the value of $p$ was not fixed to 2 and thus the relation does not hold. 

For the extent of the flattened inner region of the filament ($R_{\rm flat}$), \citet{Smith2014} find an average value of $\sim$ 0.07 pc, \citet{Kirk2015} find a range of values from 0.006 pc to 0.03 pc depending on whether the simulation included magnetic fields or not. \citet{Federrath2016} find $R_{\rm flat}$ values of 0.05 to 0.1 for non-magnetic and magnetic simulations respectively, however, note that their fits are restricted to the inner 0.2 pc of the filament profile. Looking back at Figure~\ref{fittedvalues} we see that our simulations with the highest Mach values tend to agree with the values found by \citet{Smith2014} and \citet{Federrath2016}, but are larger than those reported by \citet{Kirk2015}.

Comparing the other two Plummer profile parameters, we see that \citet{Smith2014} find an average power law index ($p$) of 2.2 while \citet{Kirk2015} quote a range of 1.3 - 2, both of which are within the range of 1.5 - 2.5 quoted by \citet{Arz2011}. \citet{Federrath2016} set their power law index value to 2 for all fits. For our simulations, we find a range between $\sim$ 2 to 4 with an overall average of 3.00. If we split the fitted filaments into those evolving in the SL and CR-only environments, we find average $p$ values of 3.10 and 2.95 respectively. Finally, looking at the central density of the filament, the fits of our simulations yield a maximum value of only $2.55\times10^{4}$ cm$^{-3}$ which is up to 3 orders of magnitude smaller than the values quoted by \citet{Smith2014} and \citet{Kirk2015} ($1.07\times10^{5}$ cm$^{-3}$ and $10^{4} - 10^{7}$ cm$^{-3}$, respectively).  \citet{Federrath2016} does not give results for this parameter. 

Looking back at Table~\ref{comp}, we also note that out of all four studies, our simulations exhibit the poorest resolution while those of \citet{Kirk2015} exhibit the best resolution with the simulations from \citet{Smith2014} and \citet{Federrath2016} falling in between. Therefore, it is not surprising that \citet{Kirk2015} simulations exhibit the smallest FWHM and $R_{flat}$ while ours exhibit some of the largest. The reason for discrepancies in the fitting parameters can be attributed to  the resolution of the simulations, the evolutionary time frame and maximum achieved densities within our simulations. That is, our simulations represent the early phases of cloud/filament formation compared to \citet{Smith2014} and \citet{Kirk2015}. 
\\

\begin{figure}
\centering
\includegraphics*[width = 0.37\textwidth, angle=-90]{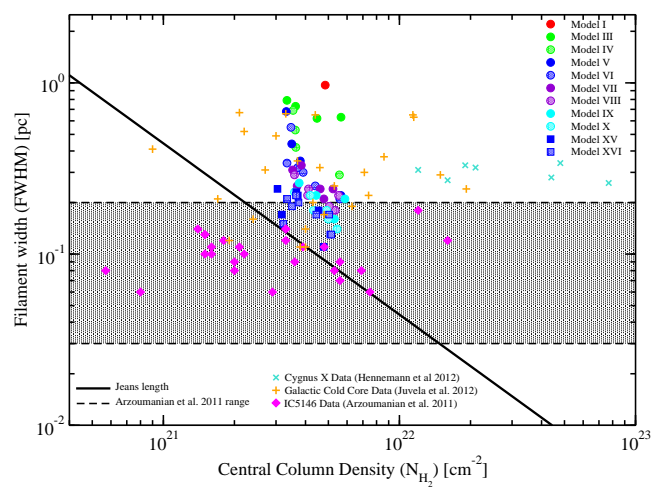}
\caption{Filament width (FWHM) as a function of central column density ($N_{H_{2}}$) for all models analysed. Colours depict the particular model as indicated by the legend. Filled symbols depict SL models while shaded symbols depict CR-only models. Circles indicate models with an initially transcritical mass-to-flux ratio ($\mu_{0} = 1.1$) while squares indicate models with and initially super-critical mass-to-flux ratio ($\mu_{0} = 2.0$). Also included are filament widths based on observations of IC5146 \citep[][diamonds]{Arz2011}, the Galactic cold core survey \citep[][pluses]{Juvela2012} and the DR21 filament within the Cygnus X region \citep[][X's]{Hennemann2012}.The solid black line depicts the computed Jeans length as a function of column density [$\lambda_{J} = c_{s}^{2}/(G\sigma_{0})]$] for a gas temperature of 10 K. The shaded area within the dashed lines indicates the region occupied by the filaments analysed in \citet{Arz2011}.  The errors in our FWHM values are too small to be seen on the plot.}
\label{fwhmvssigma}
\end{figure}

\noindent\textbf{Observational}
\\

As discussed at the beginning of Section~\ref{fits} there have been several observational studies of various regions with the intent of determining the width of the filaments found within. Figure~\ref{fwhmvssigma} shows the deconvolved filament width as a function of the central column density for our simulations combined with culled data from surveys of IC5146 \citep[][diamonds]{Arz2011}, Galactic Cold Cores \citep[][pluses]{Juvela2012} and the DR21 filament in the Cygnus X region \citep[][X's]{Hennemann2012}. The coloured symbols depict the particular models as indicated by the legends. For our model data, filled symbols depict SL models while shaded symbols depict CR-only models. Circles indicate models with an initially transcritical mass-to-flux ratio ($\mu_{0} = 1.1$) while squares indicate models with an initially super-critical mass-to-flux ratio ($\mu_{0} = 2.0$). The solid black line depicts the computed Jeans length as a function of column density [$\lambda_{J} = c_{s}^{2}/(G\sigma_{0})$] for a gas temperature of 10 K.  The shaded area within the dashed lines indicate the range of filaments widths in the Aquila and Polaris regions as depicted in \citet[][Figure 7]{Arz2011} for which no data were provided. 

From this figure we note a few interesting trends within our model data itself. First, the scatter in the measured widths tends to decrease as the central column density increases although both Models I and III show larger widths even at higher column densities. Comparing the blue coloured symbols, we see again that the filaments forming within supercritical mass-to-flux ratio regions (squares) exhibit smaller widths than those that are evolving in transcritical regions (circles). Finally, we see that the overall width tends to decrease as the degree of turbulence increases. Comparing our data to the expected Jeans length at 10 K (solid line), we see that our simulations tend to exhibit widths that are much larger than the Jeans length, however a few of the super-critical filaments do come close. Comparing our filament widths to the three observed regions, we see that our simulations tend to agree the most with those of the Galactic cold core survey \citep{Juvela2012}, however given the earlier stage of evolution that our simulations represent, no direct conclusion can be drawn. 

Looking at the figure as a whole however, we see that there exists a wide range of filament widths and corresponding central densities among the observed regions, particularly within the cold core observations of \citet{Juvela2012}. This is to be expected given the fact that these observations were performed over a wide range of points along the Galactic plane and therefore examine varying evolutionary environments between the filaments. The DR21 filament within Cygnus X is thought to be the convergence point of several filaments where high mass star formation is likely occurring \citep{Hennemann2012}. This easily accounts for the significant difference between the densities observed for this region as compared to the others regions and our simulations, all of which are assumed to be undergoing low mass star formation. Finally, the parameter space covered by the shaded area covers filaments within Aquila, IC5146 and Polaris \citep{Arz2011} and depicts regions of high, moderate and little to no star formation, respectively. Looking back at the original figure from \citet[][Figure 7]{Arz2011}, we see that there seems to be an anti-correlation between the degree of star formation and the filament width, i.e., very active star forming regions show larger filament widths than regions with little to no active star formation.  
\\

\section{Discussion}

The goal of this paper has been to determine the effect of different ionisation profiles on the formation of structures and cores within a turbulent magnetised medium. In addition, we explored the effect of varying the strength of the turbulence and the initial mass-to-flux ratio on the timescale and extent of the structures formed. To this end, we performed two sets of simulations which explored these various effects. The following subsections discuss the main results found through the analysis.

\subsection{Effect of Ionisation Profile}

The main effect explored by the simulations in this paper are those of the two ionisation profiles on the formation of structures in a turbulent medium. From our analysis, we found that the ionisation profile does not seem to matter structurally for Mach values greater than or equal to 2. As depicted in Figure~\ref{sigma}, the structures present in Models V and VI are virtually indistinguishable. Upon close inspection, the filaments in Model V (SL ionisation profile) are slightly puffed up compared to those in Model VI (CR-only ionisation profile) however the difference is too small to see without overlaying the two models. We conclude that for highly turbulent driven flows, approximate flux-freezing holds, and there is not enough time for ambipolar diffusion to affect the evolution, even in the CR ionisation cases.

Looking at the density enhancement maps for the final times as shown in Figure~\ref{sigma}, it is easy to see a pattern with increasing Mach number that seems to depend on the choice of ionisation profile. For the SL simulations, it appears that as the degree of turbulence increases, the size of the final structures formed decreases from a monolithic single filament in Model I ($v_{a}/c_{s} = 0.03$) to thinner, more fragmented filaments in Model V ($v_{a}/c_{s} = 2$). Conversely, for the CR-only simulations, the opposite trend is observed i.e., the final size of the structures increases from small core like structures in Model II to long filamentary structures in Model VI. However, when we look at the evolution of each individual model over time, such a pattern does not always hold up. Instead, all models tend to have an initial monolithic filament form at lower density values within which the final density structure forms, whether it be a continuation of a monolithic structure as in Model I, or thinner, more fragmented filamentary structures as observed in Models III - VI. In addition to the differences in the density enhancement structure, we also looked at the mass-to-flux ratio structure. Comparing the two ionisation profiles, we found that SL simulations show very little deviation from the initial mass-to-flux ratio while CR simulations allow for a more significant increase within the medium. Specifically, the nearly flux frozen state of the low density gas in SL simulations results in an amplification in the magnetic field enhancement on the outskirts of the core in comparison to the CR simulation. Although the simulations show that the different ionisation profiles do not imply structural differences for models with Mach number greater than two or a supercritical mass-to-flux ratio, this may not hold true within the kinematics of the region. These possible differences will be explored in the second paper of this series. 

\subsection{Effect of Mach Number}

The models in Set 1 (see Table~\ref{models}) explore the effect of changing the strength of the injected turbulence via the Mach number. Here we found that as the turbulent Mach number increases, the structure within the mass-to-flux ratio maps becomes more complicated. In addition, we find that the value of the Mach number has an effect on the physical properties of the structures themselves. Looking at the mass analysis, we find that as the Mach number increases, the mass contained within a particular structure decreases. Comparing corresponding models for the two ionisation profiles, we see that for Mach values greater than 2, there is not a significant difference between the derived masses of the structures.

The fitting analysis tends to agree with all of the effects mentioned above. Specifically, the trends in the FWHM and $R_{\rm flat}$ parameters shown in Figure~\ref{fittedvalues} agree with the assertion that there is little structural difference between the SL and CR simulations once the Mach number rises above 2. The fitting also shows that as the Mach number increases, the central density of the filaments also increases, however one must remember that the value of this parameter is solely a result of the fitting of Equation~(\ref{plummer}). Based on our fitting analysis, we conclude that the values of the fitted parameters could give some insight into the degree of turbulence within a region assuming that the magnetic field strength (or mass-to-flux ratio) is comparable to the value used in our simulations. Kinematics studies will be needed to better quantify the consistency of our model predictions with observations. 
 
\subsection{Effect of Initial Mass-to-Flux ratio}

The models in Set 2 in combination with Models V \& VI from Set 1 (see Table~\ref{models}) explore the effect of changing the initial mass-to-flux ratio on the formation of structures within a turbulent medium. Through these eight models, we were able to look at both a highly and moderately subcritical medium, a transcritical medium and supercritical medium evolving within either a SL or CR-only ionisation profile. The main effect of the different initial mass-to-flux ratio observed within these simulations is on the structures formed. For the subcritical simulations (Models XI to XIV), the SL ionisation profile simulations are not able to form a self-gravitating core, even with the added turbulence.  Conversely, the CR-only models are able to overcome the strong magnetic field associated with the subcritical mass-to flux ratio. This is a direct result of the low density regions of the SL simulations being held in a nearly flux frozen state within a strong magnetic field, while the ambipolar diffusion within the CR-only simulation still allows for the neutral particles to slip past the field lines even with a strong magnetic field. From these simulations, we found that the main effect of a higher mass-to-flux ratio is to form thinner filaments. This is evident in the filament profile fits performed in Section~\ref{fits}. 

Based on all of the results presented in this subsection and the previous two, we can say that the underlying ionisation and turbulent content can affect the density structure when turbulence is below Mach 2. In a future paper, the velocity field in the different models will be analysed by using the simulations to create synthetic spectra of optically thin low- and high-density tracers. In this way, we can quantify whether different ionisation profiles (i.e., different chemistries) can result in observable differences in clouds with apparently similar structures which can only be observed in dust continuum emission.

\subsection{Filaments and Fibres}

Observations of the Taurus L1495/B213 complex have shown the existence of filaments both with and without core formation \citep{Hacar2013}. Comparing to these observations, we see that the density structure of the filaments is best reproduced by our models 
with high mass-to-flux ratio ($\mu_{0} = 2.0$) and/or at least mildly supersonic turbulence. These observations have also shown the existence of fibres within the macro filamentary structure \citep{Hacar2013}. They also found that fibres could either be fertile or sterile, that is they either show evidence of core formation or they don't. Based on these observations, two possible scenarios for the formation of such structures have been proposed. First, \citet{TH2015} propose a ``Fray and Fragment'' scenario where a macro filament forms then frays into smaller fibres which then fragment into cores. The second scenario, ``Fray and Gather'', is based upon turbulent hydrodynamic simulations from \citet{Smith2016}. In their simulations, they found that the medium tends to fray into fibres rather than forming a macro filament. These fibres are then gathered into bundles by the large scale motions of the medium. Looking at the temporal evolution of our simulations (not shown), we find that they tend to support the ``Fray and Gather'' scenario, however there are a few exceptions, specifically Models I and II. In Model I, as discussed before, the medium undergoes monolithic collapse into a single large filament, which would be the first step of the ``Fray and Fragment'' scenario, however we do not see any evidence of fibres or multiple cores within the filament. Looking at the evolution of Model II, we see that it forms tenuous, large scale filaments at low densities which fragment into core-like structures with no evidence of fibres, thus skipping the ``Fray'' stage altogether. 

An interesting and somewhat unexpected result from the second set of simulations come from the two subcritical SL models (Models XI and XIII). Although these simulations did not collapse into high density filaments or cores, the evolution of the medium over the course of the simulation shows the formation of thin transient filamentary structures that are similar to those observed by \citet{Hacar2013}. Looking at the evolution of these simulations over time (not shown), we see that they form thin filaments that are gathered together by the bulk motions, as in the ``Fray and Gather'' scenario of \citet{Smith2016}, however these gatherings are transient, thus not allowing for the formation of cores or chains of cores to develop along these fibres. This could be the reason behind the fertile and sterile fibres as defined by \citet{Hacar2013}. \textit{Regions with fertile fibres likely indicate a trans- or supercritical mass-to-flux ratio within the region while the sterile fibres are likely subcritical in nature}. In addition, the evolution of our subcritical simulations indicate that these sterile fibre bundles are likely transient in nature; forming and dispersing over the course of a couple Myr. In the following kinematics paper, we will study these spatially resolved fibres to see if they will show up as different kinematic features when observed along the plane of the sheet.

\section{Summary}

Based on the preceding analysis and discussion, we summarise several of the results below:

\begin{itemize}

\item The addition of turbulence within the simulations allows for the formation of filaments within the medium. The degree of turbulence required to form filamentary structure within the simulations depends on the ionisation profile of the medium. The step-like profile requires very minimal turbulence (i.e., Mach 0.03; Model I) while the CR-only simulations require larger degrees of turbulence (e.g., Mach $\sim 1$; Model IV).  

\item Above Mach 2, the effect of the ionisation profile on the density structures formed becomes minimal. This suggests that above a certain turbulent threshold, ambipolar diffusion is not occurring sufficiently to make a dynamical difference in the magnetic forces. 

\item Analysis of the mass-to-flux ratio maps reveals a patchy distribution of high and low values within the CR-only simulations. The patchiness was found to depend on the degree of turbulence with higher turbulent velocities resulting in finer structures. 

\item Analysis of the mass-to-flux ratio structures in the vicinity of the cores reveal a unique structure with high values that coincide with the high density regions and significantly lower values just outside. This structure is elongated along the direction of the filament. Profiles across this feature are sharply peaked for the SL ionisation profile which would provide a unique signature within star forming regions. The profiles found in CR-only simulations show the same peak at the high density core however due to the lower ionisation fraction, the profiles are not as sharp as in the SL simulations.

\item Variation of the initial mass-to-flux ratio shows that the CR-only simulations are able to form self-gravitating cores within a subcritical ($\mu < 1$) region while SL simulations are unable to even with the addition of turbulence. The supercritical ($\mu > 1$) simulations yield thinner filaments than the equivalent simulation with a transcritical mass-to-flux ratio ($\mu \gtsimeq 1$) .

\item Observed filaments of \citet{Hacar2013} and \citet{TH2015} are best reproduced by models with supercritical mass-to-flux ratios and/or at least mildly supersonic turbulence.

\item Analysis of the temporal evolution within the subcritical SL simulations (Models XI and XIII) showed the formation of thin filamentary fibres within the quiescent medium that are similar to those observed by \citet{Hacar2013}. Analysis of the formation of these fibres within our simulations tend to favour the ``'Fray and Gather'' scenario of \citet{Smith2016} as opposed to the ``Fray and Fragment'' scenario proposed by \citet{TH2015}. 

\item Our simulations suggest that fertile filaments observed by \citet{Hacar2013} may exist in regions with super- or transcritical mass-to-flux ratios. Conversely, our simulations suggest that the sterile filaments observed by \citet{Hacar2013} may exist in a highly magnetised medium and that fibre bundles may be transient.  

\item Filament and core masses are highly dependent on the physical properties of the medium. In general, the CR-only simulations yield filaments with smaller masses than the SL counterparts. This is particularly evident below Mach 2. The overall mass found within the filaments decreases as the degree of turbulence and/or mass-to-flux ratio increases. This is the same for the cores embedded within the filaments. 

\item Filament fitting showed trends as a function of turbulent velocity and mass-to-flux ratio. Specifically, the Gaussian FWHM and Plummer $R_{\rm flat}$ parameters show a decreasing trend with increasing Mach number. Conversely, the central density of the filament as determined through the Plummer fits showed an increasing trend with increasing Mach number. In addition, a larger mass-to-flux ratio seems to have the same effect as an increase in turbulence for the three parameters that showed a trend. Finally, there was no discernible trend between the power law index from the Plummer fits and the degree of turbulence.

\item Our filaments tend to overlap best with the Cold Core observations \citep{Juvela2012} while exhibiting lower densities and larger widths when compared to \citet{Hennemann2012} and \citet{Arz2011} respectively, however simulations with larger degrees of turbulence and/or supercritical mass-to-flux ratios do overlap with the parameter space occupied by the parameter space defined in \citet{Arz2011}. 

\end{itemize}

Although the simulations tend to show little difference between the resulting structures formed in highly turbulent media based on the density enhancement maps, there are differences between the two ionisation profiles when other physical parameters are analysed (i.e., mass-to-flux ratio structures). As such, it is not unrealistic to expect differences to appear in other physical parameters such as the velocity structure. Analysis of such differences/similarities between the evolution of the velocity and resulting synthetic spectra for varying initial turbulent Mach numbers, initial mass-to-flux ratio values and ionisation profiles will be explored in the second paper of this series.

\begin{acknowledgements} 
SB was supported by a Natural Science and Engineering Research Council (NSERC) Discovery Grant. The research leading to these results has received funding from the European Research Council under the European Union's Seventh Framework Programme (FP/2007-2013)/ ERC Grant Agreement n. 320620-PALs
\end{acknowledgements}

\appendix
\section{Resolution Effects}

As indicated in Table~\ref{comp} and studies such as those presented in \citet{Ntormousi2016}, the resolution of the simulations can have an effect on the sizes and scale of structures produced. As such, in this Appendix, we explore the effect of resolution on our fiducial pair of simulations, namely Models V and VI. We have chosen these models as they represent the threshold degree of turbulence and mass-to-flux ratio required to produce filaments. We look at both Models V and VI to discern if either of the underlying ionisation profiles (SL or CR-only, respectively) react differently at higher resolution. For the high resolution simulations, we have doubled the number of pixels in each dimension (1024x1024) while maintaining the same physical size of the box. This results in a pixel size that is half that of the low resolution models (i.e., 0.0148 pc). 

\begin{figure*}
\centering
\includegraphics[width = 0.43\textwidth]{densitymap_t3.80000_log.eps} 
\includegraphics[width = 0.43\textwidth]{densitymap_t3.70000_log.eps}\\ 
\includegraphics[width = 0.43\textwidth]{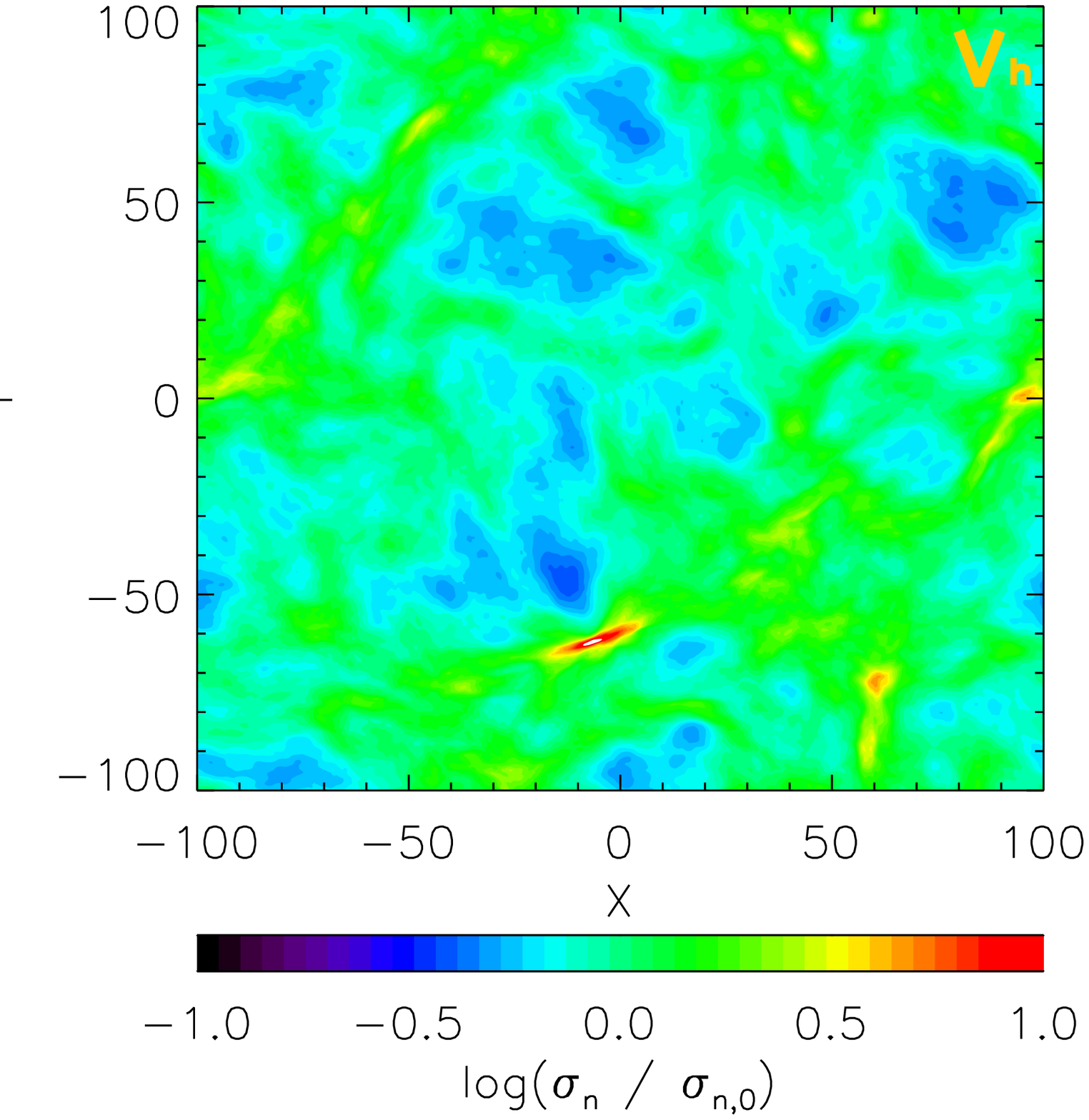} 
\includegraphics[width = 0.43\textwidth]{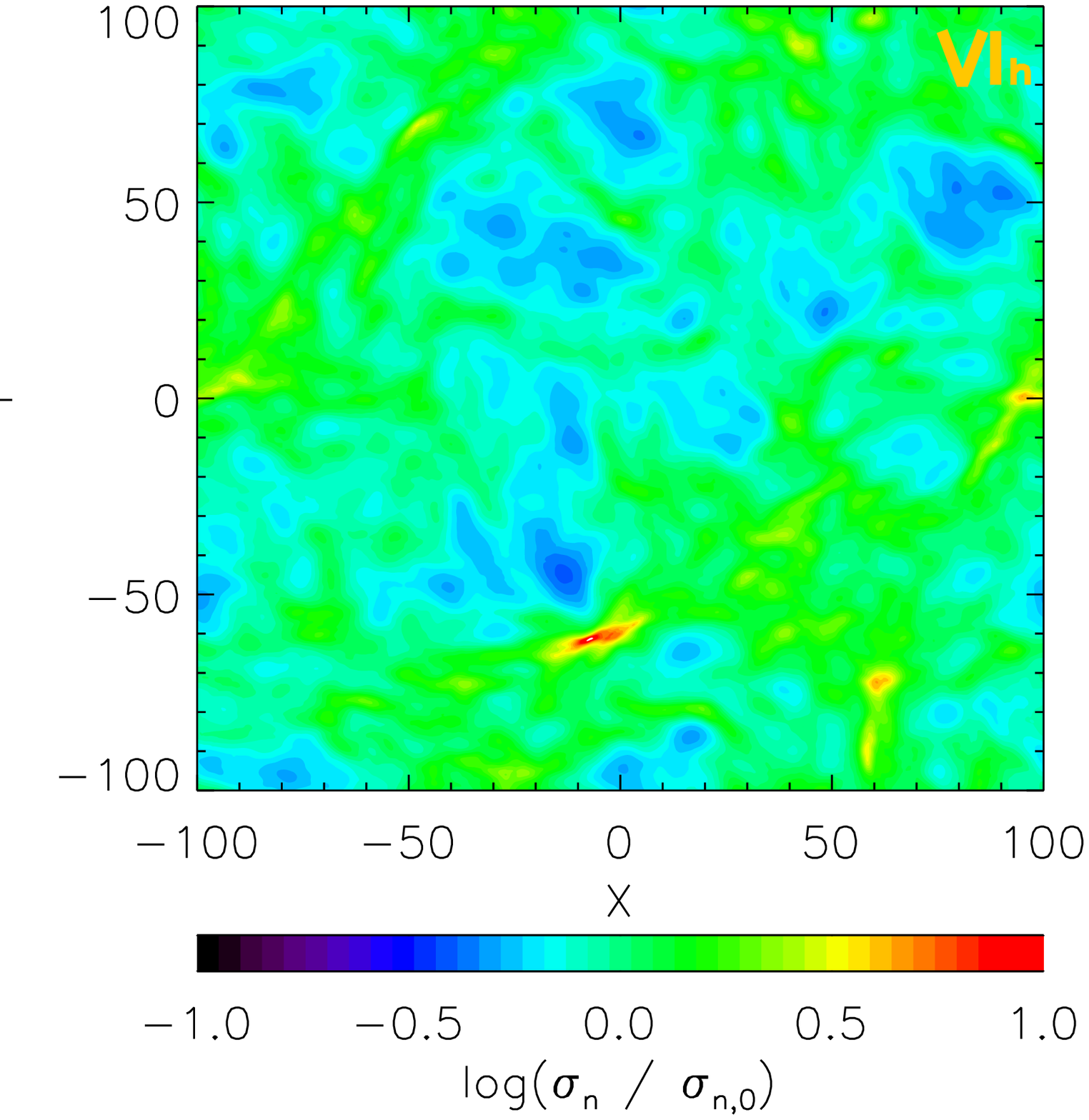}\\ 
\caption{Column density enhancement maps for high and low resolution simulations of both Step-like (SL, Model V; left column) and CR-only (CR, Model VI; right column) models. High resolution models are 1024 x 1024 pixels while low resolution models are 512 x 512 pixels. Colour bar shows the logarithm of the density enhancement. Roman numerals in the top right of each panel indicate the model number while the subscript h indicates high resolution models. }
\label{hi-lo-resmaps}
\end{figure*}

The first difference evident between the high and low resolution simulations is the run time. The original simulations for Models V and VI ran for 3.8 and 3.7 dimensionless time units, respectively, while the high resolution simulations ran until dimensionless times of 2.4 and 2.2, respectively. Figure~\ref{hi-lo-resmaps} shows a comparison of the column density enhancement maps for the low (top row) and high (bottom row) resolution simulations of Models V (left column) and VI (right column). At first look, the high and low resolution maps look different. This, however, is not surprising as there is no such thing as an identical random number field when looking at models with different resolutions, even if we use the same seed i.e., increasing the resolution increases the number of pixels and changes the arrangement of the applied random number field.

Focussing on the high resolution simulations, we see that, just as in the low resolution versions, both the SL and CR-only ionisation profiles produce similar structures. In terms of filament formation, we see that the high resolution simulations only produce one filament which conforms to our definition, namely structures with $A_{V} > 2$ however we can see that there are other regions that are in the process of forming filaments at the run termination time. High resolution runs of these models with increased stopping conditions (i.e., $\sigma/\sigma_{n,0} = 20$) show the continued formation of these filaments, however for comparison purposes, we will focus on the simulations with the same stopping conditions (i.e., $\sigma/\sigma_{n,0} = 10$). As such there is only one high density region within the high resolution simulations which we can use to compare to the low resolution runs. A further set of high/low resolution runs for Model V using different seeds for the turbulent velocity distribution (not shown) yielded simulations which ran to dimensionless times of 5.8 and 6.4, respectively and resulted in the formation of multiple filaments within the high resolution simulation. 

\begin{figure}
\centering
\includegraphics[width = 0.37\textwidth, angle = -90]{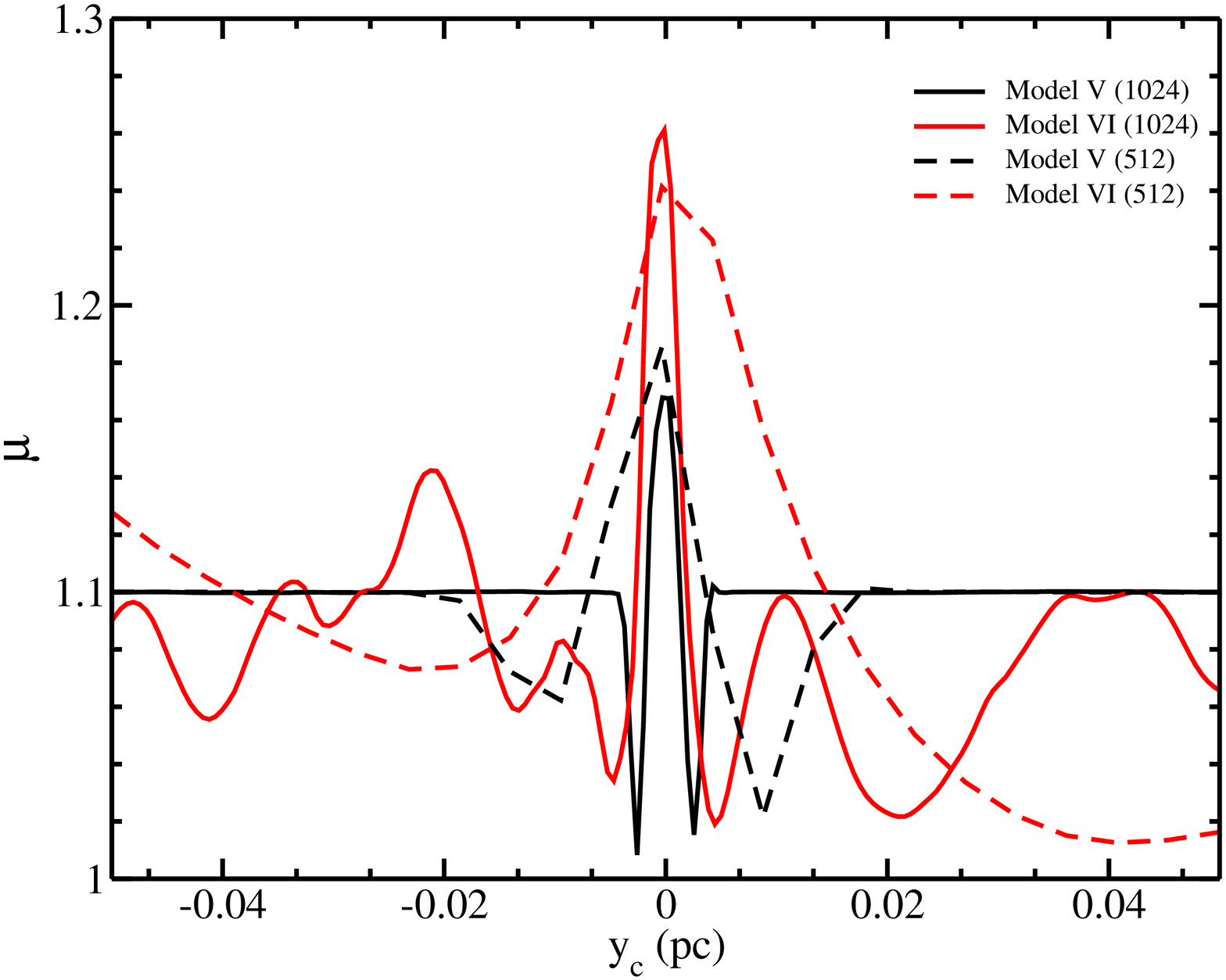} 
\caption{Mass-to-flux ratio profiles for the densest cores in the high (solid lines) and low (dashed lines) resolution cases. The black curves depict the SL simulations while the red curves depict the CR simulations. All profiles have been shifted such that the peaks coincide at y~=~0. The x axis depicts the distance from this central peak in parsecs.  }
\label{muprofilesall}
\end{figure}

Figure~\ref{muprofilesall} shows the mass-to-flux ratio profiles for the densest cores in the high (solid lines) and low (dashed lines) resolution simulations.  The black curves depict the SL models (V and V$_{h}$) while the red curves depict the CR models (VI and VI$_{h}$). Here we are comparing the only high density region within the high resolution simulation to the highest density regions within the low resolution simulations, i.e., filaments V-4 and VI-4 for Models V and VI, respectively. The curves have all been shifted such that the peaks of the profiles coincide at $y=0$. The $x$-axis depicts the distance away from the central peak in parsecs. Here, we see that the high resolution simulations again exhibit the same profile described in Section~\ref{mtfstructure}, however, the increased resolution allows for an increased degree of detail. Specifically, the peaked profile becomes more distinct in the high resolution simulations with the CR-only model again showing a wider feature with more variability in the wings compared to the SL model. 

Finally, to further compare the two resolutions, we also extracted the average profile of the high resolution filaments. As stated earlier, when assuming the same stopping conditions as the low resolution simulations, the high resolution runs only produced one filament which adheres to such criteria. Table~\ref{hi-low-fwhm-comp} shows shows a comparison between the FWHM of the single high resolution filaments and the low resolution filaments originally presented in Table~\ref{Filamentwidths} for both ionisation profiles. For both models, we see that the FWHM of the high resolution simulation falls within the range of FWHMs set by the low resolution filaments. For Model VI, we also see that the high resolution filament yields a very similar value to half of the low resolution filaments. When comparing the high resolution filament directly to the low resolution filament which contains the highest density core, we see that the high resolution simulations results in larger FWHM by 21\% and 32\% for Models V and VI, respectively. Conversely, when compared to the highest FWHM within the low resolution set, the high resolution simulations differ by 59\% and 44\% for Models V and VI, respectively. Therefore, we conclude that resolution does have an effect on the measured widths of the filaments, however it does not result in smaller filaments and on a whole the FWHMs of the high resolution filaments agree with the range of FWHMs found in the low resolution simulations.

\begin{table}
\small
\caption{Resolution comparison of filament FWHM}
\centering
\begin{tabular}{cc}
\hline\hline
Filament Designation\tablefootmark{a} & FWHM \\
\hline
V-1a & 0.35 \\
V-2  & 0.44 \\ 
V-3  & 0.68 \\ 
V-4  & 0.22 \\
V$_{h}$ & 0.28 \\
\hline
VI-1a & 0.25 \\ 
VI-1b & 0.30 \\
VI-2a & 0.34 \\ 
VI-2b & 0.32 \\
VI-3  & 0.55 \\ 
VI-4  & 0.21 \\
VI$_{h}$ & 0.31\\
\hline\hline
\end{tabular}
\tablefoot{
\tablefoottext{a}{The subscript $h$ denotes a filament in a high resolution model while the other designations denote the filaments as defined in Table~\ref{Filamentmasses}}
}

\label{hi-low-fwhm-comp}
\normalsize
\end{table}

\bibliography{thesis}{}

\begin{thebibliography}{67}
\expandafter\ifx\csname natexlab\endcsname\relax\def\natexlab#1{#1}\fi

\bibitem[{{Alves} {et~al.}(2008){Alves}, {Franco}, \& {Girart}}]{Alves2008}
{Alves}, F.~O., {Franco}, G.~A.~P., \& {Girart}, J.~M. 2008, A\&A, 486, L13

\bibitem[{{Andr{\'e}} {et~al.}(2014){Andr{\'e}}, {Di Francesco},
  {Ward-Thompson}, {Inutsuka}, {Pudritz}, \& {Pineda}}]{Andre2014}
{Andr{\'e}}, P., {Di Francesco}, J., {Ward-Thompson}, D., {et~al.} 2014,
  Protostars and Planets VI, 27

\bibitem[{{Andr{\'e}} {et~al.}(2010){Andr{\'e}}, {Men'shchikov}, {Bontemps},
  {K{\"o}nyves}, {Motte}, {Schneider}, {Didelon}, {Minier}, {Saraceno},
  {Ward-Thompson}, {di Francesco}, {White}, {Molinari}, {Testi}, {Abergel},
  {Griffin}, {Henning}, {Royer}, {Mer{\'{\i}}n}, {Vavrek}, {Attard},
  {Arzoumanian}, {Wilson}, {Ade}, {Aussel}, {Baluteau}, {Benedettini},
  {Bernard}, {Blommaert}, {Cambr{\'e}sy}, {Cox}, {di Giorgio}, {Hargrave},
  {Hennemann}, {Huang}, {Kirk}, {Krause}, {Launhardt}, {Leeks}, {Le Pennec},
  {Li}, {Martin}, {Maury}, {Olofsson}, {Omont}, {Peretto}, {Pezzuto}, {Prusti},
  {Roussel}, {Russeil}, {Sauvage}, {Sibthorpe}, {Sicilia-Aguilar}, {Spinoglio},
  {Waelkens}, {Woodcraft}, \& {Zavagno}}]{Andre2010}
{Andr{\'e}}, P., {Men'shchikov}, A., {Bontemps}, S., {et~al.} 2010, A\&A, 518,
  L102

\bibitem[{{Arzoumanian} {et~al.}(2011){Arzoumanian}, {Andr{\'e}}, {Didelon},
  {K{\"o}nyves}, {Schneider}, {Men'shchikov}, {Sousbie}, {Zavagno}, {Bontemps},
  {di Francesco}, {Griffin}, {Hennemann}, {Hill}, {Kirk}, {Martin}, {Minier},
  {Molinari}, {Motte}, {Peretto}, {Pezzuto}, {Spinoglio}, {Ward-Thompson},
  {White}, \& {Wilson}}]{Arz2011}
{Arzoumanian}, D., {Andr{\'e}}, P., {Didelon}, P., {et~al.} 2011, \aap, 529, L6

\bibitem[{{Arzoumanian} {et~al.}(2013){Arzoumanian}, {Andr{\'e}}, {Peretto}, \&
  {K{\"o}nyves}}]{Arz2013}
{Arzoumanian}, D., {Andr{\'e}}, P., {Peretto}, N., \& {K{\"o}nyves}, V. 2013,
  \aap, 553, A119

\bibitem[{{Bailey} \& {Basu}(2012)}]{BB2012}
{Bailey}, N.~D. \& {Basu}, S. 2012, ApJ, 761, 67

\bibitem[{{Bailey} \& {Basu}(2013)}]{BB2013}
{Bailey}, N.~D. \& {Basu}, S. 2013, \apj, 766, 27

\bibitem[{{Bailey} \& {Basu}(2014)}]{BB2014}
{Bailey}, N.~D. \& {Basu}, S. 2014, ApJ, 780, 40

\bibitem[{{Bailey} {et~al.}(2015){Bailey}, {Basu}, \& {Caselli}}]{BBC2015}
{Bailey}, N.~D., {Basu}, S., \& {Caselli}, P. 2015, ApJ, 798, 75

\bibitem[{{Banerjee} {et~al.}(2006){Banerjee}, {Pudritz}, \&
  {Anderson}}]{Banerjee2006}
{Banerjee}, R., {Pudritz}, R.~E., \& {Anderson}, D.~W. 2006, \mnras, 373, 1091

\bibitem[{{Basu} \& {Ciolek}(2004)}]{BC2004}
{Basu}, S. \& {Ciolek}, G.~E. 2004, ApJL, 607, L39

\bibitem[{{Basu} {et~al.}(2009{\natexlab{a}}){Basu}, {Ciolek}, {Dapp}, \&
  {Wurster}}]{BCDW2009}
{Basu}, S., {Ciolek}, G.~E., {Dapp}, W.~B., \& {Wurster}, J.
  2009{\natexlab{a}}, NewA, 14, 483

\bibitem[{{Basu} {et~al.}(2009{\natexlab{b}}){Basu}, {Ciolek}, \&
  {Wurster}}]{BCW2009}
{Basu}, S., {Ciolek}, G.~E., \& {Wurster}, J. 2009{\natexlab{b}}, NewA, 14, 221

\bibitem[{{Beuther} {et~al.}(2015{\natexlab{a}}){Beuther}, {Henning}, {Linz},
  {Feng}, {Ragan}, {Smith}, {Bihr}, {Sakai}, \& {Kuiper}}]{Beuther201509}
{Beuther}, H., {Henning}, T., {Linz}, H., {et~al.} 2015{\natexlab{a}}, \aap,
  581, A119

\bibitem[{{Beuther} {et~al.}(2015{\natexlab{b}}){Beuther}, {Ragan}, {Johnston},
  {Henning}, {Hacar}, \& {Kainulainen}}]{Beuther201512}
{Beuther}, H., {Ragan}, S.~E., {Johnston}, K., {et~al.} 2015{\natexlab{b}},
  \aap, 584, A67

\bibitem[{{Chen} \& {Ostriker}(2014)}]{CO2014}
{Chen}, C.-Y. \& {Ostriker}, E.~C. 2014, \apj, 785, 69

\bibitem[{{Ciolek} \& {Basu}(2006)}]{CB2006}
{Ciolek}, G.~E. \& {Basu}, S. 2006, ApJ, 652, 442

\bibitem[{{Dapp} \& {Basu}(2009)}]{DappBasu2009}
{Dapp}, W.~B. \& {Basu}, S. 2009, MNRAS, 395, 1092

\bibitem[{{Federrath}(2016)}]{Federrath2016}
{Federrath}, C. 2016, \mnras, 457, 375

\bibitem[{{Fiege} \& {Pudritz}(2000)}]{FP2000}
{Fiege}, J.~D. \& {Pudritz}, R.~E. 2000, \mnras, 311, 85

\bibitem[{{Franco} \& {Alves}(2015)}]{FA2015}
{Franco}, G.~A.~P. \& {Alves}, F.~O. 2015, \apj, 807, 5

\bibitem[{{Fryxell} {et~al.}(2000){Fryxell}, {Olson}, {Ricker}, {Timmes},
  {Zingale}, {Lamb}, {MacNeice}, {Rosner}, {Truran}, \& {Tufo}}]{Fryxell2000}
{Fryxell}, B., {Olson}, K., {Ricker}, P., {et~al.} 2000, \apjs, 131, 273

\bibitem[{{Glover} \& {Mac Low}(2007{\natexlab{a}})}]{GM2007a}
{Glover}, S.~C.~O. \& {Mac Low}, M.-M. 2007{\natexlab{a}}, \apjs, 169, 239

\bibitem[{{Glover} \& {Mac Low}(2007{\natexlab{b}})}]{GM2007b}
{Glover}, S.~C.~O. \& {Mac Low}, M.-M. 2007{\natexlab{b}}, \apj, 659, 1317

\bibitem[{{Hacar} {et~al.}(2013){Hacar}, {Tafalla}, {Kauffmann}, \&
  {Kov{\'a}cs}}]{Hacar2013}
{Hacar}, A., {Tafalla}, M., {Kauffmann}, J., \& {Kov{\'a}cs}, A. 2013, A\&A,
  554, A55

\bibitem[{{Heitsch}(2013)}]{Heitsch2013}
{Heitsch}, F. 2013, \apj, 769, 115

\bibitem[{{Hennebelle}(2013)}]{Hennebelle2013}
{Hennebelle}, P. 2013, \aap, 556, A153

\bibitem[{{Hennebelle} \& {Andr{\'e}}(2013)}]{HA2013}
{Hennebelle}, P. \& {Andr{\'e}}, P. 2013, \aap, 560, A68

\bibitem[{{Hennemann} {et~al.}(2012){Hennemann}, {Motte}, {Schneider},
  {Didelon}, {Hill}, {Arzoumanian}, {Bontemps}, {Csengeri}, {Andr{\'e}},
  {Konyves}, {Louvet}, {Marston}, {Men'shchikov}, {Minier}, {Nguyen Luong},
  {Palmeirim}, {Peretto}, {Sauvage}, {Zavagno}, {Anderson}, {Bernard}, {Di
  Francesco}, {Elia}, {Li}, {Martin}, {Molinari}, {Pezzuto}, {Russeil}, {Rygl},
  {Schisano}, {Spinoglio}, {Sousbie}, {Ward-Thompson}, \&
  {White}}]{Hennemann2012}
{Hennemann}, M., {Motte}, F., {Schneider}, N., {et~al.} 2012, A\&A, 543, L3

\bibitem[{{Henshaw} {et~al.}(2014){Henshaw}, {Caselli}, {Fontani},
  {Jim{\'e}nez-Serra}, \& {Tan}}]{Henshaw2014}
{Henshaw}, J.~D., {Caselli}, P., {Fontani}, F., {Jim{\'e}nez-Serra}, I., \&
  {Tan}, J.~C. 2014, MNRAS, 440, 2860

\bibitem[{{Hocuk} {et~al.}(2016){Hocuk}, {Cazaux}, {Spaans}, \&
  {Caselli}}]{Hocuk2015}
{Hocuk}, S., {Cazaux}, S., {Spaans}, M., \& {Caselli}, P. 2016, \mnras, 456,
  2586

\bibitem[{{Juvela} {et~al.}(2012){Juvela}, {Ristorcelli}, {Pagani}, {Doi},
  {Pelkonen}, {Marshall}, {Bernard}, {Falgarone}, {Malinen}, {Marton},
  {McGehee}, {Montier}, {Motte}, {Paladini}, {T{\'o}th}, {Ysard}, {Zahorecz},
  \& {Zavagno}}]{Juvela2012}
{Juvela}, M., {Ristorcelli}, I., {Pagani}, L., {et~al.} 2012, A\&A, 541, A12

\bibitem[{{Kirk} {et~al.}(2015){Kirk}, {Klassen}, {Pudritz}, \&
  {Pillsworth}}]{Kirk2015}
{Kirk}, H., {Klassen}, M., {Pudritz}, R., \& {Pillsworth}, S. 2015, \apj, 802,
  75

\bibitem[{{Kirk} {et~al.}(2013){Kirk}, {Myers}, {Bourke}, {Gutermuth},
  {Hedden}, \& {Wilson}}]{Kirk2013}
{Kirk}, H., {Myers}, P.~C., {Bourke}, T.~L., {et~al.} 2013, \apj, 766, 115

\bibitem[{{K{\"o}nyves} {et~al.}(2015){K{\"o}nyves}, {Andr{\'e}},
  {Men'shchikov}, {Palmeirim}, {Arzoumanian}, {Schneider}, {Roy}, {Didelon},
  {Maury}, {Shimajiri}, {Di Francesco}, {Bontemps}, {Peretto}, {Benedettini},
  {Bernard}, {Elia}, {Griffin}, {Hill}, {Kirk}, {Ladjelate}, {Marsh}, {Martin},
  {Motte}, {Nguy{\^e}n Luong}, {Pezzuto}, {Roussel}, {Rygl}, {Sadavoy},
  {Schisano}, {Spinoglio}, {Ward-Thompson}, \& {White}}]{Konyves2015}
{K{\"o}nyves}, V., {Andr{\'e}}, P., {Men'shchikov}, A., {et~al.} 2015, \aap,
  584, A91

\bibitem[{{Kudoh} \& {Basu}(2003)}]{kud03}
{Kudoh}, T. \& {Basu}, S. 2003, ApJ, 595, 842

\bibitem[{{Kudoh} \& {Basu}(2008)}]{KB2008}
{Kudoh}, T. \& {Basu}, S. 2008, \apjl, 679, L97

\bibitem[{{Kudoh} \& {Basu}(2011)}]{KB2011}
{Kudoh}, T. \& {Basu}, S. 2011, \apj, 728, 123

\bibitem[{{Li} {et~al.}(2015){Li}, {Yuen}, {Otto}, {Leung}, {Sridharan},
  {Zhang}, {Liu}, {Tang}, \& {Qiu}}]{Li2015}
{Li}, H.-B., {Yuen}, K.~H., {Otto}, F., {et~al.} 2015, \nat, 520, 518

\bibitem[{{Malinen} {et~al.}(2012){Malinen}, {Juvela}, {Rawlings},
  {Ward-Thompson}, {Palmeirim}, \& {Andr{\'e}}}]{Malinen2012}
{Malinen}, J., {Juvela}, M., {Rawlings}, M.~G., {et~al.} 2012, A\&A, 544, A50

\bibitem[{{McDaniel} \& {Mason}(1973)}]{MM1973}
{McDaniel}, E.~W. \& {Mason}, E.~A. 1973, The Mobility and Diffusion of Ions in
  Gases (New York: Wiley)

\bibitem[{{Men'shchikov}(2013)}]{Menshchikov2013}
{Men'shchikov}, A. 2013, \aap, 560, A63

\bibitem[{{Moeckel} \& {Burkert}(2015)}]{MB2015}
{Moeckel}, N. \& {Burkert}, A. 2015, \apj, 807, 67

\bibitem[{{Motte} {et~al.}(2010){Motte}, {Zavagno}, {Bontemps}, {Schneider},
  {Hennemann}, {di Francesco}, {Andr{\'e}}, {Saraceno}, {Griffin}, {Marston},
  {Ward-Thompson}, {White}, {Minier}, {Men'shchikov}, {Hill}, {Abergel},
  {Anderson}, {Aussel}, {Balog}, {Baluteau}, {Bernard}, {Cox}, {Csengeri},
  {Deharveng}, {Didelon}, {di Giorgio}, {Hargrave}, {Huang}, {Kirk}, {Leeks},
  {Li}, {Martin}, {Molinari}, {Nguyen-Luong}, {Olofsson}, {Persi}, {Peretto},
  {Pezzuto}, {Roussel}, {Russeil}, {Sadavoy}, {Sauvage}, {Sibthorpe},
  {Spinoglio}, {Testi}, {Teyssier}, {Vavrek}, {Wilson}, \&
  {Woodcraft}}]{Motte2010}
{Motte}, F., {Zavagno}, A., {Bontemps}, S., {et~al.} 2010, \aap, 518, L77

\bibitem[{{Mouschovias} \& {Ciolek}(1999)}]{MC1999}
{Mouschovias}, T.~Ch. \& {Ciolek}, G.~E. 1999, in NATO ASIC Proc. 540: The
  Origin of Stars and Planetary Systems, ed. C.~J. {Lada} \& N.~D. {Kylafis},
  305

\bibitem[{{Nagai} {et~al.}(1998){Nagai}, {Inutsuka}, \& {Miyama}}]{Nagai1998}
{Nagai}, T., {Inutsuka}, S.-i., \& {Miyama}, S.~M. 1998, \apj, 506, 306

\bibitem[{{Nakamura} \& {Li}(2005)}]{NL2005}
{Nakamura}, F. \& {Li}, Z.-Y. 2005, \apj, 631, 411

\bibitem[{{Nakamura} \& {Li}(2008)}]{NL2008}
{Nakamura}, F. \& {Li}, Z.-Y. 2008, \apj, 687, 354

\bibitem[{{Nelson} \& {Langer}(1997)}]{NL1997}
{Nelson}, R.~P. \& {Langer}, W.~D. 1997, \apj, 482, 796

\bibitem[{{Ntormousi} {et~al.}(2016){Ntormousi}, {Hennebelle}, {Andr{\'e}}, \&
  {Masson}}]{Ntormousi2016}
{Ntormousi}, E., {Hennebelle}, P., {Andr{\'e}}, P., \& {Masson}, J. 2016, \aap,
  589, A24

\bibitem[{{Padoan} {et~al.}(2001){Padoan}, {Juvela}, {Goodman}, \&
  {Nordlund}}]{Padoan2001}
{Padoan}, P., {Juvela}, M., {Goodman}, A.~A., \& {Nordlund}, {\AA}. 2001, \apj,
  553, 227

\bibitem[{{Palmeirim} {et~al.}(2013){Palmeirim}, {Andr{\'e}}, {Kirk},
  {Ward-Thompson}, {Arzoumanian}, {K{\"o}nyves}, {Didelon}, {Schneider},
  {Benedettini}, {Bontemps}, {Di Francesco}, {Elia}, {Griffin}, {Hennemann},
  {Hill}, {Martin}, {Men'shchikov}, {Molinari}, {Motte}, {Nguyen Luong},
  {Nutter}, {Peretto}, {Pezzuto}, {Roy}, {Rygl}, {Spinoglio}, \&
  {White}}]{Pal2013}
{Palmeirim}, P., {Andr{\'e}}, P., {Kirk}, J., {et~al.} 2013, A\&A, 550, A38

\bibitem[{{Panopoulou} {et~al.}(2014){Panopoulou}, {Tassis}, {Goldsmith}, \&
  {Heyer}}]{Panopoulou2014}
{Panopoulou}, G.~V., {Tassis}, K., {Goldsmith}, P.~F., \& {Heyer}, M.~H. 2014,
  \mnras, 444, 2507

\bibitem[{{Peretto} {et~al.}(2014){Peretto}, {Fuller}, {Andr{\'e}},
  {Arzoumanian}, {Rivilla}, {Bardeau}, {Duarte Puertas}, {Guzman Fernandez},
  {Lenfestey}, {Li}, {Olguin}, {R{\"o}ck}, {de Villiers}, \&
  {Williams}}]{Peretto2014}
{Peretto}, N., {Fuller}, G.~A., {Andr{\'e}}, P., {et~al.} 2014, \aap, 561, A83

\bibitem[{{Pineda} {et~al.}(2011){Pineda}, {Goodman}, {Arce}, {Caselli},
  {Longmore}, \& {Corder}}]{Pineda2011}
{Pineda}, J.~E., {Goodman}, A.~A., {Arce}, H.~G., {et~al.} 2011, ApJL, 739, L2

\bibitem[{{Pineda} {et~al.}(2010){Pineda}, {Goldsmith}, {Chapman}, {Snell},
  {Li}, {Cambr{\'e}sy}, \& {Brunt}}]{Pineda2010}
{Pineda}, J.~L., {Goldsmith}, P.~F., {Chapman}, N., {et~al.} 2010, ApJ, 721,
  686

\bibitem[{{Ruffle} {et~al.}(1998){Ruffle}, {Hartquist}, {Rawlings}, \&
  {Williams}}]{Ruffle1998}
{Ruffle}, D.~P., {Hartquist}, T.~W., {Rawlings}, J.~M.~C., \& {Williams}, D.~A.
  1998, A\&A, 334, 678

\bibitem[{{Sadavoy} {et~al.}(2012){Sadavoy}, {di Francesco}, {Andr{\'e}},
  {Pezzuto}, {Bernard}, {Bontemps}, {Bressert}, {Chitsazzadeh}, {Fallscheer},
  {Hennemann}, {Hill}, {Martin}, {Motte}, {Nguyn Lu'O'Ng}, {Peretto}, {Reid},
  {Schneider}, {Testi}, {White}, \& {Wilson}}]{Sadavoy2012}
{Sadavoy}, S.~I., {di Francesco}, J., {Andr{\'e}}, P., {et~al.} 2012, A\&A,
  540, A10

\bibitem[{{Schisano} {et~al.}(2014){Schisano}, {Rygl}, {Molinari}, {Busquet},
  {Elia}, {Pestalozzi}, {Polychroni}, {Billot}, {Carey}, {Paladini},
  {Noriega-Crespo}, {Moore}, {Plume}, {Glover}, \&
  {V{\'a}zquez-Semadeni}}]{Schisano2014}
{Schisano}, E., {Rygl}, K.~L.~J., {Molinari}, S., {et~al.} 2014, \apj, 791, 27

\bibitem[{{Seifried} \& {Walch}(2015)}]{SW2015}
{Seifried}, D. \& {Walch}, S. 2015, \mnras, 452, 2410

\bibitem[{{Smith} {et~al.}(2014){Smith}, {Glover}, \& {Klessen}}]{Smith2014}
{Smith}, R.~J., {Glover}, S.~C.~O., \& {Klessen}, R.~S. 2014, MNRAS, 445, 2900

\bibitem[{{Smith} {et~al.}(2016){Smith}, {Glover}, {Klessen}, \&
  {Fuller}}]{Smith2016}
{Smith}, R.~J., {Glover}, S.~C.~O., {Klessen}, R.~S., \& {Fuller}, G.~A. 2016,
  \mnras, 455, 3640

\bibitem[{{Sousbie}(2013)}]{Sousbie2013}
{Sousbie}, T. 2013, ArXiv e-prints [\eprint[arXiv]{1302.6221}]

\bibitem[{{Springel}(2010)}]{Springel2010}
{Springel}, V. 2010, \mnras, 401, 791

\bibitem[{{Tafalla} \& {Hacar}(2015)}]{TH2015}
{Tafalla}, M. \& {Hacar}, A. 2015, \aap, 574, A104

\bibitem[{{Tomisaka}(2014)}]{Tomisaka2014}
{Tomisaka}, K. 2014, \apj, 785, 24

\bibitem[{{Williams} {et~al.}(1994){Williams}, {de Geus}, \&
  {Blitz}}]{Williams1994}
{Williams}, J.~P., {de Geus}, E.~J., \& {Blitz}, L. 1994, ApJ, 428, 693

\end{thebibliography}
\bibliographystyle{aa}

\end{document}